%% file: SN2017ein-20181130.tex
\documentclass[twocolumn]{aastex62}
\pdfoutput=1 
\usepackage{amsmath,amstext}
\usepackage[T1]{fontenc}
\usepackage[figure,figure*]{hypcap}
\usepackage{graphicx}
\usepackage{epstopdf}
\usepackage{gensymb}
\usepackage{subfigure}

\graphicspath{{./}}

\shorttitle{A Type Ic Supernova \mbox{SN~2017ein}}
\shortauthors{Xiang et al.}

\begin{document}

\title{Observations of SN~2017ein Reveal Shock Breakout Emission and A Massive Progenitor Star for a Type Ic Supernova}

\author{Danfeng Xiang}
\author{Xiaofeng Wang}
\author{Jun Mo}
\affiliation{Physics Department and Tsinghua Center for Astrophysics (THCA), Tsinghua University, Beijing, 100084, China}
\author{Lingjun Wang}
\affiliation{Astroparticle Physics, Institute of High Energy Physics, Chinese Academy of Sciences, Beijing 100049, China}
\author{Stephen Smartt}
\affiliation{Astrophysics Research Centre, School of Mathematics and Physics, Queens University Belfast, Belfast BT7 1NN, UK}
\author{Morgan Fraser}
\affiliation{School of Physics, University College Dublin, Belfield, Dublin 4, Ireland}
\author{Shuhrat A. Ehgamberdiev}
\affiliation{Ulugh Beg Astronomical Institute, Uzbekistan Academy of Sciences, Uzbekistan, Tashkent, 100052, Uzbekistan}
\author{Davron Mirzaqulov}
\affiliation{Ulugh Beg Astronomical Institute, Uzbekistan Academy of Sciences, Uzbekistan, Tashkent, 100052, Uzbekistan}
\author{Jujia Zhang}
\affiliation{Yunnan Observatories, Chinese Academy of Sciences, Kunming 650216, China}
\affiliation{Key Laboratory for the Structure and Evolution of Celestial Objects, Chinese Academy of Sciences, Kunming 650216, China}
\affiliation{Center for Astronomical Mega-Science, Chinese Academy of Sciences, 20A Datun Road, Chaoyang District, Beijing, 100012, China}
\author{Tianmeng Zhang}
\affiliation{National Astronomical Observatory of China, Chinese Academy of Sciences, Beijing, 100012, China}
\author{Jozsef Vinko}
\affiliation{Konkoly Observatory, MTA CSFK, Konkoly Thege M. ut 15-17, Budapest, 1121, Hungary}
\affiliation{Department of Optics \& Quantum Electronics, University of Szeged, Dom ter 9, Szeged, 6720 Hungary}
\affiliation{Department of Astronomy, University of Texas at Austin, Austin, TX, 78712, USA}
\author{J. Craig Wheeler}
\affiliation{Department of Astronomy, University of Texas at Austin, Austin, TX, 78712, USA}
\author{Griffin Hosseinzadeh}
\affiliation{Las Cumbres Observatory, 6740 Cortona Drive, Suite 102, Goleta, CA 93117-5575, USA}
\affiliation{Department of Physics, University of California, Santa Barbara, CA 93106-9530, USA}
\author{D. Andrew Howell}
\affiliation{Las Cumbres Observatory, 6740 Cortona Drive, Suite 102, Goleta, CA 93117-5575, USA}
\affiliation{Department of Physics, University of California, Santa Barbara, CA 93106-9530, USA}
\author{Curtis McCully}
\affiliation{Las Cumbres Observatory, 6740 Cortona Drive, Suite 102, Goleta, CA 93117-5575, USA}
\author{James M DerKacy}
\author{E. Baron}
\affiliation{Homer L. Dodge Department of Physics and Astronomy, University of Oklahoma, Norman, OK}
\author{Peter Brown}
\affiliation{George P. and CynthiaWoods Mitchell Institute for Fundamental Physics \& Astronomy, Texas A. \& M. University,
Department of Physics and Astronomy, 4242 TAMU, College Station, TX 77843, USA}
\author{Xianfei Zhang}
\author{Shaolan Bi}
\affiliation{Department of Astronomy, Beijing Normal University, Beijing 100875, China}
\author{Hao Song}
\author{Kaicheng Zhang}
\affiliation{Physics Department and Tsinghua Center for Astrophysics (THCA), Tsinghua University, Beijing, 100084, China}
\author{A.~Rest} 
\affiliation{Space Telescope Science Institute, 3700 San Martin Drive, Baltimore, MD 21218, USA}
\affiliation{Department of Physics and Astronomy, Johns Hopkins University, Baltimore, MD 21218, USA}
\author{Ken'ichi Nomoto}
\author{Alexey Tolstov}
\affiliation{Kavli Institute for the Physics and Mathematics of the Universe (WPI), The University of Tokyo Institutes for Advanced Study, The University of Tokyo, 5-1-5 Kashiwanoha, Kashiwa, Chiba 277-8583, Japan}
\author{Sergei Blinnikov}
\affiliation{Institute for Theoretical and Experimental Physics (ITEP), 117218, Moscow, Russia}
\affiliation{All-Russia Research Institute of Automatics (VNIIA), 127055, Moscow, Russia}
\affiliation{Kavli Institute for the Physics and Mathematics of the Universe (WPI), The University of Tokyo Institutes for Advanced Study, The University of Tokyo, 5-1-5 Kashiwanoha, Kashiwa, Chiba 277-8583, Japan}

\begin{abstract}
We present optical and ultraviolet observations of nearby type Ic supernova SN 2017ein as well as detailed analysis of its progenitor properties from both the early-time observations and the prediscovery Hubble Space Telescope (HST) images. The optical light curves started from within one day to $\sim$275 days after explosion, and optical spectra range from $\sim$2 days to $\sim$90 days after explosion. Compared to other normal SNe Ic like SN 2007gr and SN 2013ge, \mbox{SN 2017ein} seems to have more prominent C{\footnotesize II} absorption and higher expansion velocities in early phases, suggestive of relatively lower ejecta mass.
The earliest photometry obtained for \mbox{SN 2017ein} show indications of shock cooling. The best-fit obtained by including a shock cooling component gives an estimate of the envelope mass as $\sim$0.02 M$_{\odot}$ and stellar radius as 8$\pm$4 R$_{\odot}$. Examining the pre-explosion images taken with the HST WFPC2, we find that the SN position coincides with a luminous and blue point-like source, with an extinction-corrected absolute magnitude of M$_V$$\sim$$-$8.2 mag and M$_I$$\sim$$-$7.7 mag. 
Comparisons of the observations to the theoretical models indicate that the counterpart source was either a single WR star or a binary with whose members had high initial masses, or a young compact star cluster.
To further distinguish between different scenarios requires revisiting the site of the progenitor with HST after the SN fades away.
\end{abstract}


\section{Introduction}\label{sec:intro}

Observationally, type Ic supernovae (SNe Ic) \mbox{represent} a subclass of core-collapse SNe (CCSNe) that do not show either hydrogen or helium features in their spectra.
These events, together with SNe IIb and Ib, are all called stripped-envelope (SE) SNe \citep{1996ApJ...462..462C} due to the fact that all or part of the outer H/He envelopes have been stripped from their progenitors prior to explosion.
The absence of helium absorptions in the spectra further distinguishes SNe Ic from SNe Ib. However, recent studies suggest that residual helium seems to exist in the former subclass, e.g., SN 2007gr \citep{2014ApJ...790..120C}, \mbox{SN~2016coi} \mbox{\citep{2018MNRAS.478.4162P, 2017ApJ...837....1Y}}, and \mbox{recent} statistical analysis from \mbox{\cite{2018A&A...618A..37F}}, suggestive of a continuous distribution between SNe Ic and SNe Ib.
Modeling by \cite{2012MNRAS.424.2139D} \mbox{indicates} that nickel mixing and hidden He during the explosion can make differences to the appearance of He in SNe spectra, and progenitors with similar structures can produce both type Ib and Ic.
While there are also some studies arguing for little or no helium in SNe Ic \citep{2013ApJ...773L...7F, 2016ApJ...827...90L}.
Classification of SE-SNe is revisited in \cite{2017MNRAS.469.2672P} based on  the strength of He lines in the SNe spectra.

Among SNe Ic, a subgroup called broad-lined supernovae (SNe Ic-BL) gained more attention due to their broad spectral line profiles, relativistic ejecta velocities, and higher luminosities.
Some SNe Ic-BL are found to be associated with long gamma-ray bursts (\mbox{LGRBs}), called GRB-SNe, e.g., SN 1998bw \citep{1998Natur.395..670G} and SN 2002ap \citep{2002MNRAS.332L..73G}.
And those that are not associated with GRBs are thought to be due to an artifact of viewing angle, as GRB radiation is strongly beamed.
While late-time radio observations of some SNe Ic-BL, e.g. SN 2002ap \citep{2002ApJ...577L...5B}, SN 2010ah \citep{2011ApJ...741...76C}, exclude relativistic jets accompanying these supernovae.
Consequently, not \mbox{all} SNe Ic-BL that not associated with GRB are due to beaming effects. Nevertheless, progenitors of these two subclasses of SNe Ic-BL are found to differ in metallicity \citep{2011ApJ...731L...4M, 2008AJ....135.1136M}.
For instance, GRB-SNe  occur preferentially in metal-poor galaxies relative to SNe Ic-BL and other SNe Ic.
These facts demonstrate that \mbox{SNe Ic-BL} may have different origins and/or explosion mechanisms compared with normal SNe Ic.
Moreover, GRB-SNe may be used as standard candles to measure cosmic distances and cosmological parameters \citep{2014ApJ...796L...4L, 2014ApJ...794..121C}.

The progenitors of SNe Ib/c have been thought to be Wolf-Rayet (WR) stars with high initial masses (M$_{\mathrm{ZAMS}}\gtrsim25$ M$_{\odot}$) \citep{2007ARA&A..45..177C}.
\mbox{Before} core-collapse, these stars usually have experienced severe mass loss through strong stellar winds or due to interaction with companion stars \citep{2015ApJS..220...15P, 2006A&A...458..453V}.
As the evolution of massive stars is usually dominated by binary evolution \citep{2003ApJ...591..288H}, and also depends largely on metallicity and rotation etc. \citep{2013A&A...558A.103G, 2012A&A...542A..29G, 2003ApJ...591..288H}, this makes the direct identification of their progenitors complicated \citep{2015PASA...32...16S}.
However, there are increasing studies suggesting that a lower mass binary scenario is more favorable for most SNe Ib/c, considering the measured low ejecta masses \citep{2013MNRAS.436..774E, 2016MNRAS.457..328L}. And the H/He envelopes of the progenitor stars are stripped by binary interaction.
There are many detections of progenitor stars for SNe II.
For example, most SNe IIP are found to originate from red supergiants \citep{2009MNRAS.395.1409S}, while \mbox{SNe IIL} are typically from progenitors with somewhat warmer colors \citep[see][for a review]{2015PASA...32...16S}, and \mbox{SNe IIb} are from those with higher effective temperatures such as yellow supergiants which have had their H/He envelopes partially stripped through binary interaction \cite[e.g. \mbox{SN 1993J;}][]{2014ApJ...790...17F, 2004Natur.427..129M, 1993Natur.364..509P}.
Until recently, there has been \mbox{only} one report of the possible identification of progenitor star for \mbox{SNe Ib}, namely \mbox{iPTF 13bvn}, which was proposed to spatially coincide with a single WR-like star identified on the pre-explosion Hubble \mbox{Space} Telescope (HST) images \citep{2013A&A...558L...1G, 2013ApJ...775L...7C}.
But such an identification is still controversial \citep[e.g.][]{2014AJ....148...68B, 2016MNRAS.461L.117E, 2015MNRAS.446.2689E, 2014A&A...565A.114F}.
Direct detection of progenitor stars is still elusive for SNe Ic, which prevents us from further testing the theoretical evolution of massive stars \citep{2013MNRAS.436..774E}.

During the core collapse explosions of massive stars, a ``breakout'' may occur when the shock breaks through the stellar surface or circumstellar medium (CSM).
This process should be accompanied by an X-ray and/or \mbox{UV} flash on a timescale of a few minutes to 1 -- 2 hours due to high temperatures
and is then followed by UV/optical ``cooling'' emission from the expanding envelope\footnote{The word ``envelope'' is usually used to represent the H or He layers of a star during its evolution. While in shock breakout theory, ``envelope'' is used as an indication of a thin layer of mass in which the shock breaks out. In this paper, to distinguish between the two interpretations, we use the following rules: ``envelope'' appears together with ``H'' or ``He'' for the former case; while the others for the latter case. In particular, ``stellar envelope'' represents the envelope at the surface of the star.}, where both the duration and strength of the cooling depend on the size and density of the stellar envelope or the CSM \citep{2016arXiv160701293W}.
Thus, the shock breakout and the cooling observations can provide information on the progenitor stars or circumstellar environment.
Owing to a very short timescale, the shock breakout phenomenon has been possibly detected for very few events, including type Ib SN 2008D \citep{2009ApJ...702..226M,2008Sci...321.1185M}, the GRB-SN SN 2006aj \citep{2006Natur.442.1008C,2007ApJ...667..351W}, and type IIb SN 2016gkg \citep{2018Natur.554..497B}.
By comparison, the cooling phase can last for a few days and can be more easily captured, e.g. \mbox{\cite{2017ApJ...837L...2A}} for SN 2016gkg.
For normal SNe Ic, the shock breakout cooling might have been observed in \mbox{only} one object, i.e., LSQ14efd \citep{2017MNRAS.471.2463B}.

\begin{figure}[ht!]
\centering
\includegraphics[width=1.0\linewidth]{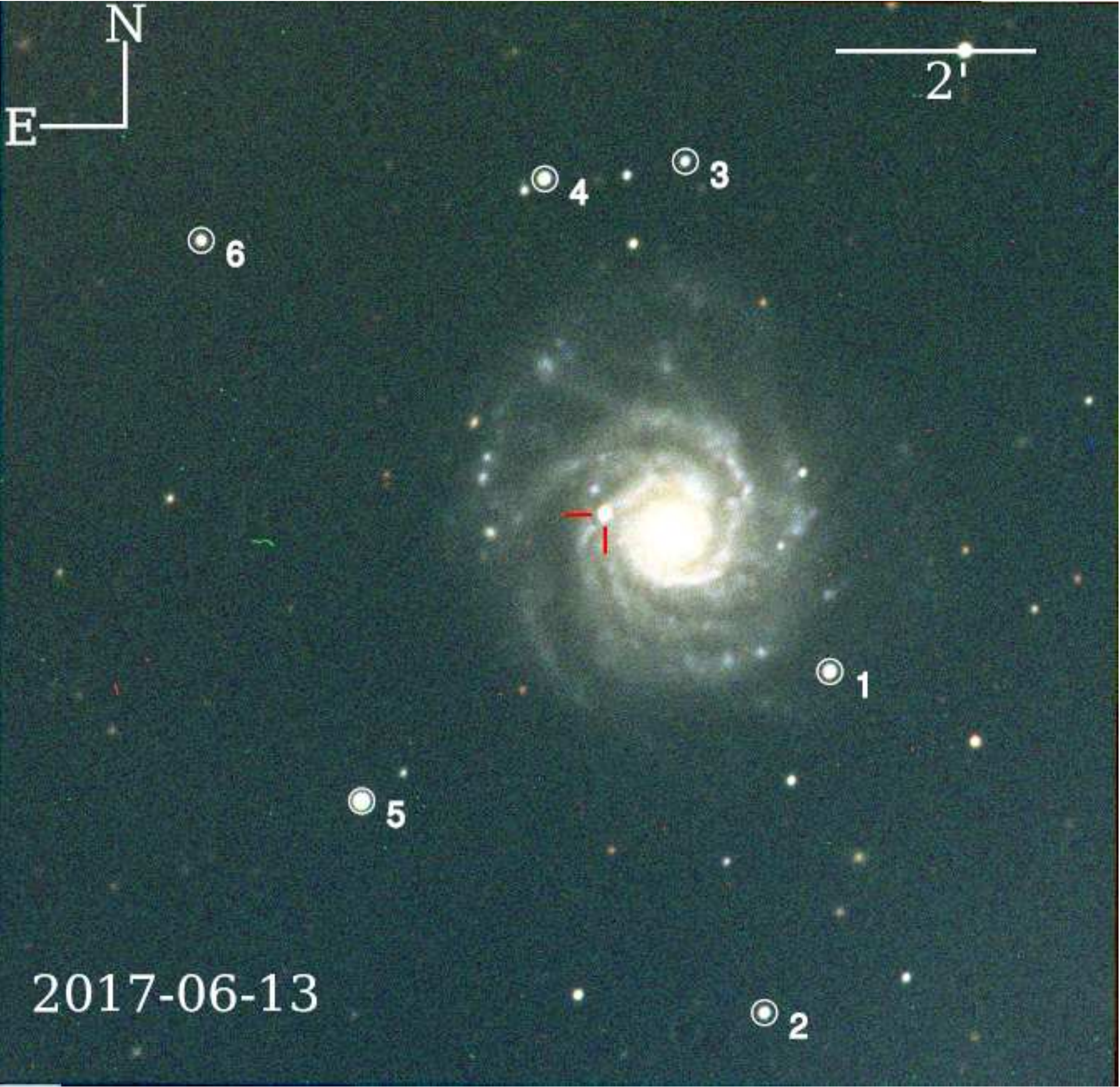}
\caption{A pseudo-color finder chart of \mbox{SN~2017ein} and its host galaxy NGC 3938, produced from the $BVRI$-band images obtained with the Tsinghua-NAOC 0.8-m telescope on 2017 June 13.
\mbox{SN~2017ein} is marked by red ticks.
The comparison stars used for flux calibration are marked by white circles.
North is up and west is right. \label{fig:SNimage}}
\end{figure}

\mbox{SN~2017ein} provides us a rare opportunity to constrain the progenitors of SNe Ib/c, as it has both the pre-discovery HST images and very early-time photometry obtained within a fraction of a day after explosion.
\mbox{SN~2017ein} was discovered in NGC 3938 by \mbox{Ron Arbour} on 2017 May 25.97 (UT dates are used throughout this paper), with a magnitude of 17.6 mag in a clear filter.
NGC 3938 is a very nearby type Sc, face-on spiral galaxy at a distance of $\sim$17 Mpc, which has produced SN 1961U (type II), SN 1964L (type Ic), and \mbox{SN 2005ay} (type II) before \mbox{SN~2017ein}.
Figure \ref{fig:SNimage} shows the finder chart of \mbox{SN~2017ein} which has coordinates of R.A.=11h52m53s.25, Decl.=+44\degree07$^{\prime}$26$^{\prime\prime}$.20 (J2000).
A~spectrum was obtained for \mbox{SN~2017ein} half a day after its discovery, and it was classified as a type Ic supernova near maximum light or one that suffered significant reddening \citep{2017ATel10434....1X}.
Follow-up photometry indicated that this supernova showed a rapid rise in the brightness, revealing that it was discovered at a very young phase \citep{2017ATel10481....1I}.
Moreover, the pre-discovery photometric data extracted from the Asteroid Terrestrial-impact Last Alert System \cite[ATLAS;][]{2018PASP..130f4505T} survey images indicate that the observations of \mbox{SN~2017ein} can be traced back to a phase within a fraction of one day after explosion, representing one of the earliest detection for a type Ic supernova.

In this paper we present extensive optical and ultraviolet (UV) observations of \mbox{SN~2017ein}, including light curves covering from $\sim$1 -- 275 days after explosion and spectra covering from $\sim$1.8 -- 90 days.
These datasets are used to constrain the explosion parameters and determine the properties of the  progenitor star for a normal SN Ic.
Moreover, we also analyze the prediscovery HST images to further constrain the progenitor star of \mbox{SN~2017ein}.
The paper is structured as follows.
In section \ref{sec:obs}, our photometric and spectroscopic observations of \mbox{SN~2017ein} are presented.
In section \ref{sec:lc}, the light curves are described and analyzed.
In section \ref{sec:dist_extinc} we discuss the distance, foreground extinction to the SN and the absolute magnitudes of \mbox{SN~2017ein}.
The spectral evolution of \mbox{SN~2017ein} is presented in section \ref{sec:spec}.
We model the light curves to place constraints on the progenitor in section \ref{sec:shockfitting}.
In section \ref{sec:HST}, we perform the analysis of the properties of stars at and near the SN position on the pre-discovery HST images.
Discussions of our results and conclusions are given in section \ref{sec:summary}.

\section{Observations and Data Reduction}\label{sec:obs}
As \mbox{SN~2017ein} was discovered at a young phase and it is located at a relatively close distance, we triggered followup observations immediately after its discovery.
This supernova was observed in both optical and ultraviolet (UV) bands with the \textit{Swift}-UVOT \citep{2005SSRv..120...95R, 2004ApJ...611.1005G}.

\subsection{Optical and ultraviolet photometric obervations}\label{sec:obs_phot}
Photometric observations of \mbox{SN~2017ein} were collected with several telescopes, including the 0.8-m \mbox{Tsinghua} University-NAOC telescope \citep[TNT,][]{2012RAA....12.1585H} at Xinglong Observatory of NAOC, the AZT-22 \mbox{1.5-m} telescope (hereafter AZT) at Maidanak Astronomical \mbox{Observatory} \citep{2018NatAs...2..349E}.
The TNT and AZT observations were obtained in standard Johnson-Cousin $BVRI$ \mbox{bands}.
These observations covered the phases from MJD 57904 (2017 May 31) to MJD 58170 (2018 Feb. 22).
At late times, the supernova is only visible on the $R$- and $I$-band images on MJD 58107 (2017 Dec. 20) and becomes undetectable after that.

\begin{figure*}
    \begin{minipage}{0.32\linewidth}
    \includegraphics[width=1.0\linewidth]{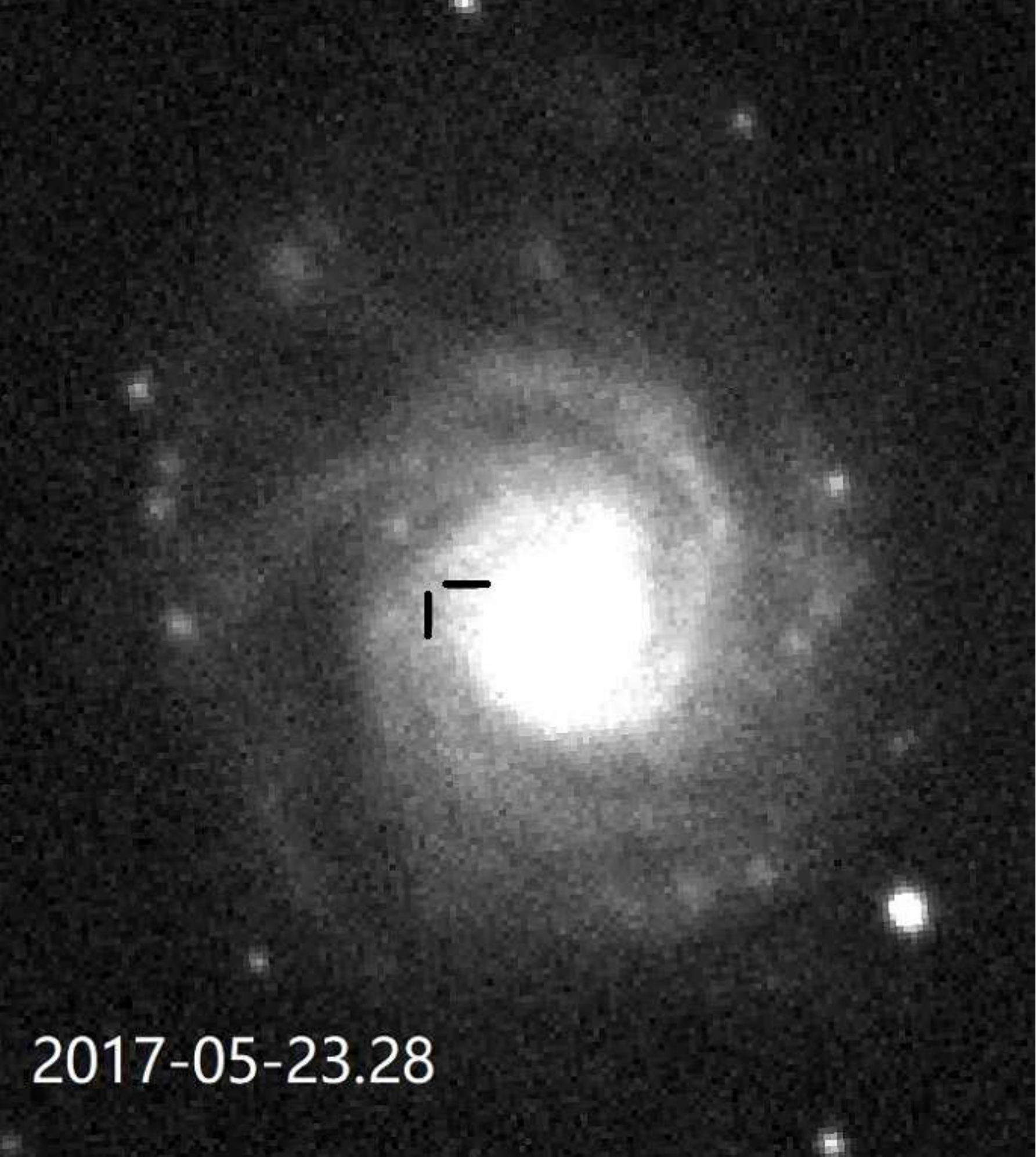}
    \end{minipage}
    \begin{minipage}{0.32\linewidth}
    \includegraphics[width=1.0\linewidth]{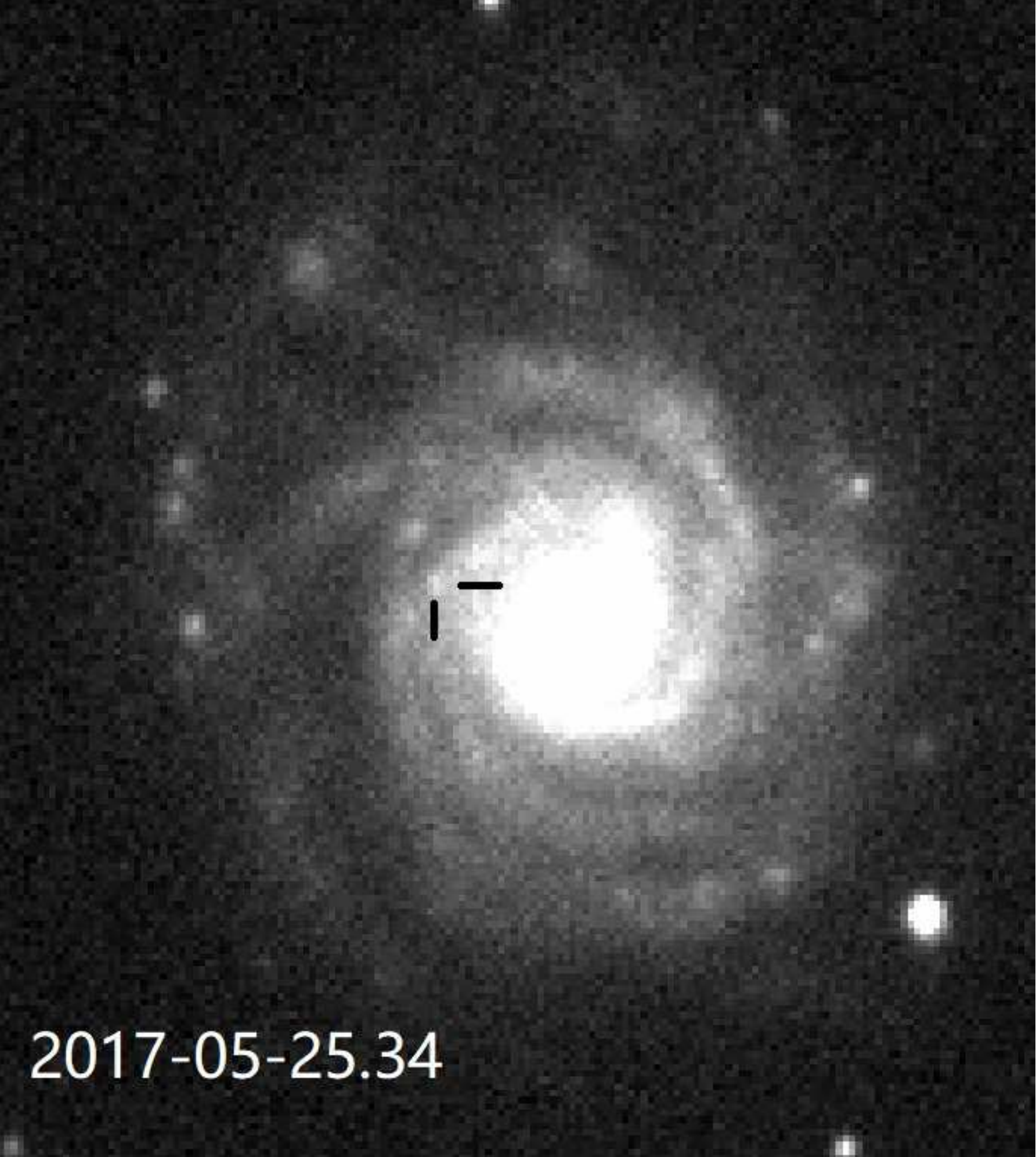}
    \end{minipage}
    \begin{minipage}{0.32\linewidth}
    \includegraphics[width=1.0\linewidth]{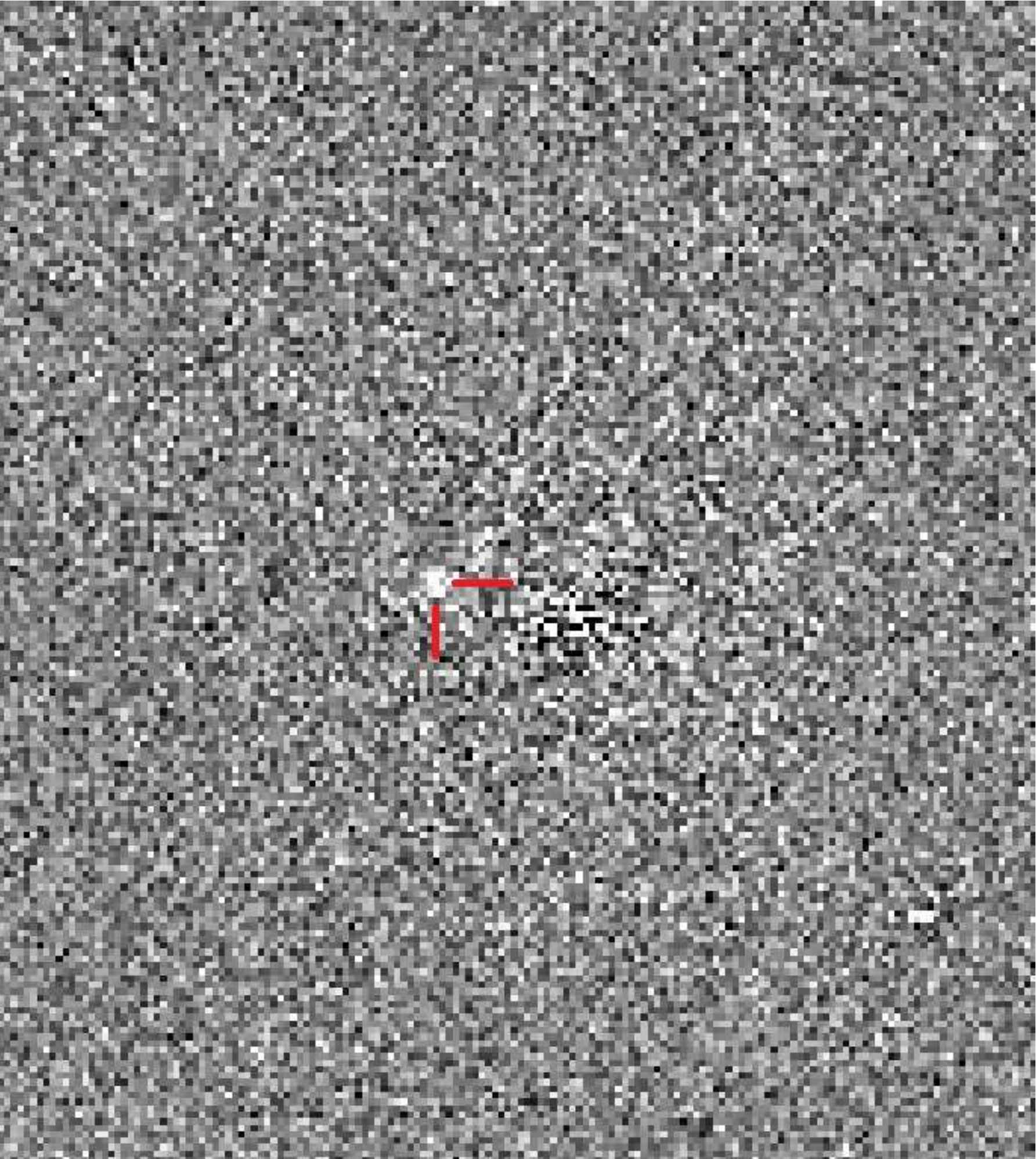}
    \end{minipage}
\caption{The ATLAS images centering at the galaxy NGC 3938, north is up and west is right. \textit{left}: The latest non-detection image taken on 2017 May 23.28 in \textit{o}-band. The AB magnitude at the SN position is fainter than 20 mag (after subtracting template); \textit{middle}: the first detection image of \mbox{SN~2017ein} on 2017 May 25.34 in \textit{c}-band; \textit{right}: residual image obtained by subtracting the template from the first-detection image, with the position of the SN marked. \label{fig:cyanimg}}
\end{figure*}

ATLAS is a twin 0.5-m telescope system that surveys in cyan (\textit{c}) and orange (\textit{o}) filters \citep{2018PASP..130f4505T}.
The typical sky coverage is once every 2 \mbox{nights}.
\mbox{ATLAS} monitored the field of \mbox{SN~2017ein} from 2017 Apr. 26, and it did not detect any flux excess relative to the background until 2017 May 25.33 when the SN became detectable at an AB magnitude of \mbox{18.43$\pm$0.10 mag} in the \textit{c}-band, which is about 0.64 day before the first discovery.
The SN was also detectable in the \textit{o}-band image taken on 2017 May 27.25.
Figure \ref{fig:cyanimg} shows the ATLAS $c$-band prediscovery and the earliest detection images of \mbox{SN~2017ein}.
The ATLAS observations of \mbox{SN~2017ein} lasted until 2017 July 03.
The ATLAS data are reduced and calibrated automatically as described in \citep{2018PASP..130f4505T}.
The photometry was done on the \mbox{difference} images by subtracting pre-explosion templates.
A model of the point-spread-function (PSF) was created and fitted to the excess flux sources in the images \citep[as described in][]{2018PASP..130f4505T}.
The photometric calibration was achieved with a custom-built catalogue based on the Pan-STARRS reference catalogues \mbox{\citep{2016arXiv161205242M}}, and the magnitudes are in the AB system.

All $BVRI$ CCD images were pre-processed using \mbox{standard} \textsc{IRAF}\footnote{IRAF is distributed by the National Optical Astronomy Observatories, which are operated by the Association of Universities for Research in Astronomy, Inc., under cooperative agreement with the National Science Foundation (NSF).} routines, which includes \mbox{corrections} for bias, flat field, and removal of cosmic rays.
As \mbox{SN~2017ein} exploded near a spiral arm of the host galaxy, the late-time photometry may be influenced by contamination from the galaxy.
We thus use the template images taken on 2018 Jun. 16 and 18 (which correspond to $\sim+$390 days from the discovery) to get better photometry for the AZT images taken at late phases when the supernova was relatively faint.
We did not apply template subtraction to the TNT images as the SN is bright relative to the galaxy background when these images were taken.
The instrumental magnitudes of both the SN and the reference stars were then measured using the standard point spread function (PSF).
These magnitudes were converted to those of the standard Johnson system using the Pan-STARRS reference catalogue\footnote{\url{http://pswww.ifa.hawaii.edu/pswww/}}.
The magnitudes of the comparison stars used for flux calibration, as marked in Figure \ref{fig:SNimage}, are listed in Table \ref{tab:comp_stars}.
Although the supernova is not visible in images taken after MJD 58107, we also measured the magnitudes at the supernova site.
These measurements provide upper limits of the late-time light curves.
The final flux-calibrated magnitudes (Vega) of \mbox{SN~2017ein} from TNT and AZT are shown in Table \ref{tab:lcBVRI}, and the photometric results (AB magnitudes) from ATLAS are reported in Table \ref{tab:atlas}.

\mbox{SN~2017ein} was also observed with the Ultraviolet/Optical Telescope \cite[UVOT;][]{2005SSRv..120...95R} onboard the Neil Gehrels \textit{Swift} Observatory \citep{2004ApJ...611.1005G}.
The observations were obtained in uvw1, uvm2, uvw2, $u$, $b$, and $v$ filters, from 2017 May 27 until 2017 July 21.
The Swift/UVOT data reduction is based on that of the Swift Optical Ultraviolet Supernova Archive \cite[SOUSA;][]{2014Ap&SS.354...89B}.
A 3\arcsec\ aperture is used to measure the source counts with an aperture correction based on an average PSF.
Magnitudes are computed using the zeropoints from \cite{2011AIPC.1358..373B} for the UV and \cite{2008MNRAS.383..627P} for the optical and the 2015 redetermination of the temporal sensitivity loss.
The photometry is performed on the images subtracted by the template that was obtained on 2018 Jun. 02.
The supernova was visible only in uvw1 band but not in uvw2 and uvm2 bands, likely due to the significant reddening that it suffered from the host galaxy (see discussion in Section \ref{sec:dist_extinc}).
The \textit{Swift}-UVOT photometry is listed in Table \ref{tab:swift}.
The UVOT $b$- and $v$-band photometry are comparable to those of the ground-based observations around peak in $B$ and $V$ bands, respectively, while the UVOT photometry suffered large uncertainties at later phases.

\begin{deluxetable*}{ccccccc}
\tablecaption{Magnitudes of Comparison Stars in Field of \mbox{SN~2017ein}. \label{tab:comp_stars}}
\tablehead{
\colhead{Star No.} &\colhead{R.A.} &\colhead{Decl.}&\colhead{\textit{B}} &\colhead{\textit{V}} &\colhead{\textit{R}} &\colhead{\textit{I}}}
\startdata
1   &11h52m40s.50   &44\degree05$^{\prime}$49$^{\prime\prime}$.41   &15.848$\pm$0.037   &15.067$\pm$0.013   &14.611$\pm$0.015   &14.194$\pm$0.016\\
2   &11h52m44s.22   &44\degree02$^{\prime}$20$^{\prime\prime}$.76   &16.490$\pm$0.035   &15.774$\pm$0.012   &15.354$\pm$0.015   &14.927$\pm$0.016\\
3   &11h52m48s.69   &44\degree11$^{\prime}$01$^{\prime\prime}$.51   &16.879$\pm$0.037   &16.442$\pm$0.013   &16.173$\pm$0.015   &15.865$\pm$0.016\\
4   &11h52m56s.69   &44\degree10$^{\prime}$50$^{\prime\prime}$.83   &15.874$\pm$0.036   &15.166$\pm$0.013   &14.750$\pm$0.015   &14.337$\pm$0.016\\
5   &11h53m07s.08   &44\degree04$^{\prime}$30$^{\prime\prime}$.23   &14.386$\pm$0.036   &13.843$\pm$0.013   &13.516$\pm$0.015   &13.149$\pm$0.016\\
6   &11h53m16s.23   &44\degree10$^{\prime}$13$^{\prime\prime}$.08   &16.880$\pm$0.039   &16.009$\pm$0.016   &15.504$\pm$0.017   &15.052$\pm$0.017\\
\enddata
\end{deluxetable*}

\subsection{Optical spectroscopy}\label{sec:obs_spec}
Our first spectrum of \mbox{SN~2017ein} was obtained on UT 2017 May 26.57 with the Xinglong 2.16-m telescope of NAOC (hereafter XLT), which is at $\sim$2.5 days after explosion, assuming that the explosion date is 2017 May 24.0 according to the analysis conducted in Section \ref{sec:shockfitting}.
A sequence of 9 low-resolution spectra were collected with different telescopes including the XLT, the Lijiang 2.4-m telescope (hereafter LJT), the ARC 3.5-m telescope at Apache Point Observatory (hereafter \mbox{APO~3.5-m}), the 2-m Faulkes Telescope North (FTN) of the Las Cumbres Observatory network, and the \mbox{9.2-m} Hobby-Eberly Telescope (HET).
The details of the spectroscopic \mbox{observations} are listed in Table \ref{tab:spec}.

All spectra were reduced using the standard IRAF routines, which involves corrections for bias, flat field, and removal of cosmic rays.
The Fe/Ar and Fe/Ne \mbox{arc} lamp spectra obtained during the observation night are used to calibrate the wavelength of the spectra, and \mbox{standard} stars observed on the same night at similar airmasses as the supernova were used to calibrate the flux of spectra.
The spectra were further corrected for continuum atmospheric extinction during flux calibration, using mean extinction curves obtained at Xinglong Observatory, Lijiang Observatory, Apache Point Observatory, and Haleakala Observatory in Hawaii; moreover, telluric lines were removed from the data.

\begin{deluxetable*}{lcccccc}
\tablecaption{Journal of Optical Spectroscopy of \mbox{SN~2017ein}. \label{tab:spec}}
\tablehead{
\colhead{UT} &\colhead{MJD} &\colhead{$t_e$\tablenotemark{a}} &\colhead{$t_V$\tablenotemark{b}}
&\colhead{Telescope} &\colhead{Instrument} &\colhead{Exposure time (s)}
}
\startdata
2017/05/26.57 &57899.57 &2.6 &$-13.9$ &BAO 2.16~m &BFOSC+G4 &3600\\
2017/05/30.63 &57903.63 &6.6 &$-9.9$  &BAO 2.16~m &BFOSC+G4&3000\\
2017/05/31.60 &57904.60 &7.6 &$-8.9$  &BAO 2.16~m &BFOSC+G4 &3000\\
2017/06/15.15 &57919.15 &21.7&$+5.2$  &HET 9.2~m  &LRS2     &1500\\
2017/06/20.56 &57924.56 &27.5&$+11.0$ &BAO 2.16~m &BFOSC+G4 &2700\\
2017/06/23.29 &57927.29 &30.3&$+13.8$ &FTN 2.0~m  &FLOYDS   &2700\\
2017/06/28.28 &57932.28 &35.3&$+18.8$ &FTN 2.0~m  &FLOYDS   &2700\\
2017/07/29.55 &57963.55 &66.5&$+50.1$ &LJO 2.4~m  &YFOSC+G10&2100\\
2017/08/21.04 &57987.04 &90.0&$+73.6$ &APO 3.5~m  &DIS+B300\&R400&1800\\
\enddata
\tablenotetext{a}{Days after explosion.}
\tablenotetext{b}{Phase relative to the \textit{V}-band maximum.}
\end{deluxetable*}

\section{Photometric Properties of \mbox{SN~2017ein}}\label{sec:lc}
\subsection{Optical and ultraviolet light curves}\label{sec:OUV-lc}
\begin{figure*}[ht!]
\centering
\includegraphics[width=0.8\linewidth]{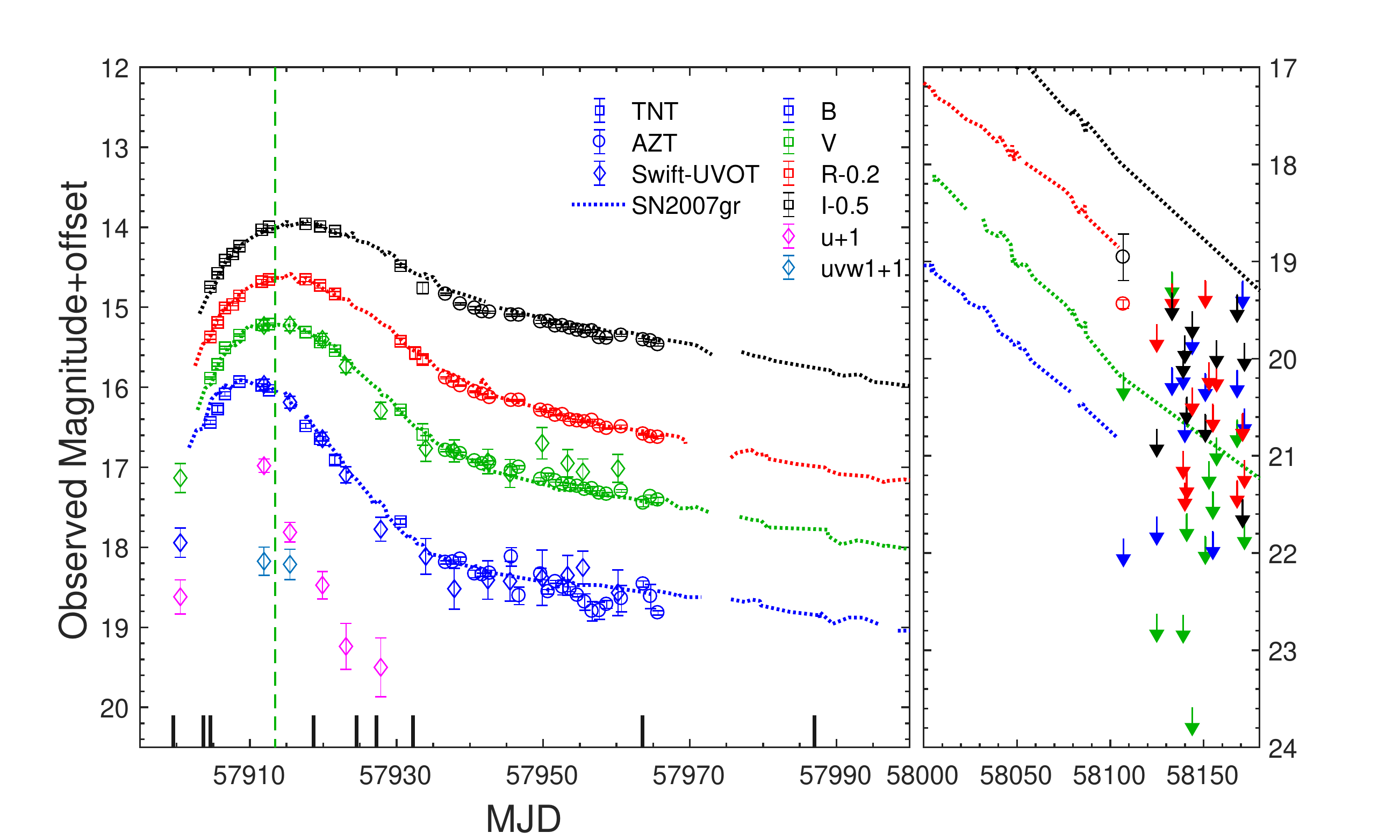}
\caption{$BVRI$ light curves of \mbox{SN~2017ein}, obtained with the 0.8-m Tsinghua-NAOC telescope, AZT-22 1.5-m telescope, and Swift UVOT. The vertical dashed line marks the maximum time in the $V$ band. The ticks at the bottom mark the epoches with spectroscopic observations. Also plotted are the $BVRI$ light curves of SN 2007gr \citep{2014ApJ...790..120C}. The light curves of SN 2007gr are shifted in time and magnitudes to match the corresponding light curve peaks of \mbox{SN~2017ein}.\label{fig:lc-BVRI}}
\end{figure*}

Figure \ref{fig:lc-BVRI} shows the AZT and TNT \textit{BVRI} light curves of \mbox{SN~2017ein}, in comparison with those of SN 2007gr \citep{2014ApJ...790..120C}.
The \textit{Swift} UVOT $uvw1$- and $ubv$-band light curves are also overplotted.
\mbox{Applying} polynomial fits to the data around maximum light, we found that \mbox{SN~2017ein} reached $B$-band peak of m$_{B}$(peak)=15.91$\pm$0.04 mag on MJD 57909.86, and $V$-band peak of m$_{V}$(peak)=15.20$\pm$0.02 mag at $\sim$3.6 days later.
This gives an estimate of $B_{\mathrm{max}}-V_{\mathrm{max}}$ color as 0.71$\pm$0.05 mag for \mbox{SN~2017ein}, which is much redder than the typical value for normal SNe Ic.
Given the \mbox{small} reddening of E$(B-V)_{\mathrm{MW}}$=0.019 mag due to the Milky Way \citep{2011ApJ...737..103S} and a typical $B-V$ color of 0.4 mag for normal SNe Ic near the maximum light \citep{2011ApJ...741...97D}, the red color observed for \mbox{SN~2017ein} indicates that it suffered significant reddening due to the host galaxy (see also discussion in Section \ref{sec:dist_extinc}).

\begin{figure}[ht!]
\centering
\includegraphics[width=1.\linewidth]{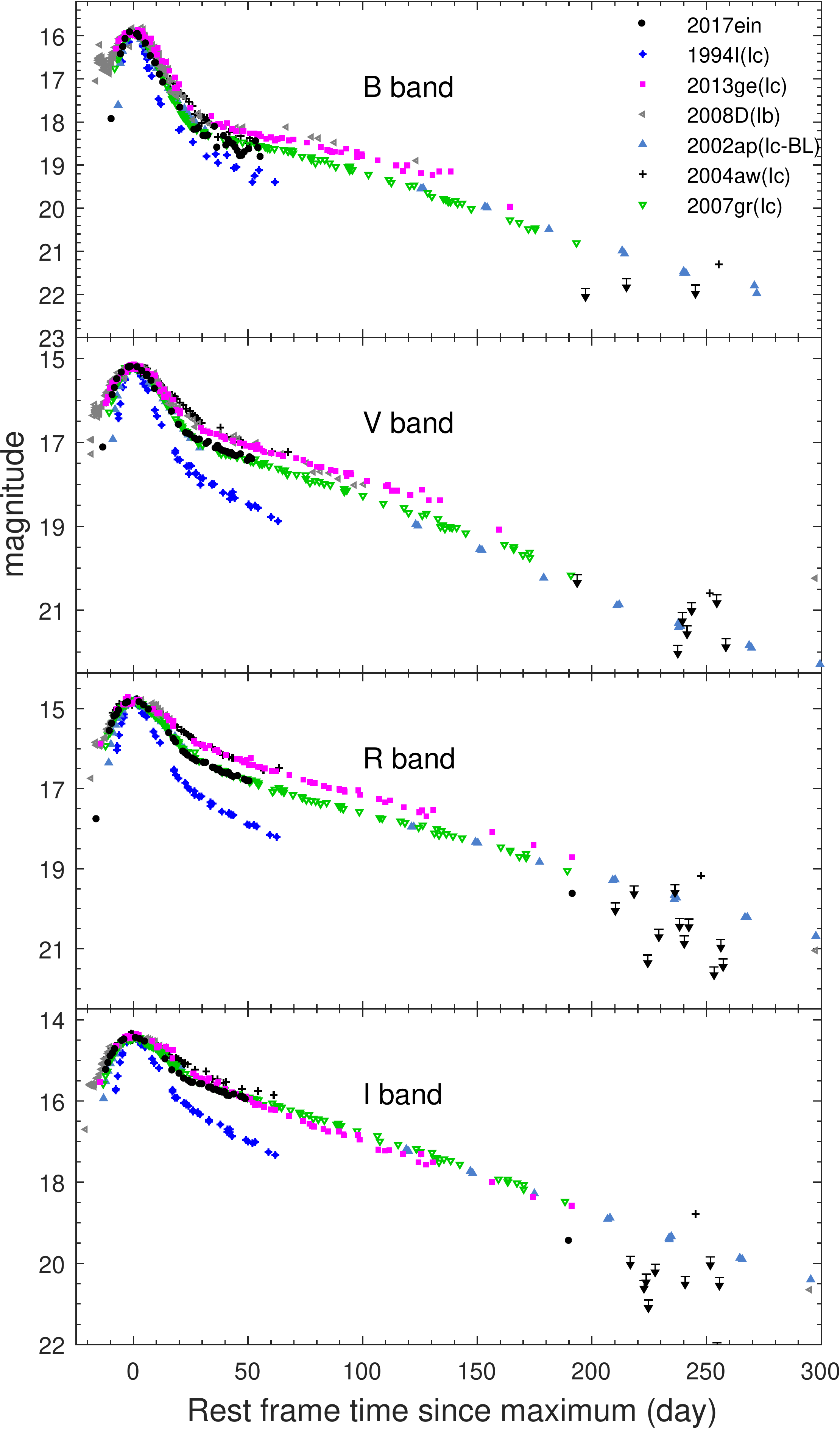}
\caption{Comparison of $BVRI$ light curves of \mbox{SN~2017ein} with other SNe Ib/c: SN 1994I, SN 2002ap, 2004aw, 2007gr, SN 2008D and SN 2013ge. The light curves are shifted in time and magnitude to match the peaks of \mbox{SN~2017ein}. \mbox{Data} references: SN 1994I \citep{1996AJ....111..327R, 1994PASJ...46L.191Y}, SN 2002ap \citep{2003PASP..115.1220F}, SN 2004aw \mbox{\citep{2006MNRAS.371.1459T}}, SN 2007gr \citep{2014ApJ...790..120C}, SN 2008D \citep{2008Sci...321.1185M, 2009ApJ...702..226M, 2014ApJS..213...19B, 2014Ap&SS.354...89B}, and SN 2013ge \citep{2016ApJ...821...57D}.
\label{fig:lc_comp}}
\end{figure}

In Figure \ref{fig:lc_comp}, we compare the $BVRI$-band light curves of \mbox{SN~2017ein} to those of other well observed SNe Ib/c. \mbox{SN~2017ein} shows light curves that are very similar to SN 2007gr in all four $BVRI$ bands before t$\sim$+60 days from the peaks.
\mbox{Adopting} the explosion date as MJD 57897.0 obtained in Section \ref{sec:shockfitting}, we find a rise time of $\sim$13.2 days in the $B$ band and $\sim$16.6 days in the $V$ band for \mbox{SN~2017ein}.
This rise time is longer than that of \mbox{SN 1994I} and \mbox{SN 2004aw}, but similar to that of \mbox{SN 2007gr} and \mbox{SN 2013ge}.
Moreover, we also measured the post-peak magnitude decline-rate parameter $\Delta m_{15}$, which is derived as 1.33$\pm$0.07 mag in $B$ and 0.93$\pm$0.04 mag in $V$ respectively for \mbox{SN~2017ein}.
These values are nearly equal to those measured for \mbox{SN 2007gr} \cite[$\Delta m_{15}(B)$=1.31 mag;][]{2014ApJ...790..120C} and \mbox{slightly} larger than those for SN 2013ge \cite[$\Delta m_{15}(B)$=1.11 mag;][]{2016ApJ...821...57D}.
The full light-curve parameters in different bands are listed in Table \ref{tab:lcsum}.
Our \mbox{observations} did not cover the period from t$\sim$60 -- 190 days after $V$-band maximum due to solar conjunction. The later observations restarted on MJD 58107, corresponding to a phase of t$\sim+$193 d with respect to the $V$ maximum.
The supernova became too faint to be detectable except in $R$- and $I$-bands, where \mbox{SN~2017ein} is found to be apparently fainter than SN 2007gr at this phase.
Due to the absence of observations during t$\sim+$60 -- 190d, it is unclear when the light curves of \mbox{SN~2017ein} started to deviate from those of SN 2007gr.

Figure \ref{fig:lc_comp} shows the comparison of \mbox{SN~2017ein} with some well-observed SNe Ibc including SNe 1994I, 2002ap, 2004aw, 2007gr, 2008D, and 2013ge.
The \mbox{upper} limits showing obvious large deviations from the evolutionary trend are due to improper subtraction and bad weather and are not plotted.
The overall light curve evolution of \mbox{SN~2017ein} shows close resemblance to that of SN 2007gr at t $<$ +60 days from the peak, while in the nebular phases it clearly shows a much faster decline than SN 2007gr and other SNe Ib/c with similar light curve shapes near maximum light.
On day MJD 58107, \mbox{SN~2017ein} is found to be fainter than \mbox{SN 2007gr} by $\sim0.9$ mag in $I$-band and $\sim0.4$ mag in $R$-band, \mbox{respectively}.
The relatively slower decline in the $R$ band is probably due to a contribution of O~{\footnotesize I}, O~{\footnotesize II}, and Ca~{\footnotesize II} emission in the nebular phases.
For SN 2007gr, \cite{2014ApJ...790..120C} found that it dimmed at a pace faster than the $^{56}$Co decay in all bands after $t\sim+80$ d from the peak, which can be explained with an increase of escape rate of $\gamma$-rays at this phase; and its $R$-band light curve also showed a relatively slower evolution than other bands.
Thus, the faster decay seen for the late-time light curves of \mbox{SN~2017ein} could be due to larger escape rate of the $\gamma$ photons, suggestive of a low ejecta mass for \mbox{SN~2017ein} (see also discussion in Section \ref{sec:shockfitting}).
Alternatively, the newly formed dust in the cooling ejecta may also shed light from the supernova.

\begin{deluxetable*}{ccccc}
\tablecaption{Light-curve Parameters Obtained for \mbox{SN~2017ein}. \label{tab:lcsum}}
\tablehead{
\colhead{band} &\colhead{peak date (MJD)} &\colhead{peak obs. mag.} &\colhead{peak abs. mag.} &\colhead{$\Delta m_{15}$(mag)}}
\startdata
$B$            &57909.86$\pm$0.24        &15.91$\pm$0.04    &$-17.20\pm0.38$         &1.33$\pm$0.07\\
$V$            &57913.49$\pm$0.20        &15.20$\pm$0.02    &$-17.47\pm0.35$         &0.93$\pm$0.04\\
$R$            &57914.67$\pm$0.28        &14.81$\pm$0.05    &$-17.54\pm0.33$         &0.75$\pm$0.09\\
$I$            &57916.32$\pm$0.30        &14.43$\pm$0.04    &$-16.93\pm0.32$         &0.61$\pm$0.09\\
\enddata
\end{deluxetable*}

\begin{figure*}[ht!]
    \begin{minipage}{0.5\linewidth}
    \includegraphics[width=1.\linewidth]{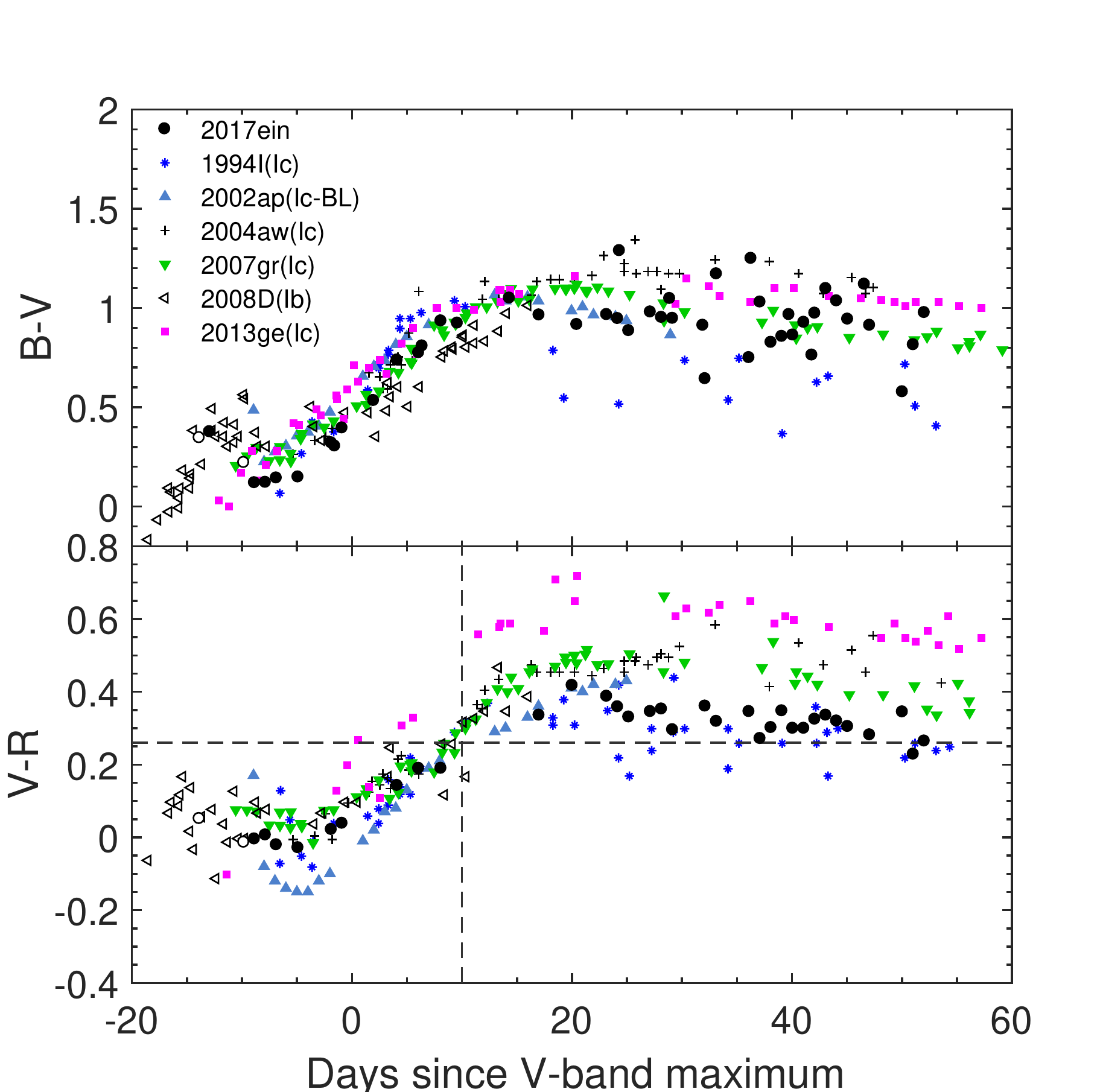}
    \end{minipage}
    \begin{minipage}{0.5\linewidth}
    \includegraphics[width=1.\linewidth]{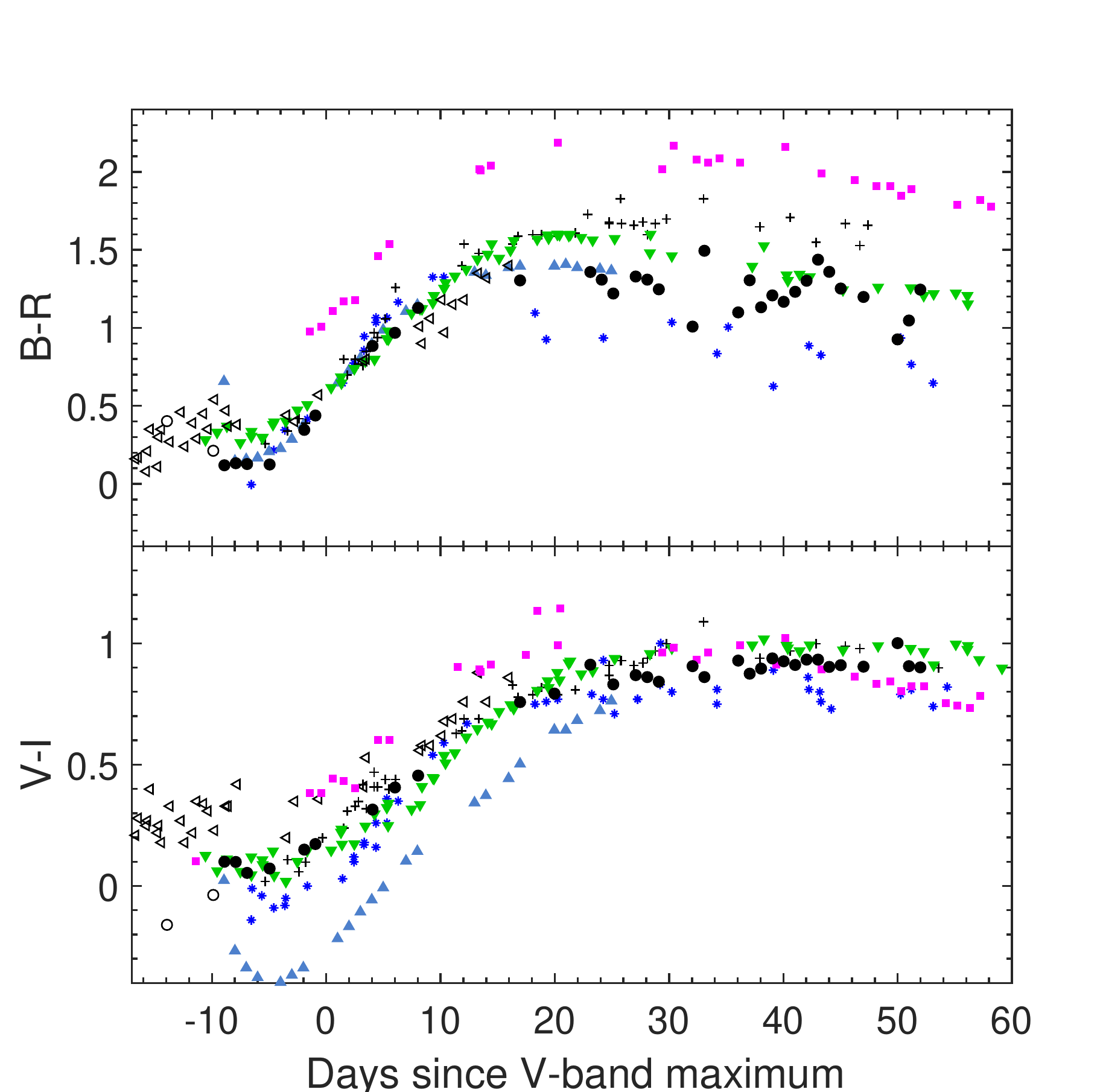}
    \end{minipage}
\caption{Color evolution of \mbox{SN~2017ein} and some well-observed SNe Ic, all corrected for the Galactic and host-galaxy reddening. The color curve of SN 2008D (SN Ib) is overplotted for comparison. The open circles represent the synthetic color derived from the spectra. The dashed lines mark that the \vr\ color tends to have a uniform value of $\sim0.26$ in \cite{2011ApJ...741...97D} at t$\sim$+10 days from the $V$-band maximum. Data references are the same as in Figure \ref{fig:color}. }
\label{fig:color}
\end{figure*}

The color evolution of \mbox{SN~2017ein} and the comparison sample is presented in Figure \ref{fig:color}, where all color curves have been corrected for the total reddening to the supernova.
The colors derived from the earliest two spectra are also included in the plot.
For \mbox{SN~2017ein}, a total reddening of E$(B - V)$$\sim$0.42 mag derived in the next section is removed from the color curves. As can be seen, all color curves seem to follow a similar trend, evolving bluewards from t$\sim$2 weeks to a few days before the $V$-band maximum.
Note that the $B - V$ and $B - R$ colors of \mbox{SN~2008D} are very blue at t $\lesssim$ 2 weeks before $V$-band maximum and they evolve redwards quickly, suggestive of an initial high temperature that can been interpreted as shock heating or interaction \citep{2009ApJ...702..226M, 2008Sci...321.1185M}.
This feature is also observed in the color evolution of \mbox{SN~2017ein}, suggesting that its early-time emission may have also experienced a cooling stage of shock breakout emission.
We will further investigate this feature  in Section \ref{sec:shockfitting}.

After reaching the bluest color, all color curves become progressively red until 3 weeks after maximum, followed by a plateau phase.
The $B - V$ color curves of different SNe Ib/c tend to converge at $\sim$0.5 mag near maximum light, though they show relatively large scatter before and after the peak.
The $V - R$ color shows a similar trend, and it tends to converge at $\sim$0.2--0.3 mag at about 10 days from $V$ band maximum as suggested by \cite{2011ApJ...741...97D}.
\mbox{SN~2017ein} shows color evolution that overall is quite similar to SN 2007gr.

\subsection{The early light curves from ATLAS}\label{sec:atlaslc}
\begin{figure}[ht!]
\includegraphics[width=0.95\linewidth]{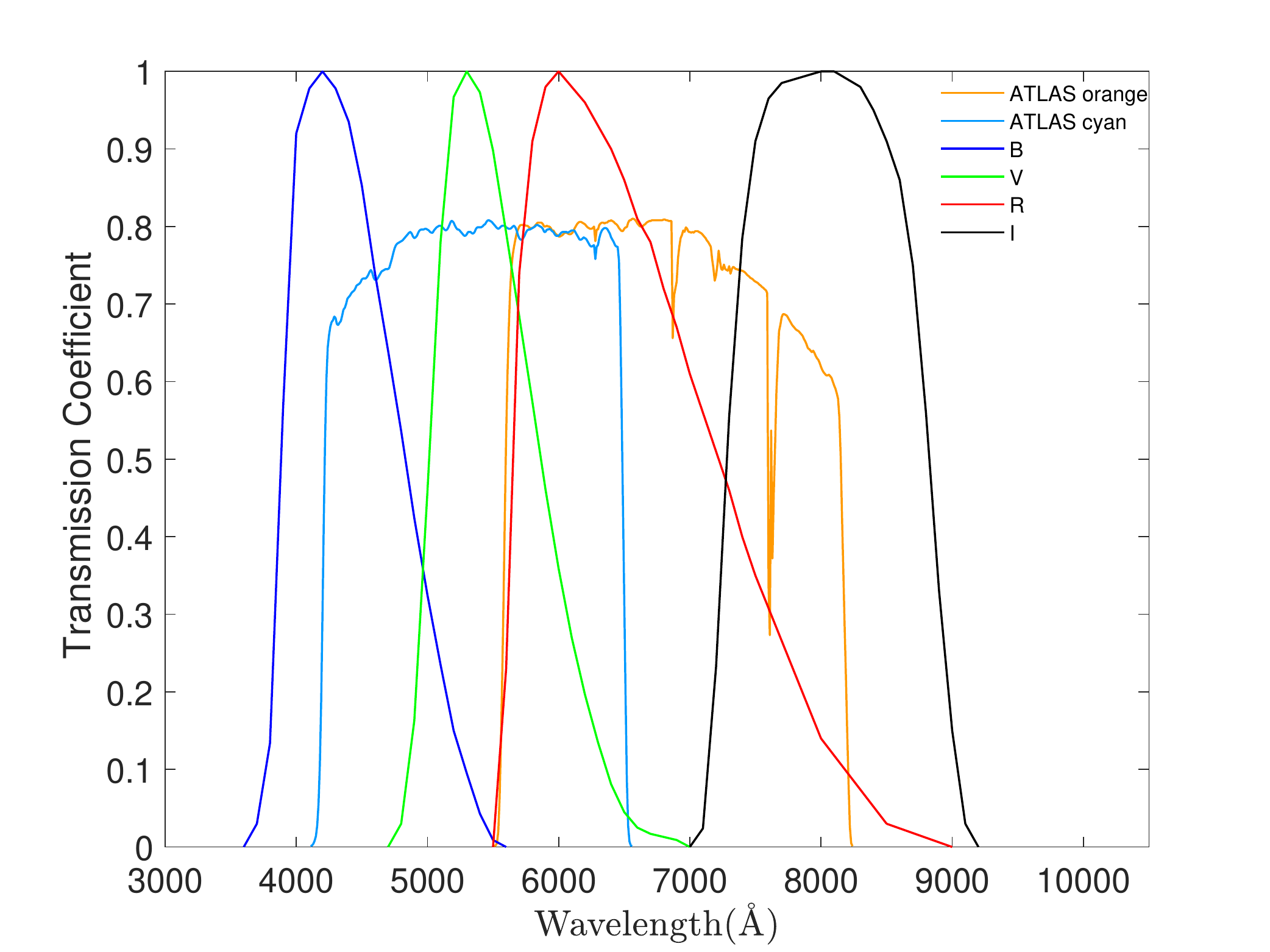}
\caption{The transmission functions of the two ATLAS filters, compared with the Johnson \textit{BVRI} filters. \label{fig:filtertran}}
\end{figure}

\begin{figure*}[ht!]
    \begin{minipage}{0.33\linewidth}
    \includegraphics[width=1.0\linewidth]{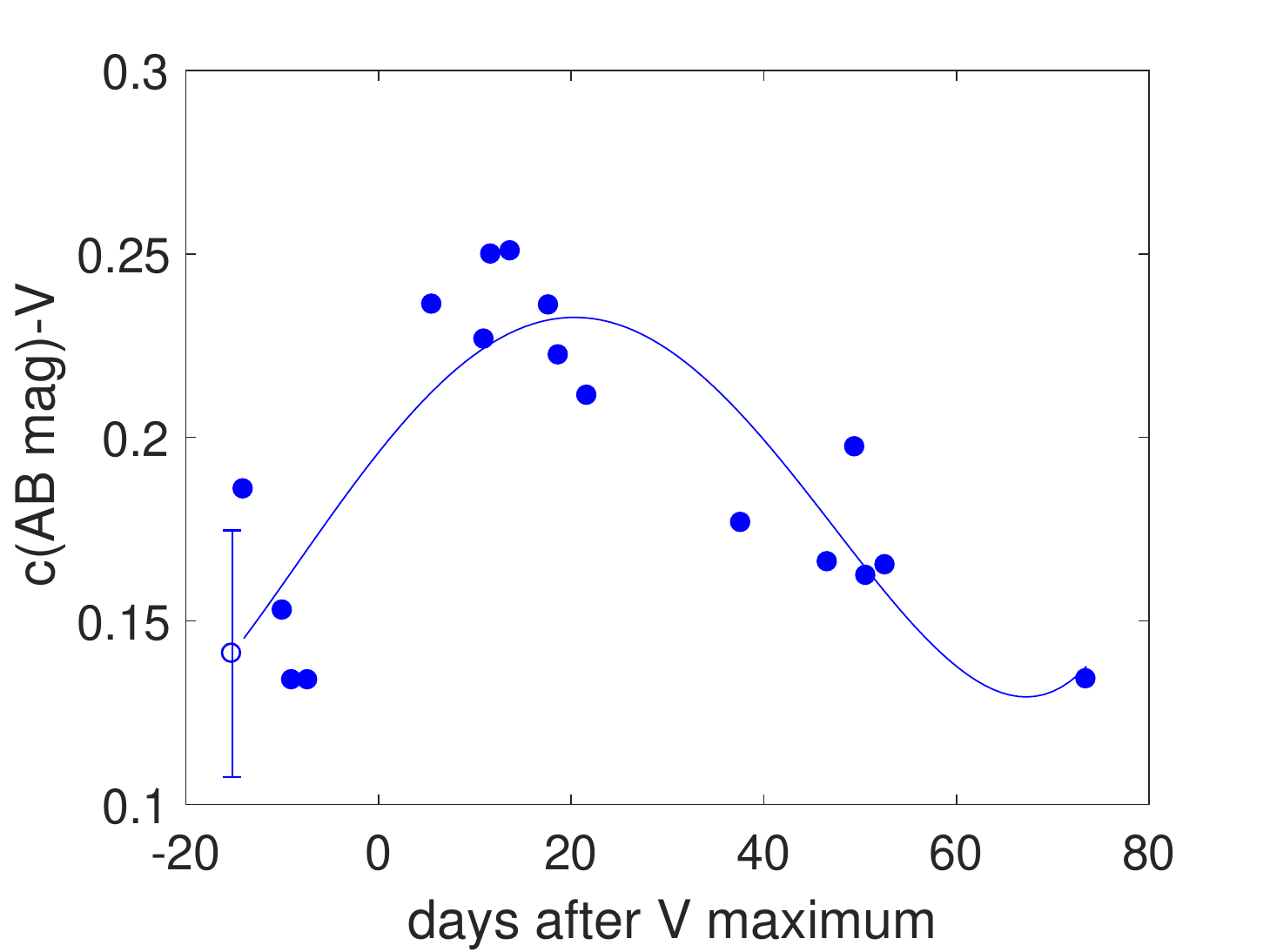}
    \end{minipage}
    \begin{minipage}{0.33\linewidth}
    \includegraphics[width=1.0\linewidth]{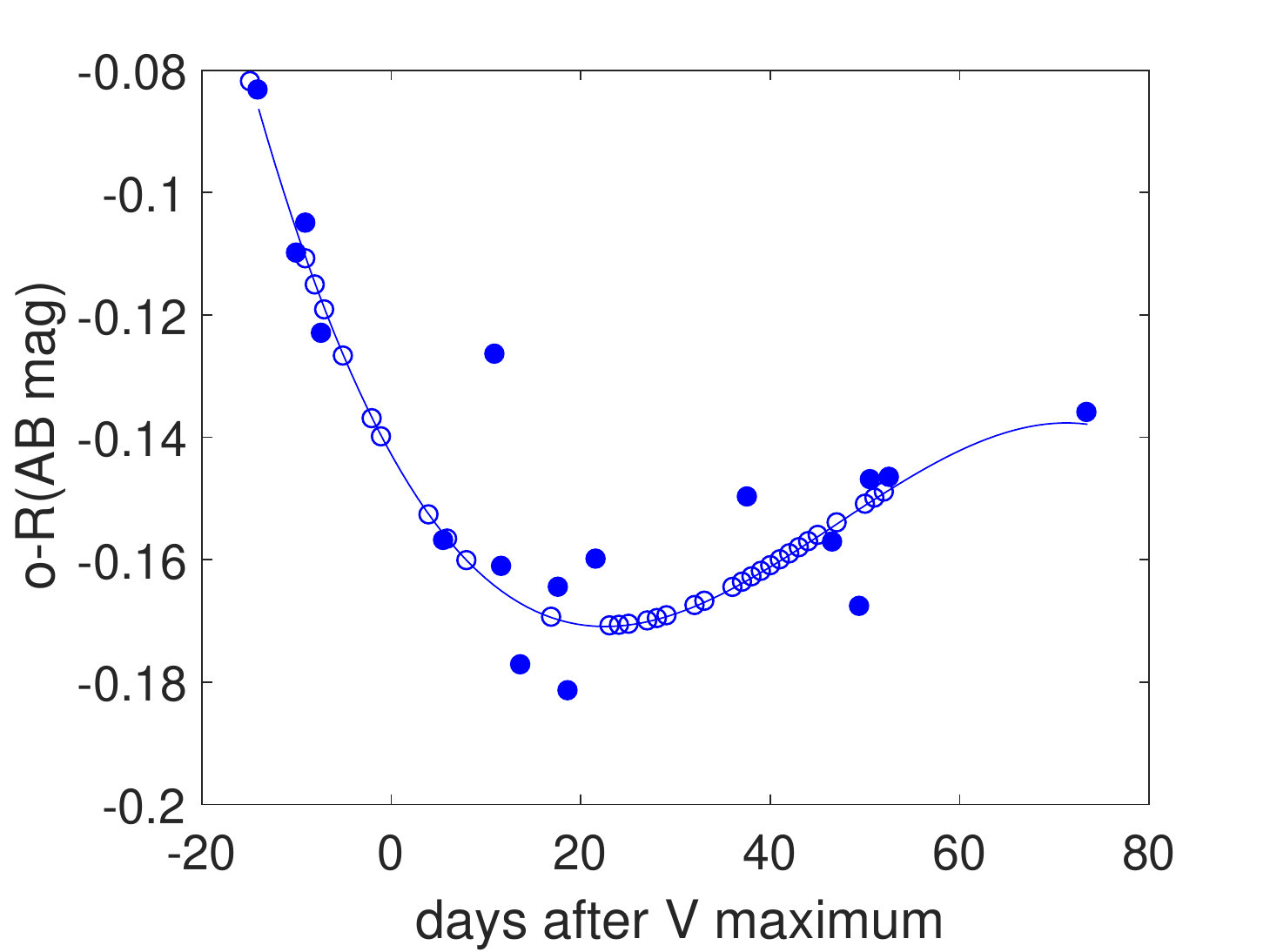}
    \end{minipage}
    \begin{minipage}{0.33\linewidth}
    \includegraphics[width=1.0\linewidth]{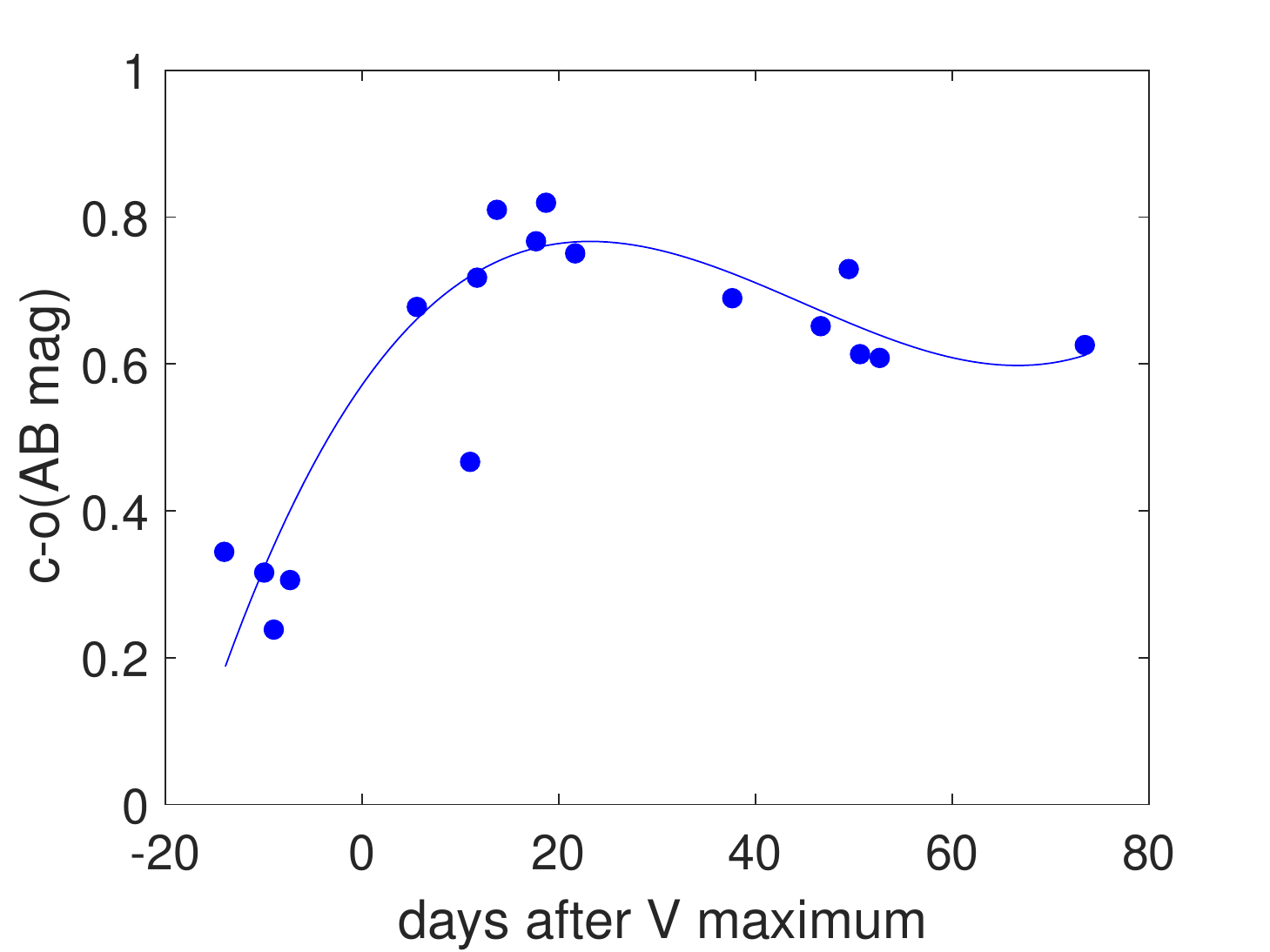}
    \end{minipage}
\caption{\textit{left}: The $c-V$ (left), $o-R$ (middle), and  $c-o$ (right) color evolutions of \mbox{SN~2017ein}, obtained using the supernova spectra presented in this paper and those taken from \cite{2018ApJ...860...90V}. The solid lines in each panel are the lower-order polynomial fits to the observed data. The open circle with an error bar is the $c-V$ color at the ATLAS-$c$ photometry epoch which is used to transform the $c$ magnitude to $V$ magnitude. The open circles in the middle panels represent the fitting values at the $R$ band photometry epochs, which are used to convert the $R$ band magnitudes to $o$ magnitudes. \label{fig:c-o-R_evol}}
\end{figure*}

The survey observation of ATLAS is operated in \mbox{two} specifically designed bands, ATLAS-cyan ($c$) and ATLAS-orange ($o$) \citep{2018PASP..130f4505T}. Figure~\ref{fig:filtertran} shows the transmission curves of these two filters and the broadband $BVRI$-band filters.
One can see that the ATLAS $o$ and $c$ filters have similar effective wavelengths to the broadband $R$ and $V$ filters, respectively.
To examine the difference between them, we calculate the synthetic $c-V$ and $o-R$ colors using the spectra of \mbox{SN~2017ein} presented in this paper, and those from \cite{2018ApJ...860...90V}.
The results are shown in Figure \ref{fig:c-o-R_evol}.
The $V$-band magnitudes appear slightly brighter than those in $c$-band by $\sim$0.2 mag, and this \mbox{difference} tends to be smaller in the early and late phases (see the left panel of Figure \ref{fig:c-o-R_evol}).
The ATLAS-$o$ magnitude differs from the $R$-band value at the same phase by $\sim$0.15 mag, as shown in the middle panel of Figure \ref{fig:c-o-R_evol}; and in very early phase this difference is less than 0.1 mag.
In the right panel of Figure \ref{fig:c-o-R_evol}, we show the evolution of the ATLAS $c-o$ color for \mbox{SN~2017ein}, with the $c$-band magnitudes being fainter than the $o$-band by 0.2--0.8 mag.
A low-order polynomial fit is applied to the $c-V$, $o-R$, and $c-o$ color evolutions, and the best-fit result is then used to interpolate the magnitudes in different bands when necessary.

\begin{figure}[ht!]
\centering
\includegraphics[width=1.07\linewidth]{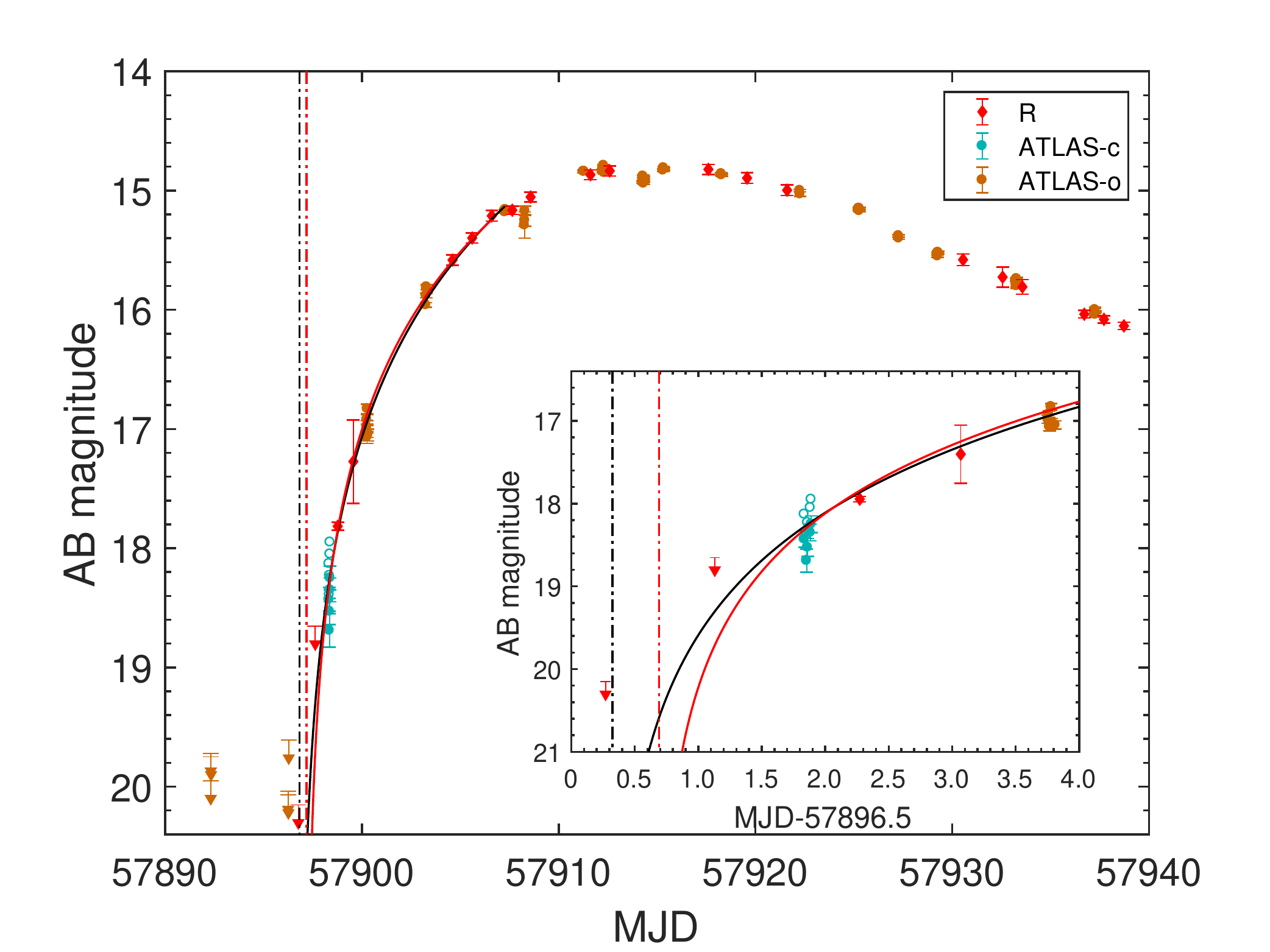}
\caption{The ATLAS $c$ and $o$-band light curves of \mbox{SN~2017ein}. The $R$-band light curves from the TNT and AZT transferred to $o$-band light curves using the relation derived in Figure \ref{fig:c-o-R_evol} are also overplotted. The detection limits from ATLAS and the report in \cite{2017ATel10481....1I} are shown by red downward arrows. The red line shows the power-law fit ($f\propto t^n$) to the pre-maximum $R$-band data, while the black line represents the fit to the combined data from \mbox{ATLAS}, AZT, and TNT. The corresponding best-fit explosion dates are shown by the vertical dot-dashed lines. The empty symbols in blue show the estimated magnitudes in $o$ band at epochs with $c$-band observations. The embedded plot shows the zoomed-in early-time light curve and the fitting. \label{fig:atlas}}
\end{figure}

Figure \ref{fig:atlas} shows the ATLAS $o$-band light curves (AB magnitude) of \mbox{SN~2017ein}, together with those converted from the AZT and TNT $R$-band photometry using the $o-R$ relationship shown in Figure \ref{fig:c-o-R_evol}.
In the conversion, we added a systematic error of 0.03 mag to the converted magnitude errors. And the converted light curve is shifted by 0.13 mag to match the peak of the $o$-band light curve.
The earliest prediscovery detection from the ATLAS $c$-band observation, taken on 2017 May 25, is \mbox{also} plotted for comparison.
In this plot, we \mbox{also} include the magnitude derived from our first spectrum taken on 2017 May 26.57.
This spectrum was flux-calibrated \mbox{using} the first $R$-band photometry from AZT (taken on 2017 May 25.77) and the earliest \mbox{ATLAS} $o$-band photometry (taken on 2017 May 27.26).
The earliest $R$-band photometry was reported in \cite{2017ATel10481....1I} with a magnitude of 17.7 mag on MJD 57898.77.
We obtained the raw images through private communication and reduced the data with our pipeline \mbox{ZrutyPhot} (Mo et al. in prep.) and obtain a magnitude of \mbox{17.78$\pm$0.01 mag}.
This measurement is converted to the $o$-band magnitude and included in our analysis as well.
The latest non-detection is from the 1-m telescope at Deok-Heong \mbox{Optical} Astronomy Observatory (DOAO) in South \mbox{Korea}, which gives an $R$-band limit of \mbox{$>$18.4 mag} from images taken on 2017 May 24.63 \citep{2017ATel10481....1I}, which is transformed to the AB system and also plotted in Figure \ref{fig:atlas}.

These early-time data can be used to constrain the time of first-light for \mbox{SN~2017ein}.
Using the data points before MJD 57907 and adopting a function of luminosity evolution as $L(\lambda)\propto t^n$ in the fit, we find the explosion date of MJD 57897.2$\pm$0.6 by fitting the early the \mbox{TNT} and AZT $R$-band data.
To get better data sample, especially at early times, we also combine the converted $R$-band data with the ATLAS $o$-band data together to infer the explosion time as MJD 57896.8$\pm$0.3 (see Figure \ref{fig:atlas}).
However, we caution that such a combined analysis may suffer some uncertainty due to that the conversion between these bands is not very accurate.
These estimated dates are very close to the non-detection date MJD 57896.77 presented by \cite{2017ATel10481....1I}, with $R$>19.9 mag.
Thus, the explosion date should be near MJD=57896.9, with a 1-$\sigma$ error of 0.3 day.
The corresponding power-law index for the above fits is $n$=1.36 and 1.51, respectively, which are smaller than the typical value of $n=2$ for the fireball model.
A more detailed analysis of the explosion date is presented in Section \ref{sec:shockfitting}, which makes use of the whole range of the light curves.

As \mbox{SN~2017ein} tends to have relatively bluer $c-o$ color in the early phase after explosion (see Figure \ref{fig:c-o-R_evol}), we assume a $c-o$ color of $\sim$0.3 mag according to the $c-o$ evolution (Figure \ref{fig:c-o-R_evol}).
The $o$-band magnitudes are thus inferred from the corresponding $c$-band photometry and are shown as empty circles in Figure \ref{fig:atlas}.
When the supernova was first detected in $c$ band at t$\sim-$15.15 days from the maximum light, it showed an excess emission beyond the power-law model.
This serves as an indication for possible detection of shock breakout cooling as addressed in section \ref{sec:shockfitting}.

\section{Distance, Reddening and Absolute Magnitude}\label{sec:dist_extinc}
The distance to the host galaxy of NGC 3938 (and hence \mbox{SN~2017ein}) was estimated by three \mbox{methods}.
One is based on the standardization of SNe II through an empirical relation between photospheric magnitude and color \citep{2014AJ....148..107R}, which is calibrated \mbox{using} SNe II with Cepheid distances.
With the observed parameters of the SN IIP exploded in \mbox{NGC 3938}, \mbox{SN~2005ay}, a distance of 31.75$\pm$0.24 mag and 31.70$\pm$0.23 mag can be derived for this galaxy, by using the $V$ and $I$-band data respectively.
While a smaller value of 31.27$\pm$0.13 mag was obtained when using correlation between luminosity and photospheric velocity of SNe IIP \citep{2002ApJ...566L..63H, 2009ApJ...694.1067P}.
A smaller value of 31.15$\pm$0.40 mag can be inferred from the Tully-Fisher relation if the Hubble constant is adopted as 75 km s$^{-1}$ Mpc$^{-1}$ \citep{1988ngc..book.....T}.
In contrast to adopting a distance modulus in a range from 31.15 to 31.75 mag \cite[see][]{2018ApJ...860...90V}, we adopt an average (not weighted) value of 31.38$\pm$0.30 mag for the distance modulus throughout this paper.

The red peak $B - V$ color and the red spectra at early epoches both suggest that \mbox{SN~2017ein} suffered severe reddening from the host galaxy.
We estimated the exact reddening through the Na {\footnotesize I} D absorption detected in the spectra.
A moderately strong Na{\footnotesize I}D absorption with an equivalent width (EW) of 2.48$\pm$0.35\AA\ can be inferred from the spectra of \mbox{SN~2017ein} taken by 2-m FTN telescope with good signal-to-noise ratio (SNR).
Based on the relation of E$(B - V)$ and EW of Na {\footnotesize I} D \citep{2003fthp.conf..200T}, i.e., \mbox{E$(B-V)$=0.16$\times$EW(Na {\footnotesize I} D)}, the host-galaxy reddening is thus estimated to be \mbox{E$(B-V)_{\mathrm{host}} = 0.40\pm0.06$ mag}.

On the other hand, a systematic statistical study of SNe Ib/c light curves \citep{2011ApJ...741...97D} concludes that the $V - R$ color of SNe Ib/c clusters at $\sim$0.26$\pm$0.06 mag, as shown by the dashed lines in the $V - R$ color plot of Figure \ref{fig:color}.
To match this trend, the observed $V - R$ color of \mbox{SN~2017ein} has to be shifted by an \mbox{amount} of about 0.4 mag, which is consistent with that derived from the Na {\footnotesize I} D absorption.
\mbox{Adopting} the extinction \mbox{coefficient} R$_{V}$ = 3.1, the total extinction is estimated to be A$_{V}$=1.29$\pm$0.18 mag for \mbox{SN~2017ein}.
\mbox{With} the distance and extinction values derived above, we estimate the absolute $V$-band peak magnitude of \mbox{SN~2017ein} as $-$17.47$\pm$0.35 mag.
This value is within the range of normal SNe Ib/c, similar to \mbox{SN 2007gr}, \mbox{SN 2013ge}, \mbox{SN 1994I} etc., but fainter than \mbox{SN 1998bw}, \mbox{SN 2004aw} and \mbox{LSQ14efd}.
In the comparison sample, SN 1998bw is the brightest, while SN 2008D is the faintest.

\section{Optical Spectral Properties}\label{sec:spec}
\subsection{Spectral classification}\label{sec:classify}
\begin{figure}[ht!]
\includegraphics[width=1.1\linewidth]{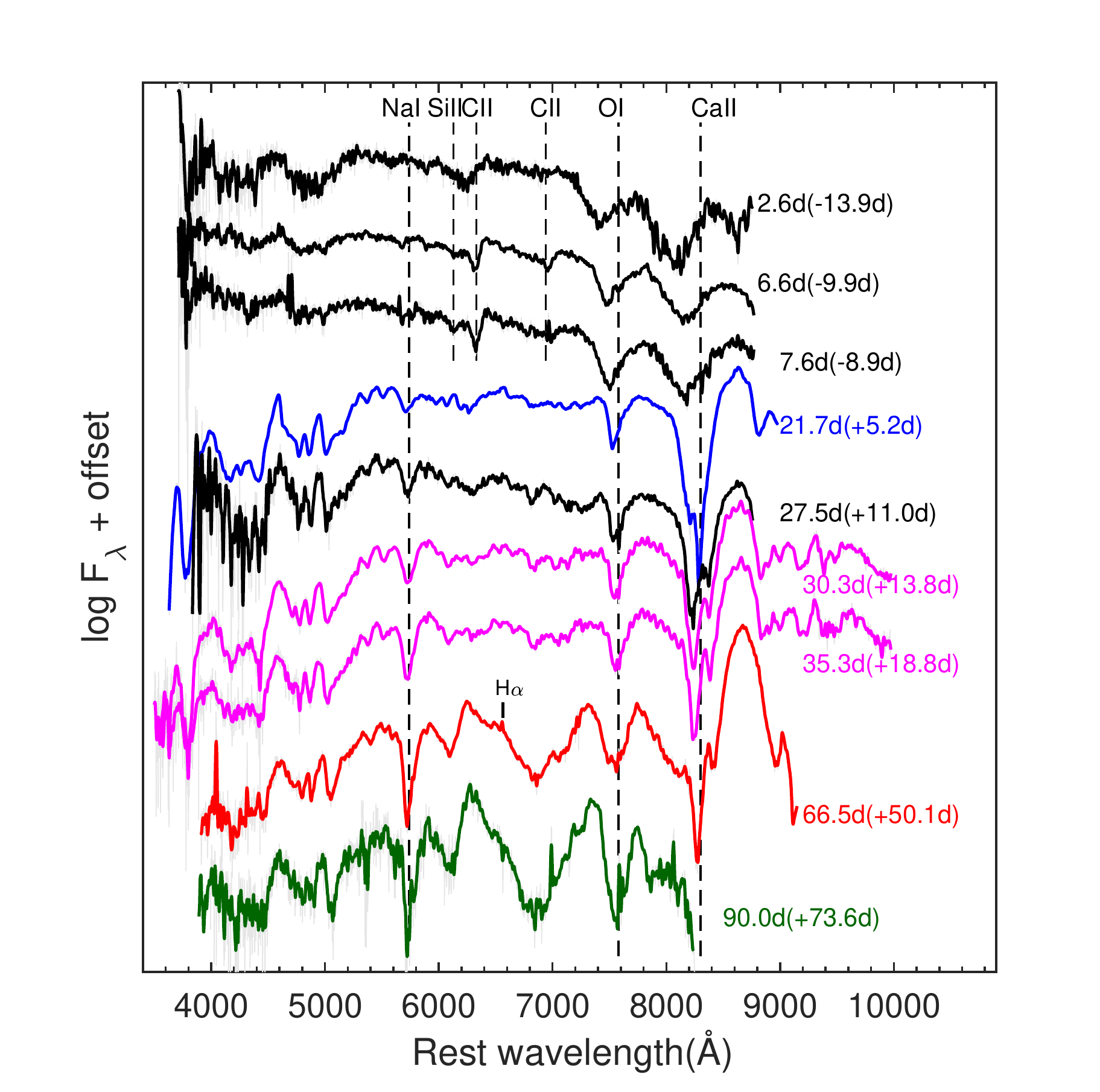}
\caption{Optical spectra of \mbox{SN~2017ein}. Different colors indicate spectra taken by different telescopes: XLT (black), HET (blue), FTN (magenta), LJT (red), and APO \mbox{3.5-m} telescope (dark green). The spectral wavelength is in the rest-frame of the host galaxy (corrected for a redshift z=0.0027), and the flux is plotted in logarithmic scale and shifted vertically for better display. The numbers beside each spectrum shows the day after explosion and numbers in bracket represent the phase in days relative to the $V$-band maximum. The positions of some noticeable absorption features are marked by dashed lines. The H$\alpha$ line from the host galaxy is marked out.\label{fig:spec}}
\end{figure}

The complete spectral evolution of \mbox{SN~2017ein} is shown in Figure \ref{fig:spec}, spanning the phase from t$\sim-$13.9 days to t$\sim$+73.6 days relative to the $V$-band maximum light.
The earliest spectrum of \mbox{SN~2017ein} is characterized by a prominent broad absorption line near 6200\AA\, which showed rapid evolution and split into two absorption features at 6100\AA\ and 6250\AA\ a few days later.
This feature in earliest spectrum can be due to a combination of Si{\footnotesize II} $\lambda$6355 and C{\footnotesize II} $\lambda$6580 absorptions.
Prominent broad absorption features can also be seen \mbox{near} 7400\AA\ and 8100\AA\, which can be attributed to the O{\footnotesize I} triplet and the Ca{\footnotesize II} NIR triplet, respectively.
The weak absorption near 6950\AA\ could also be due to C{\footnotesize II} $\lambda$7235 instead of He{\footnotesize I} $\lambda$7267, as it disappeared near maximum light.
For a SN Ib, the intensity of the main helium features will increase with time.
Thus, the \mbox{identification} of the above features give us confidence to classify \mbox{SN~2017ein} as a type Ic supernova.
This classification is further justified by the fact that it closely resembles two typical SNe Ic, SN 2007gr and \mbox{SN 2013ge}, as presented in Section \ref{sec:spec_evol}.

\subsection{Spectral evolution}\label{sec:spec_evol}
In this subsection, we examine the spectral evolution of \mbox{SN~2017ein} in comparison with other SNe Ib/c at several phases.
Most of the spectra of the comparison SNe are obtained via the Open Supernova Catalog\footnote{\url{https://sne.space}} \citep[OSC;][]{2017ApJ...835...64G}.
All spectra have been corrected for host-galaxy redshift and the total line-of-sight extinction using the extinction law of the Milky Way with R$_V$=3.1 \citep{2003ApJ...594..279G}.

\begin{figure}[ht!]
\centering
\includegraphics[width=1.05\linewidth]{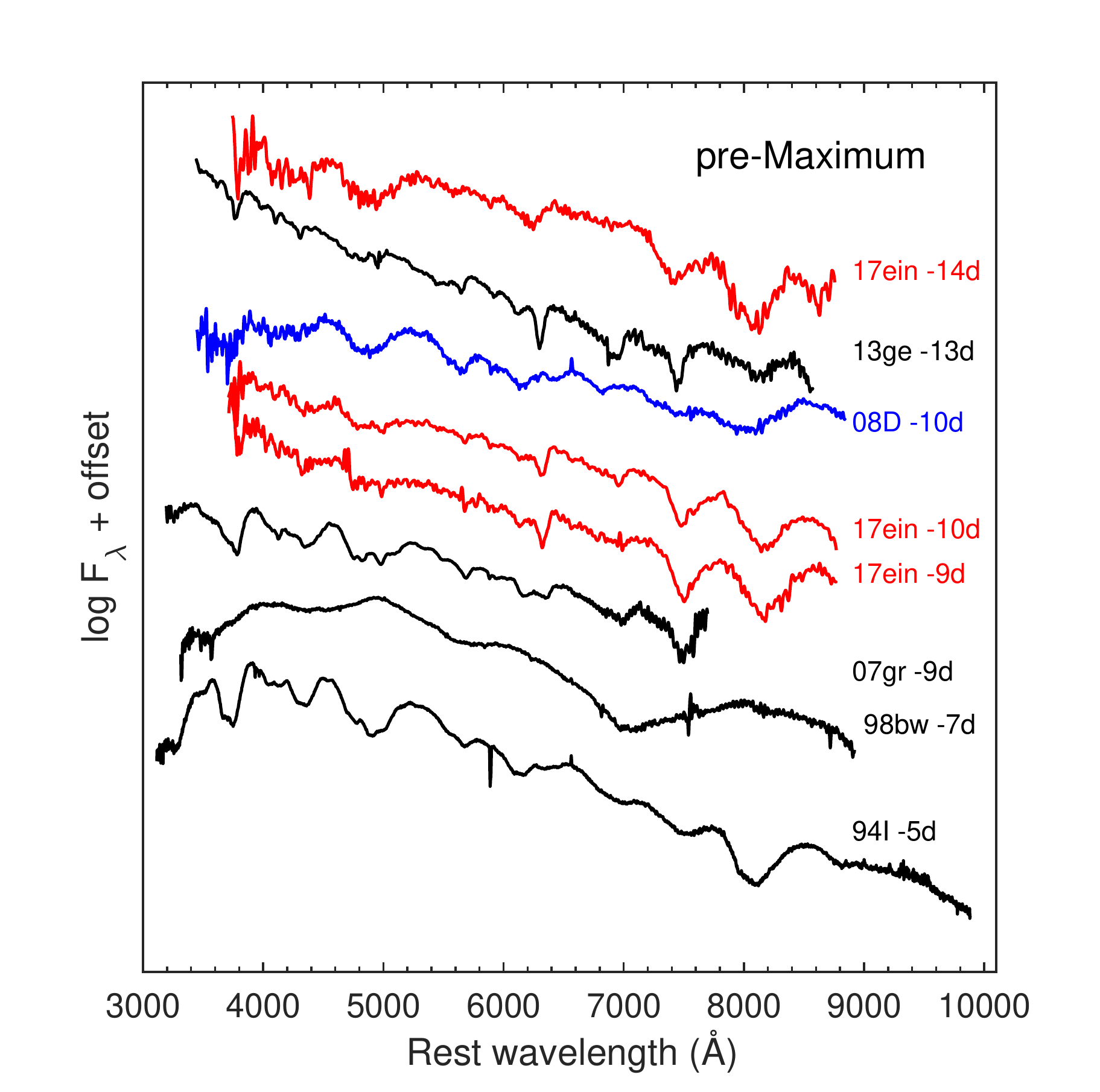}
\caption{The early-phase spectra of \mbox{SN~2017ein}, compared with those of \mbox{SN 1994I} \citep{1995ApJ...450L..11F}, \mbox{SN 1998bw} \citep{2001ApJ...555..900P}, SN 2007gr \citep{2008ApJ...673L.155V}, SN 2008D \citep{2009ApJ...702..226M}, and SN 2013ge \citep{2016ApJ...821...57D}. Spectra of SN 2017ein are plotted with red color. Different colors represent different SN types: black for Ic, blue for Ib, and magenta for IIb.
\label{fig:spectra_comppre}}
\end{figure}

\begin{figure}[ht!]
\centering
\includegraphics[width=1.05\linewidth]{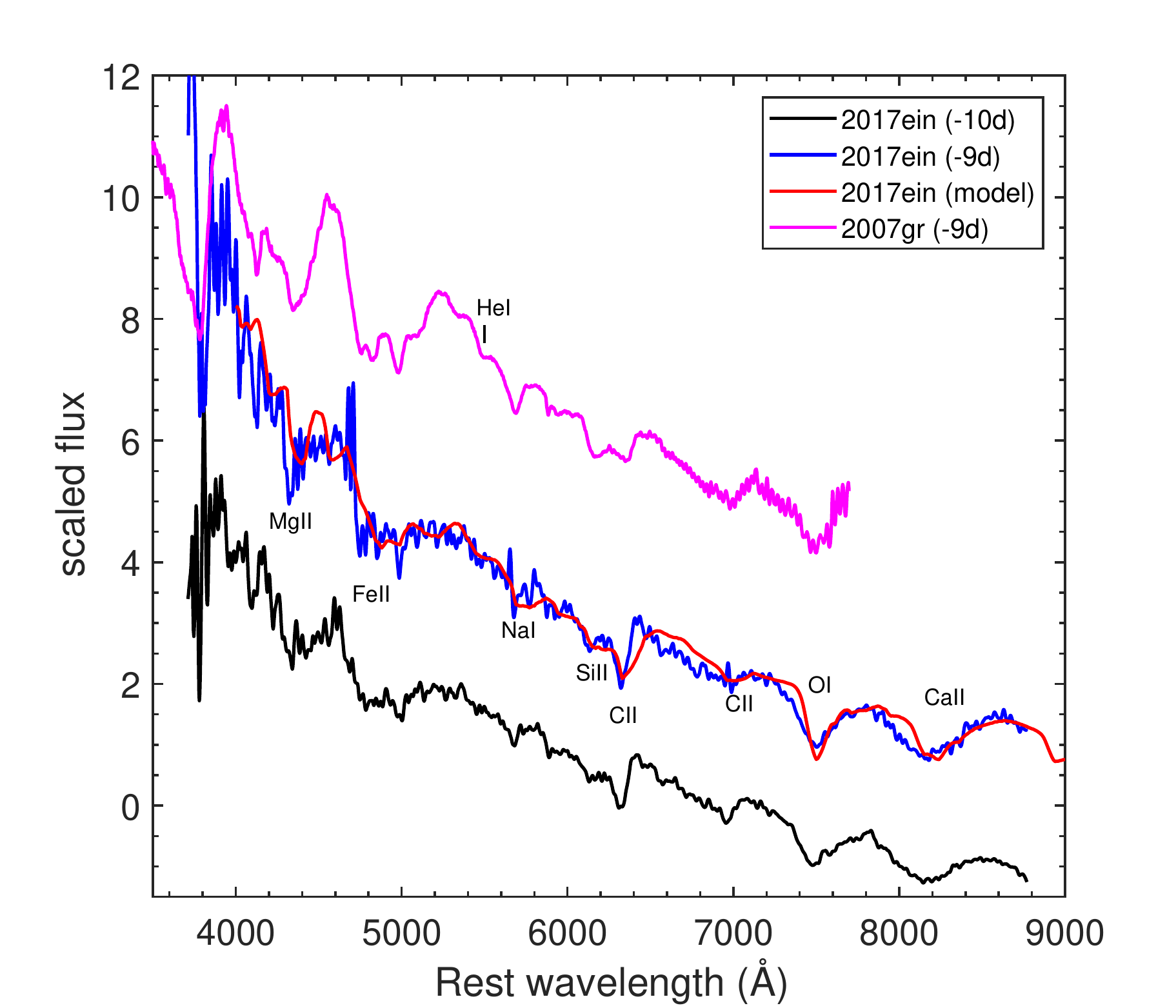}
\caption{SYNAPPS fit to the spectrum of \mbox{SN~2017ein} at t = $-$9d relative to $V$-band maximum. Also plotted the spectrum at -10d and spectrum of SN 2007gr at t = $-$9d relative to $V$-band peak \citep{2008ApJ...673L.155V}. The He{\footnotesize I} feature in the spectrum of SN 2007gr proposed by \cite{2014ApJ...790..120C} is marked out.
\label{fig:specfit}}
\end{figure}

Figure \ref{fig:spectra_comppre} shows the comparison of pre-maximum spectra.
The broad absorption feature near 6200\AA\ at t$\sim-$13.9d evolved into two separate narrow absorption features at t$\sim-$10d.
Assuming that these two features correspond to Si{\footnotesize II} $\mathrm{\lambda}$6355 and C{\footnotesize II} $\mathrm{\lambda}$6580 absorptions, they have an expansion velocity of about 10,500 km s$^{-1}$ and $12,000\mathrm{\ km\ s^{-1}}$, respectively.
The identification of prominent C{\footnotesize II} $\mathrm{\lambda}$6580 absorption in the t$\sim-$10 day spectrum is supported by the presence of another absorption feature near 6950\AA\ that can be attributed to C{\footnotesize II} $\mathrm{\lambda}$7235 with a similar blueshifted velocity.
This is also confirmed by the SYNAPPS fit\footnote{SYNAPPS \citep{2011PASP..123..237T} is a spectrum fitting code based on the  supernova synthetic-spectrum code SYN++, which is rewrite of the original SYNOW code in modern C++. }, with input of C{\footnotesize II}, O{\footnotesize I}, Na{\footnotesize I}, Mg{\footnotesize II}, Si{\footnotesize II}, Ca{\footnotesize II}, and the singly-ionized iron group elements.
The fitting result is shown in Figure \ref{fig:specfit}.
Thus in the t$\sim-$13.9 day spectrum, the Si{\footnotesize II} and C{\footnotesize II} lines likely blended together.
Applying a double gaussian fit to decompose these two features, they are found to have velocities of $\sim$14,000 km s$^{-1}$ (Si{\footnotesize II} $\lambda$6355) and $\sim$16,000 km s$^{-1}$ (C{\footnotesize II} $\lambda$6580), respectively.

The pre-maximum spectra features of \mbox{SN~2017ein} closely resembles those of SN 2007gr and SN 2013ge, which are all characterized by prominent, narrow C{\footnotesize II} $\mathrm{\lambda}$6580 absorption and weak Si{\footnotesize II} $\lambda$6355 absorption compared with SN 1994I.
The presence of high-velocity carbon in \mbox{SN~2017ein} indicates that the outer part of its progenitor star should have abundant carbon, consistent with a WC-type Wolf-Rayet progenitor star that has lost its H and He envelopes and remains a bare CO core before explosion.

\begin{figure}[ht!]
\centering
\includegraphics[width=1.05\linewidth]{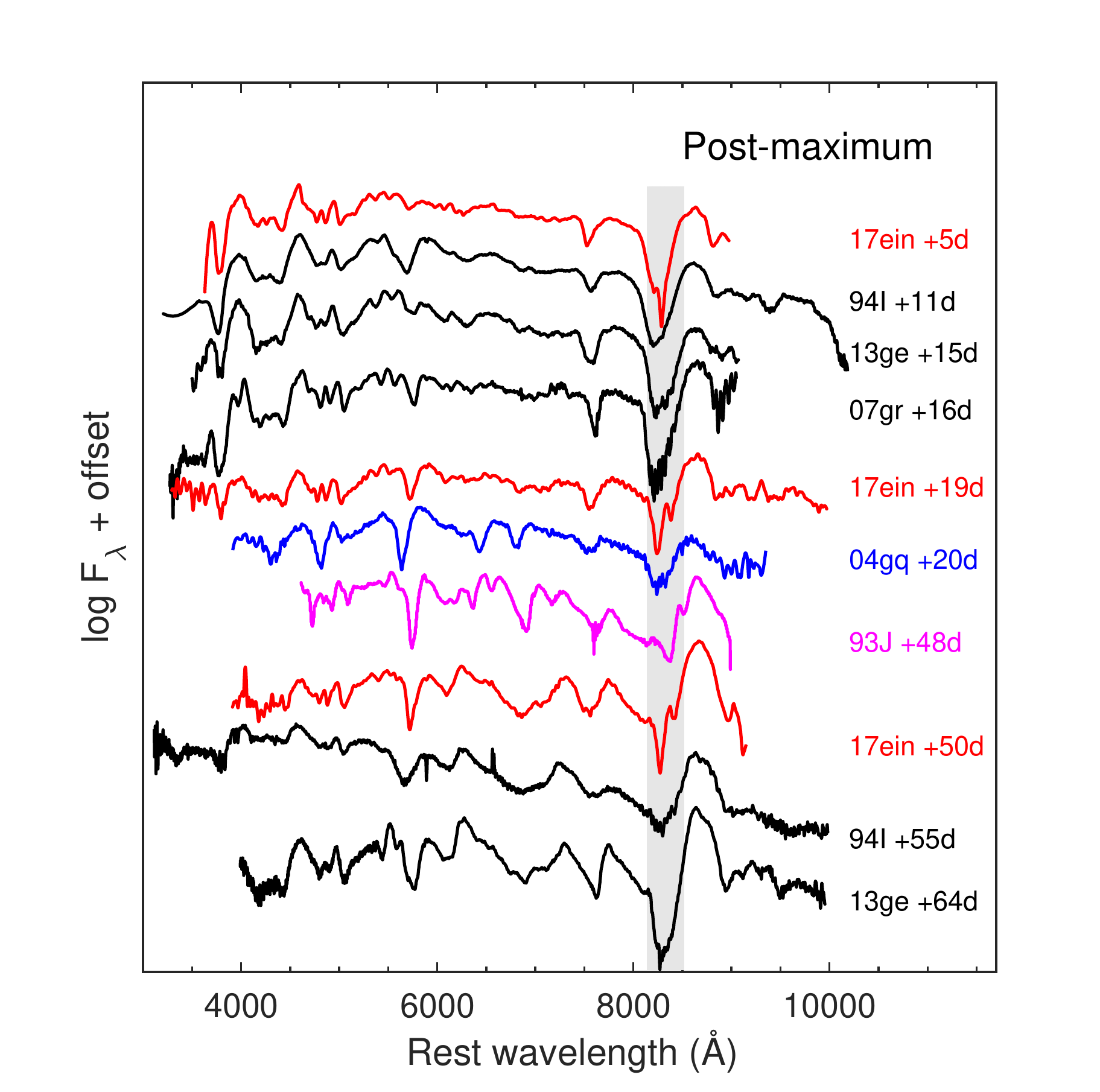}
\caption{Comparison of the post-peak spectra between \mbox{SN~2017ein} and other SNe Ib/c. The Ca{\footnotesize II} NIR triplet feature is marked by grey shaded area.
The line colors are the same as in Figure \ref{fig:spectra_comppre}. Data references: SN 1993J \citep{1995A&AS..110..513B}, SN 1994I \citep{1995ApJ...450L..11F}, SN 2004gq \citep{2014AJ....147...99M}, SN 2007gr \citep{2008ApJ...673L.155V}, and SN 2013ge \citep{2016ApJ...821...57D}.
\label{fig:spectra_comppost}}
\end{figure}

The comparison of the post-maximum spectra is shown in Figure \ref{fig:spectra_comppost}.
A late-time spectrum of type IIb \mbox{SN 1993J} is also overplotted.
As one can see, \mbox{SN 1993J} has a strong $\mathrm{H\alpha}$ line near 6150\AA, and the type Ib \mbox{SN 2004gq} and \mbox{SN 2008D} have several significant lines at 4300\AA, 5600\AA, 6500\AA\ and 6800\AA\, which can be attributed to He{\footnotesize I}.
All these features are absent in \mbox{SN~2017ein}.
\mbox{After} maximum light, the spectra of \mbox{SN~2017ein} are also similar to those of SN 2007gr and \mbox{SN 2013ge} except for the Ca{\footnotesize II} NIR triplet.
As can be seen in Figure \ref{fig:spectra_comppost}, the Ca{\footnotesize II} absorption feature in \mbox{SN~2017ein} splits into two narrow lines since t$\sim$$+$19d, which is not seen in other comparison SNe Ib/c.
This is also seen in the late spectra of SN 1993J \citep{1995A&AS..110..513B}.
In previous studies, these two lines are both attributed to the Ca{\footnotesize II} NIR triplet.
In most cases, the SN photosphere is optically thick to absorption lines, and thus the adjacent lines are likely to blend and become unresolved.
One example is the Ca{\footnotesize II} $\mathrm{\lambda\lambda\lambda}$8498, 8542, 8662, which usually blend together into one broad line.
These lines get narrower when the photosphere becomes more transparent
and recedes to lower velocities.
In this case, the line Ca{\footnotesize II} $\mathrm{\lambda}$8662 splits from the other \mbox{two} Ca{\footnotesize II} NIR lines.
Another understanding of the split feature is that the absorption line on the right is not related to Ca{\footnotesize II} NIR triplet but be due to C{\footnotesize I} $\lambda$8727 as the neutral carbon line may become stronger as a result of photospheric cooling.

\subsection{Velocity evolution}\label{sec:velocity}
\begin{figure}[ht!]
\centering
\includegraphics[width=0.95\linewidth]{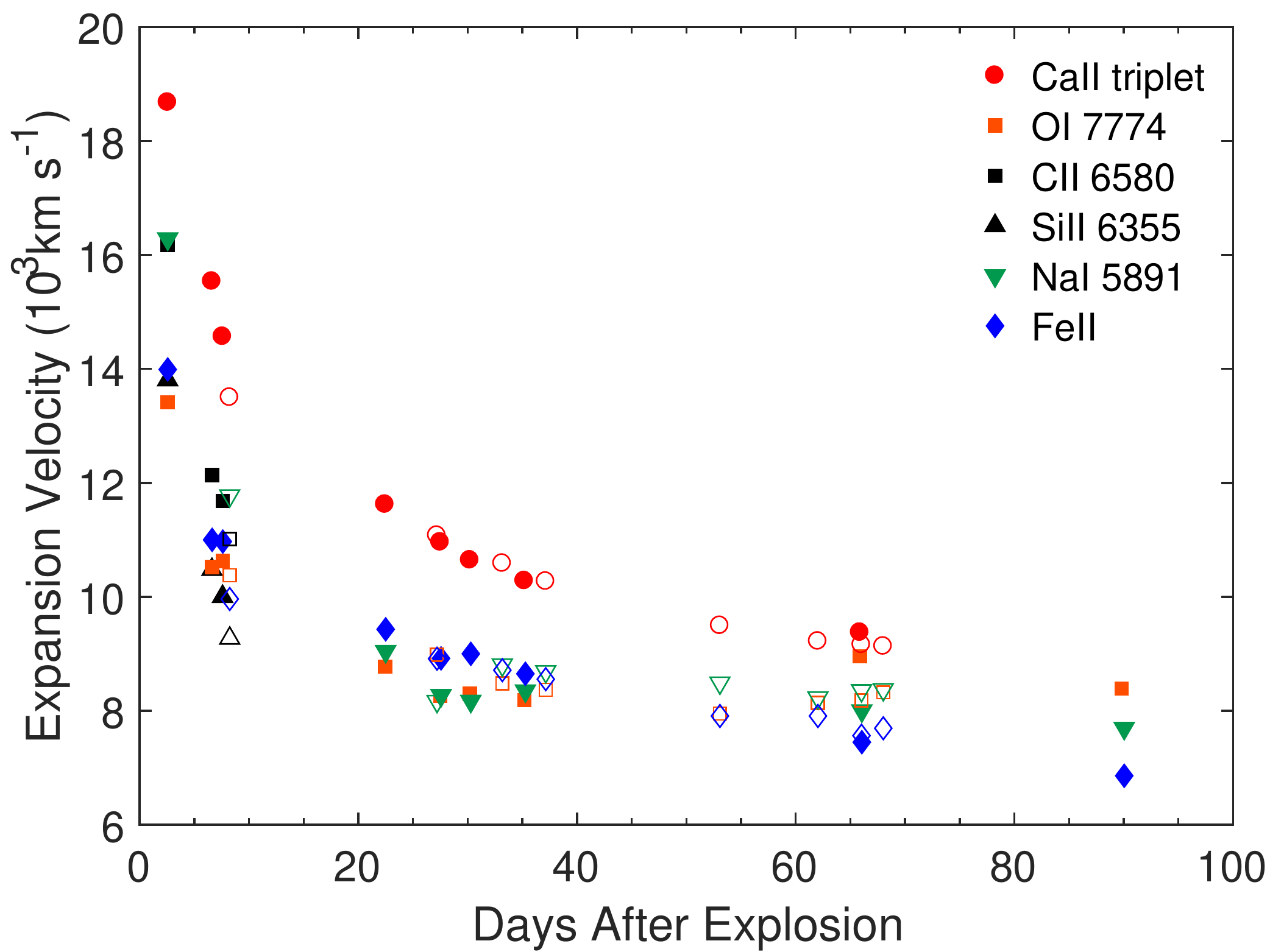}
\caption{Expansion velocity inferred from absorption lines in the spectra of \mbox{SN~2017ein}. The open symbols represent the data measured from the spectra in \cite{2018ApJ...860...90V}. \label{fig:velocity}}
\end{figure}

We measured the ejecta velocity inferred from different absorption lines in the spectra of \mbox{SN~2017ein}.
Figure \ref{fig:velocity} shows the velocity evolution of Fe{\footnotesize II} $\lambda$5169, \mbox{Si{\footnotesize II} $\lambda$6355}, C{\footnotesize II} $\lambda$6580, O{\footnotesize I} $\lambda$7774, and Ca{\footnotesize II} NIR triplet lines.
The velocity of Ca{\footnotesize II} NIR triplet is obviously higher than other lines, which is common for SNe since calcium has a larger opacity at the same radius and Ca{\footnotesize II} absorption lines thus traces the outer layers of the ejecta at higher \mbox{velocity}.

\begin{figure*}[ht!]
    \begin{minipage}{0.34\linewidth}
    \centering
    \includegraphics[width=1.0\linewidth]{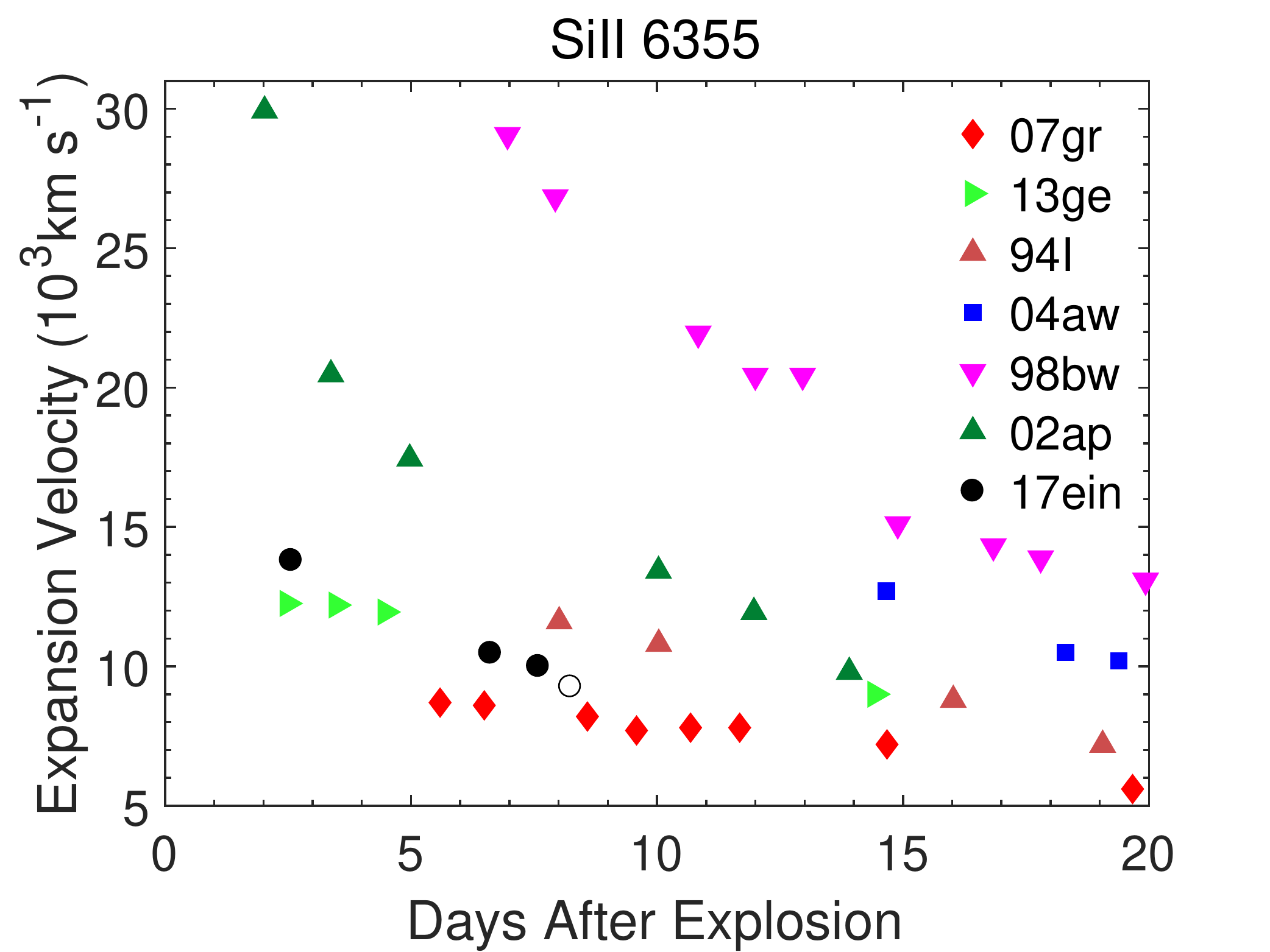}
    \end{minipage}
    \begin{minipage}{0.34\linewidth}
    \centering
    \includegraphics[width=1.0\linewidth]{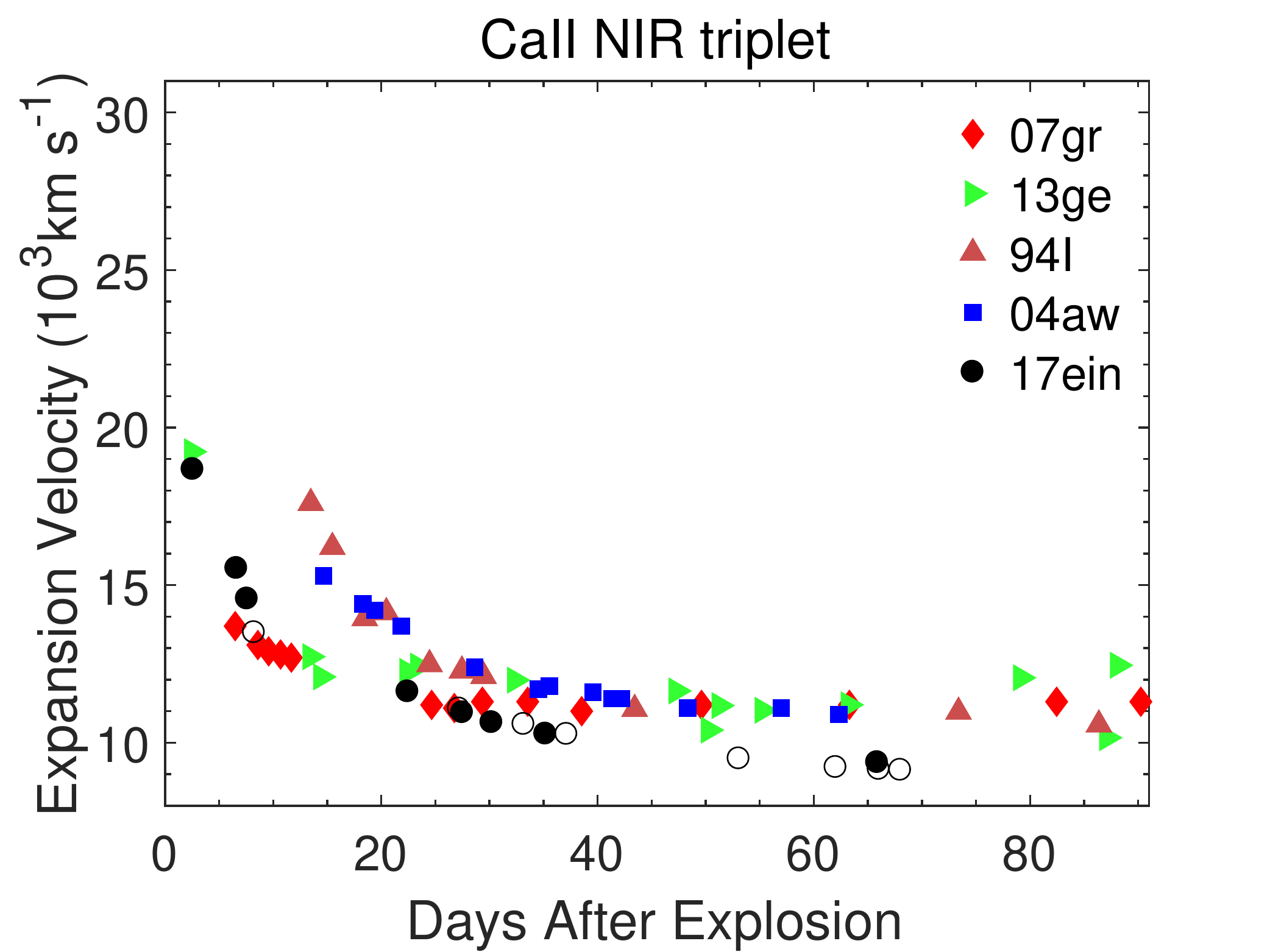}
    \end{minipage}
    \begin{minipage}{0.34\linewidth}
    \centering
    \includegraphics[width=1.0\linewidth]{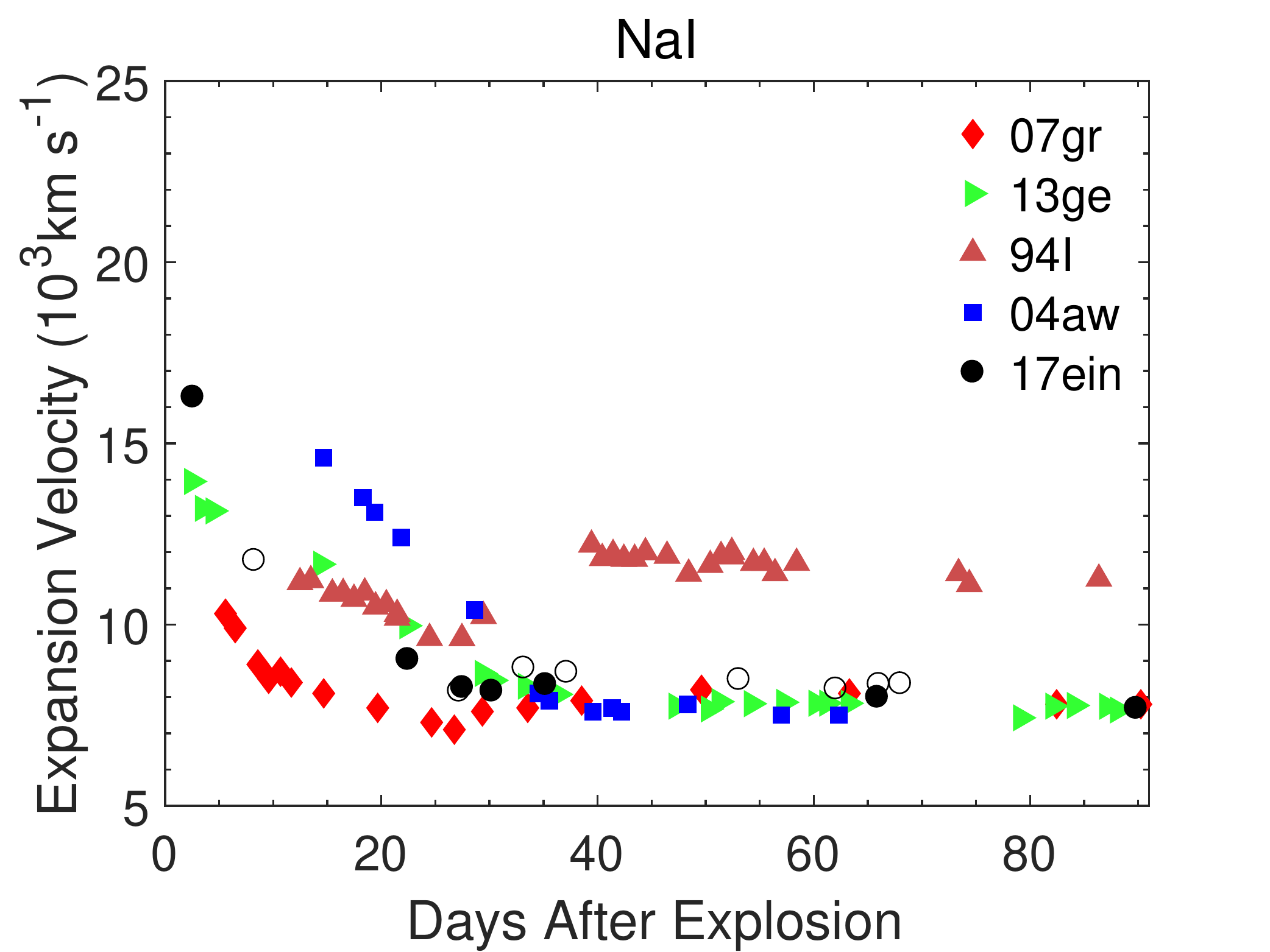}
    \end{minipage}
\caption{Evolution of the ejecta velocity inferred from Si{\footnotesize II} $\lambda$6355 (left), Ca{\footnotesize II} NIR triplet $\lambda$8571 (middle), and Na{\footnotesize I}D $\lambda$5891 (right) in the spectra of \mbox{SN~2017ein} and some well-observed SNe Ic. The open circles represent the measurement for \mbox{SN~2017ein} using spectra from \cite{2018ApJ...860...90V}.
\label{fig:vel_com}}
\end{figure*}

Figure \ref{fig:vel_com} compares line velocities with some representative SNe Ic, including SNe 1994I, 1998bw, 2002ap, 2004aw, 2007gr, and 2013ge.
It is apparent that \mbox{SNe Ic-BL} have much higher Si{\footnotesize II} velocity from very early phases until maximum light.
The ejecta velocity of \mbox{SN~2017ein} shows similar evolution to other normal SNe~Ic like SN 2007gr and SN 2013ge, which drops rapidly in about 10 days from explosion and then remains roughly constant in later phases.
Among SN 2013ge, SN 2007gr and \mbox{SN~2017ein}, SN 2007gr has the lowest Si{\footnotesize II} and Na{\footnotesize I} velocity in the early fast-decreasing stage.
But the Na{\footnotesize I} velocity of SN 2007gr shows a re-increasing trend at later epochs.
In comparison with other SNe Ic of our sample, \mbox{SN~2017ein} has a lower Ca{\footnotesize II} velocity, e.g., $\sim$12,000 km s$^{-1}$ versus $\sim$14,000 km s$^{-1}$ at around maximum light.
This difference may be caused in part by a different method of measuring the line velocity.
In \mbox{SN~2017ein}, the Ca{\footnotesize II} absorption splits into two narrower lines, and we measured the velocity from Ca{\footnotesize II} $\lambda$8538 absorption instead of treating it as a blended trough of $\lambda$8571.
Whereas we can not perform such a measurement for other SNe Ic of our sample because their Ca{\footnotesize II}~NIR features blended into one broad absorption feature.

It is noteworthy that the velocity of the Na {\footnotesize I} D line of SN 1994I showed an obvious increase from t$\sim$27 days to t$\sim$40 days from the explosion and reached a velocity plateau after that.
The Na{\footnotesize I} velocity inferred for SN 2007gr seems to have a similar trend of evolution, but with much lower plateau velocity.
As discussed in \cite{2009A&A...508..371H}, this is probably related to the combination of Na{\footnotesize II} to Na{\footnotesize I} or the departure from LTE to NLTE, although this trend could be also caused by contamination with He{\footnotesize I} $\lambda$5876.
The latter scenario gets support from recent studies that residual He may exist in some SNe Ic \cite[e.g.][]{2014ApJ...790..120C, 2018MNRAS.478.4162P}.
While \cite{2014ApJ...790..120C} suggested that in the early-time spectra of SN 2007gr, the feature near 5500\AA, \mbox{bluewards} of Na~{\footnotesize I} ~D line, should be attributed to He{\footnotesize I} (see their Figure 7), our fitting of the spectrum shows that this feature of \mbox{SN~2017ein} can be produced without adding a contribution from He (see Figure \ref{fig:specfit}).

\section{Shock Breakout Cooling}\label{sec:shockfitting}
In this section, we perform a quantitative analysis of the overall evolution of the light curves to investigate the explosion parameters of the supernova and properties of its progenitor.
The observation data include the $BVR$-bands light curve from AZT, Swift UVOT, and \mbox{TNT}, and the earliest detection from ATLAS in $c$-band.
The light curves of SNe Ic are usually believed to be powered by radioactive decay of $^{56}$Ni and $^{56}$Co \citep{1982ApJ...253..785A}, though the magnetar model has recently been proposed to account for some subclasses of SNe Ic such as \mbox{SNe Ic-BL} \citep[e.g.][]{2017ApJ...837..128W, 2016ApJ...821...22W, 2010ApJ...717..245K} and super luminous supernovae (SLSNe).
We note the possible excess emission compared to the power-law fit in the early light curves of \mbox{SN~2017ein} (see Section \ref{sec:atlaslc}), and the similar color evolution with SN 2008D during the first one or two days after explosion (see Figure \ref{fig:color}).
Given the short duration of these features, it is possible that the early light curve of \mbox{SN~2017ein} is related to the cooling of the shock breakout.
So we model the light curves by including a component due to shock cooling.

During the cooling phase, the light curve is dominated by cooling emission from shock breakout \citep{2015ApJ...808L..51P, 2013ApJ...769...67P}.
This shock cooling model considers a thermodynamical solution of the variation of the luminosity, which is related to energy, mass, and radius of the surrounding material, and expanding velocity of the shocked envelope.
The analytical model given by \mbox{\cite{1982ApJ...253..785A}} can be utilized to compute the bolometric luminosity.
Due to the lack of good UV and IR light curve samples, we try to fit the multiband light curves instead of bolometric light curve.
To fit the multiband data, we assume a blackbody SED.
Combined with the \mbox{calculated} photospheric radius, the effective temperature of the photosphere can be determined, by which the flux density at every band can be calculated.
Then we convert the flux density into absolute magnitude.

Based on the Arnett $^{56}$Ni-decay and shock cooling models and methods described above, we applied the MCMC method \citep{2017ApJ...837..128W} to fit the observed $BVR$ multi-band light curves of \mbox{SN~2017ein}.
The \mbox{observational} data include \textit{Swift} UVOT, AZT, and \mbox{TNT} light curves, and the magnitudes derived from the first spectrum.
We also convert the earliest ATLAS-\textit{c} photometry to the corresponding value in the $V$-band and include it in the $V$-band light curve.
Before that, the five earliest ATLAS-$c$ band data were averaged as they were taken within 1--2 hours.
To transform the $c$-band magnitude to the $V$-band, we use the $c-V$ color relationship described in Section \ref{sec:atlaslc}.
At the epoch when ATLAS-$c$ photometry was taken, the $c-V$ value is estimated to be 0.14$\pm$0.04 mag, from a polynomial fit to the $c-V$ evolution shown in Figure \ref{fig:c-o-R_evol}.

\begin{figure}
    \begin{minipage}{1.05\linewidth}
    \includegraphics[width=1.\linewidth]{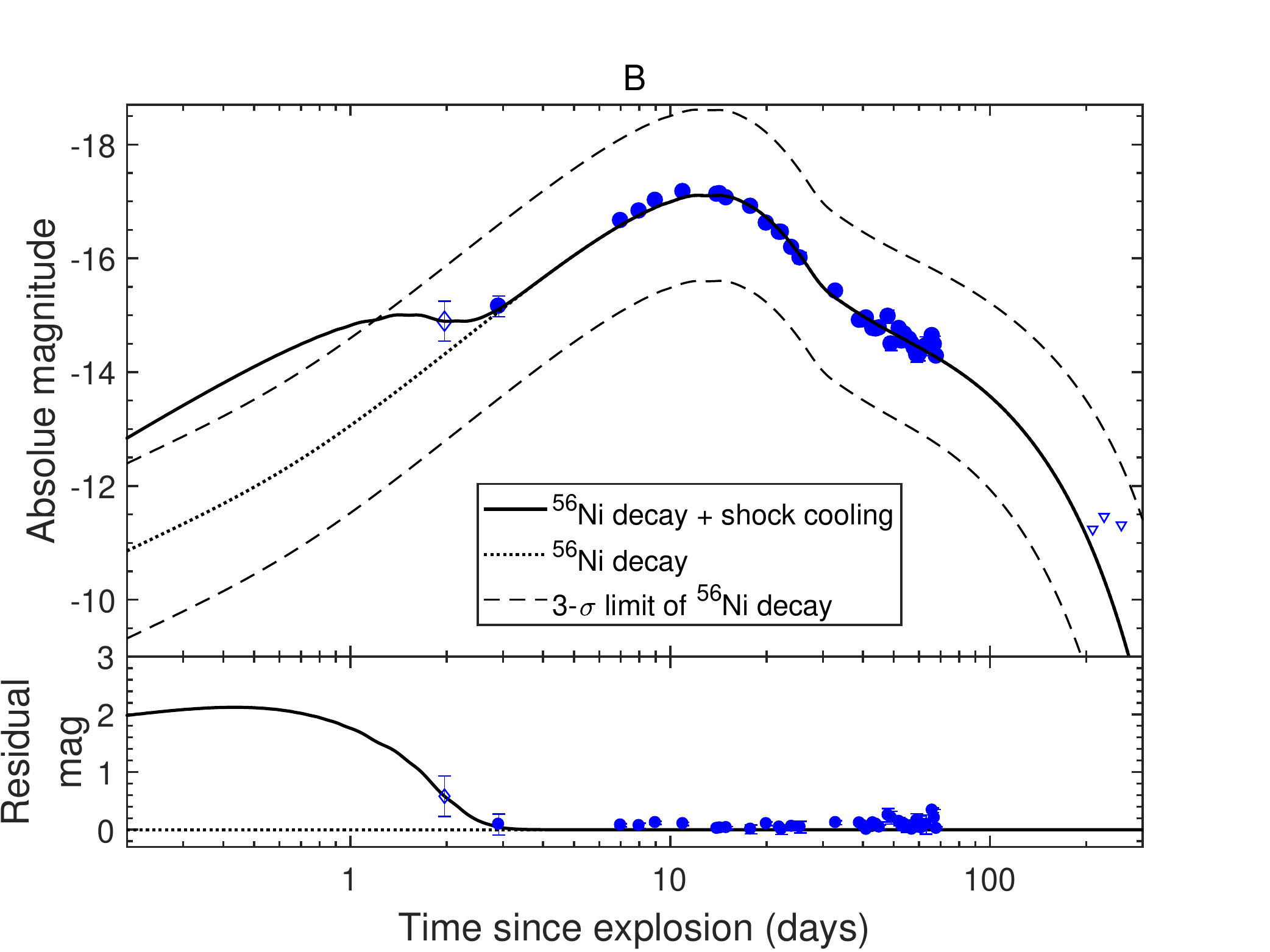}
    \end{minipage}
    \begin{minipage}{1.05\linewidth}
    \includegraphics[width=1.\linewidth]{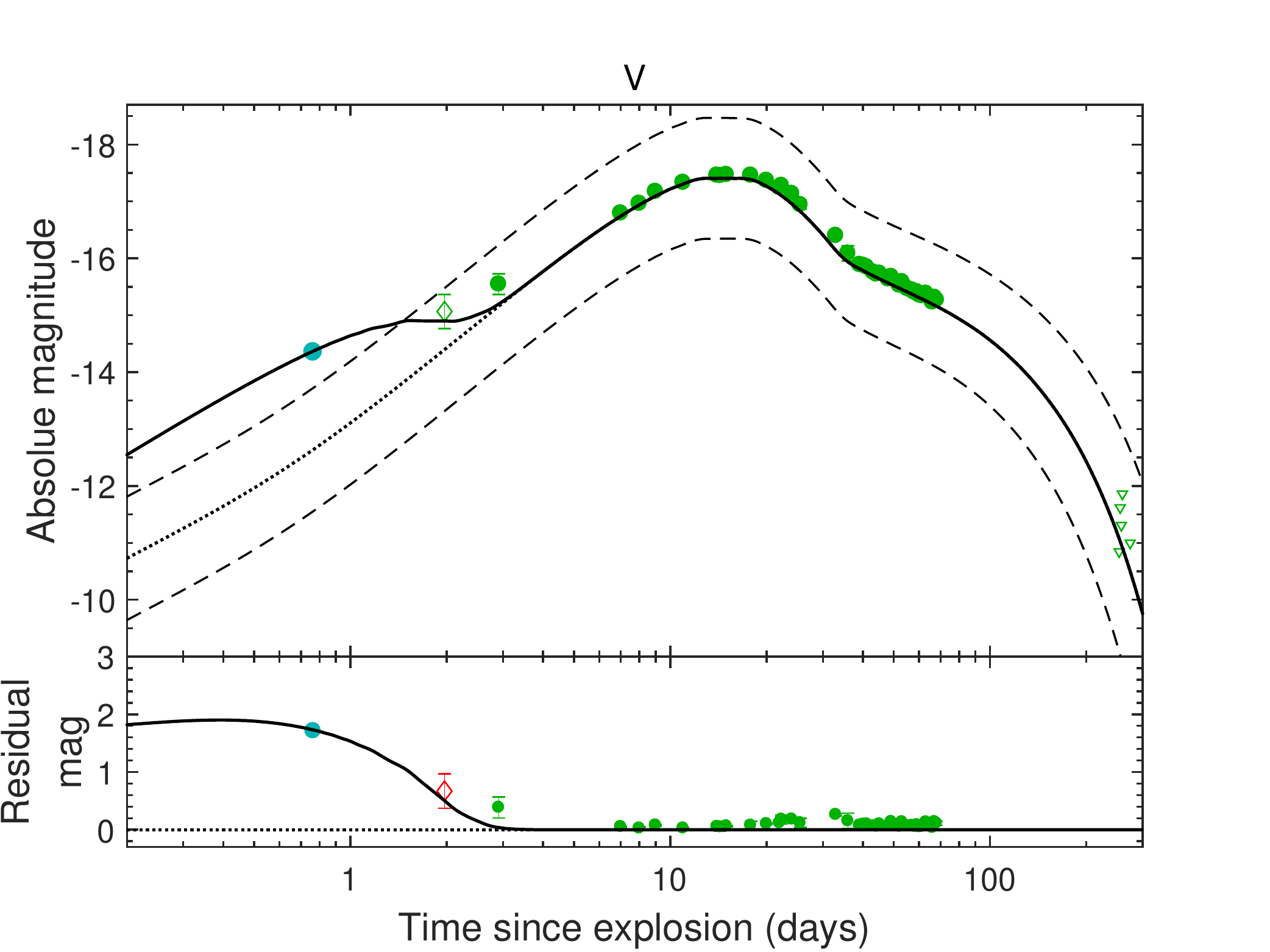}
    \end{minipage}
    \begin{minipage}{1.05\linewidth}
    \includegraphics[width=1.\linewidth]{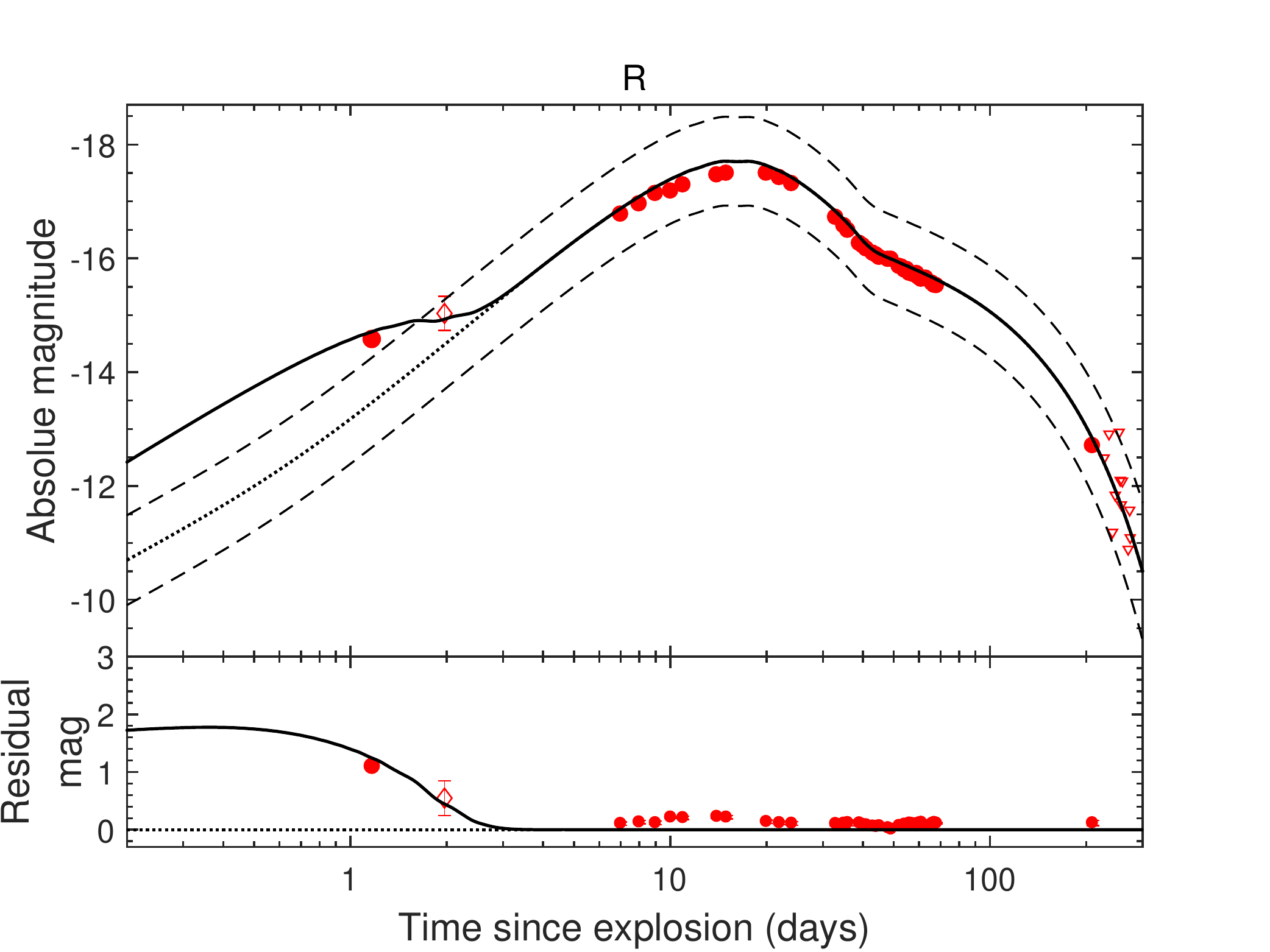}
    \end{minipage}
\caption{The fitting results of the $BVR$ light curves of SN 2017ein. Each subplot shows one band.
The absolute magnitudes from ATLAS, \textit{Swift} UVOT, AZT and TNT data are shown by solid symbols.
The open diamond represents the magnitude derived from the first spectrum taken on 2017 May 26.57. The data point in cyan represent the magnitude converted from ATLAS $c$-band magnitudes in the $V$-band figure (see text for details).
The lower panel of each plot shows the residual of the shock cooling model (solid line) and observations (data points) relative to non-cooling model. \label{fig:modelfitV}}
\end{figure}

In the fitting, the opacity to the $^{56}$Ni and $^{56}$Co decay photons is adopted as $\kappa_{\gamma, \mathrm{Ni}}=0.027\ \mathrm{cm^2\ g^{-1}}$, $\kappa_{\mathrm{opt}}=0.1\ \mathrm{cm^2\ g^{-1}}$, \mbox{respectively}, which are typical values for normal SNe Ic.
The fitting result including only the contribution of $^{56}$Ni decay is shown by the dotted lines in Figure \ref{fig:modelfitV}, and one can see that it can not explain the early ``bump'' in the light curves despite that it fits almost the whole light curve in $V$- and $R$-bands.
By fitting the light curves we also find that the photosphere cooled to a constant effective temperature of 3000$\sim$4000~K \mbox{since} t$\sim$40 days after explosion.
To account for this phenomenon, we apply a technique similar to \cite{2017ApJ...850...55N} (see their Equation 9) so that we can also fit the late-time light curve.

The best-fit light curves are shown in Figure \ref{fig:modelfitV}, and the resultant parameters of the supernova and the progenitor star are reported in Table \ref{tab:fit_result}.
We do not show the $I$-band light curve because it can not be fit very well, possibly due to a larger discrepancy between the blackbody assumption and the actual SED of the supernova in the $I$ band, where the spectra of SN are dominated with strong O and Ca absorption lines.
The fitted parameter $M_{\mathrm{env}}$ is the mass of the outermost shell of the star, from the stellar surface to position where the optical depth is $\sim v/c$ ($v$ is the shock velocity and $c$ the light speed).
Note that $M_{\mathrm{ej}}$ is the rest of the ejected mass of the explosion, while the total ejecta mass is $M=M_{\mathrm{env}}+M_{\mathrm{ej}}$.
So the total ejecta mass is $0.92\pm0.10$ M$_{\odot}$.
Our fitting gives the explosion time as MJD 57897.6$\pm$0.6, which is consistent with that derived in Section \ref{sec:atlaslc}.
We thus adopt the explosion time as the weighted average value of these two estimates, i.e. MJD 57897.0$\pm$0.3.

\begin{deluxetable*}{ccccccc}
\tablecaption{The explosion and progenitor parameters determined from the MCMC fit to the observed $BVR$ light curves of \mbox{SN~2017ein}. \label{tab:fit_result}}
\tablehead{
\colhead{$M_{\mathrm{ej}}$} &\colhead{$v_\mathrm{ph}$}&\colhead{$E_K$} &\colhead{$M_{\mathrm{Ni}}$} &\colhead{$t_{\mathrm{exp}}$} &\colhead{$R_{\mathrm{env}}$} &\colhead{$M_{\mathrm{env}}$}\\
\colhead{M$_{\odot}$} &\colhead{$\mathrm{km\ s^{-1}}$} &\colhead{$10^{51}\mathrm{erg}$} &\colhead{M$_{\odot}$} &\colhead{MJD} &\colhead{R$_{\odot}$} &\colhead{M$_{\odot}$}
}
\startdata
$0.9\pm0.1$ &9300$\pm$120 &$\sim$0.5  &$0.13\pm0.01$ &57897.6$\pm$0.6 &8$\pm$4 &$0.020\pm0.004$
\enddata
\end{deluxetable*}

As shown in Figure \ref{fig:modelfitV}, the early-time $V$ and $R$-band light curves of \mbox{SN~2017ein} can not be explained by $^{56}$Ni heating alone.
Adding the contribution from shock cooling improves the fit significantly.
The first spectrum was taken at the end of the cooling tail.
While the shock-cooling feature in the $B$-band is not so convincing, because there are no observations at earlier phases in this waveband when shock-cooling emission was stronger.
The emission of the cooling component rises to a peak at about 0.7 day after explosion, and then fades away $\sim$3 days after explosion.
After the cooling phase, the Ni-decay model well reproduces the observations.
The model fit with shock cooling gives a stellar envelope radius as $R_{\mathrm{env}}$=8$\pm$4\ R$_{\odot}$, which is consistent with the radius inferred from analysis of the pre-explosion images (see Section \ref{sec:HST}).
This also suggests that the cooling is from the stellar envelope but not the \mbox{CSM} around the progenitor star.
In the latter case, the distance to the outer region of the CSM should be much larger than a few R$_{\odot}$, and the cooling process will have a longer timescale and produce a more luminous cooling tail.
Thus, the early excess emission detected in \mbox{SN~2017ein} is more likely related to shock breakout at the surface of the progenitor star.

Although the ``bump'' feature in the early light curve seems to be well fit by including a shock cooling contribution, it is still possible that the excess emission was caused by other processes.
One popular interpretation is the mixing of $^{56}$Ni in the outer layers of the SN ejecta, which is successful in modeling the light curves of some hypernovae like SN 1997ef \citep{2000ApJ...534..660I}, SN 1998bw \citep{1998Natur.395..672I, 2001ApJ...550..991N}, and normal type Ic SN 2013ge \citep{2016ApJ...821...57D} and iPTF15dtg \citep{2016A&A...592A..89T}.
Mixing of nickel into the outer layers of SN ejecta could produce a fast rise and an extra peak or ``bump'' in the early phases, as well as broader shaped light curves.
However, the nickel-mixing model usually has a longer timescale.
For example, \mbox{SN~2013ge} has an extra ``bump'' peaking at about four days after explosion in the UV bands \citep{2016ApJ...821...57D}.
While in \mbox{SN~2017ein}, the early-time flux peak appears within one day after explosion, suggestive of a lower possibility for nickel mixing on the surface of the exploding star.
Recently \cite{2018arXiv181003108Y} discussed the effect of Ni-mixing on the early-time photometric evolution of SNe Ib/c, and found that the color curves will become progressively red rather than showing an early peak as seen in SN 2008D and \mbox{SN 2017ein} (see Figure \ref{fig:color}).
We thus prefer the shock cooling model instead of nickel mixing for \mbox{SN 2017ein}.

The ejecta mass we estimated here as $M\sim1\mathrm{M_{\odot}}$ is at the lower end of SNe Ib/c \citep{2016MNRAS.457..328L}.
For comparison, the ejecta mass of SN 2007gr has been estimated in the literature, i.e., 2.0--3.5 M$_{\odot}$ by \cite{2009A&A...508..371H} and $1.8^{+0.6}_{-0.4}$ M$_{\odot}$ by \cite{2016MNRAS.457..328L}. \cite{2009A&A...508..371H} did not describe the details of their fitting.
While \cite{2016MNRAS.457..328L} adopts an opacity of $\kappa_{\mathrm{opt}}=0.06\ \mathrm{cm^2\ g^{-1}}$ in their analysis, which is smaller than the value adopted here.
In the photospheric phase, the \mbox{Arnett} light curve model is described by two parameters: $M_{\mathrm{Ni}}$ and $\tau_m$, the latter is given by $\tau_m^2=\frac{2\kappa_{\mathrm{opt}}M}{\beta cv_{\mathrm{sc}}}$, where $\beta=13.8$ is a constant, $M$ is the total ejecta mass, $v_{\mathrm{sc}}$ is the \textit{scale velocity} of the SN, which is observationally set as the photospheric velocity at maximum light ($v_{\mathrm{ph}}$).
In our model, $v_{\mathrm{ph}}$ is given by the fit with input of the Fe {\footnotesize II} line velocities measured from the spectra.
\cite{2009A&A...508..371H} found $v_{\mathrm{ph}}\approx$ 6,700 $\mathrm{km\ s^{-1}}$ from Si {\footnotesize II} line, but \cite{2016MNRAS.457..328L} uses $v_{\mathrm{ph}}\approx$ 10,000 $\mathrm{km\ s^{-1}}$ from Fe {\footnotesize II} line.
The velocity evolution of SN 2007gr and \mbox{SN~2017ein} shows that the velocity used by \cite{2016MNRAS.457..328L} may be overestimated (see Figure \ref{fig:vel_com}).
Using the \mbox{results} from \cite{2016MNRAS.457..328L}, we find $\tau_m\approx12$ d.
And we obtain a similar value for \mbox{SN~2017ein}. Therefore, the different $M_{\mathrm{ej}}$ values inferred for SN 2007gr and \mbox{SN~2017ein} are mainly a result of different opacity values adopted in the analysis.
Moreover, the discrepancy in the ejected mass could also be related to the fitting methods and wavebands of the light curve used in the analysis.

\mbox{SN~2017ein} is one of the two SNe Ic showing possible evidence for early shock breakout cooling.
\mbox{Another} candidate is supernova LSQ14efd, with its $V$-band light curve showing a bump feature at early times, which was also interpreted as cooling of the shocked stellar envelope \mbox{\citep{2017MNRAS.471.2463B}}.
In comparison with \mbox{SN~2017ein}, \mbox{LSQ14efd} is more luminous and energetic, with an $R$-band peak absolute magnitude of $-$18.14 mag, kinematic energy $E_{\mathrm{kin}}=$5.3 B, and ejecta mass $M_{\mathrm{ej}}=$ 6.3 M$_{\odot}$, respectively, making it stand between normal \mbox{SNe Ic} and SNe Ic-BL.

\section{HST Progenitor Observation}\label{sec:HST}
\subsection{Pre- and post-explosion HST images and data reduction}\label{sec:HSTdata}
We searched the pre-explosion HST images from Mikulski Archive for Space Telescopes (MAST)\footnote{\url{http://archive.stsci.edu/}} and the Hubble Legacy Archive (HLA)\footnote{\url{http://hla.stsci.edu/}}, and found publicly available F555W and F814W WFPC2 images taken on 2007 Dec. 11 (PI: W. Li). A F438W-band image at the location of \mbox{SN~2017ein} was taken by HST on \mbox{2017 Jun. 12} with the WFC3/UVIS configuration (PI: \mbox{van Dyk}).
This image was released and publicly available immediately after the observation.
We use this image as a post-explosion image to identify the position of the SN progenitor. Table \ref{tab:hst-obs} lists the information of these images.

\begin{figure*}[ht!]
\begin{minipage}{1.0\linewidth}
\centering
\includegraphics[width=0.6\linewidth]{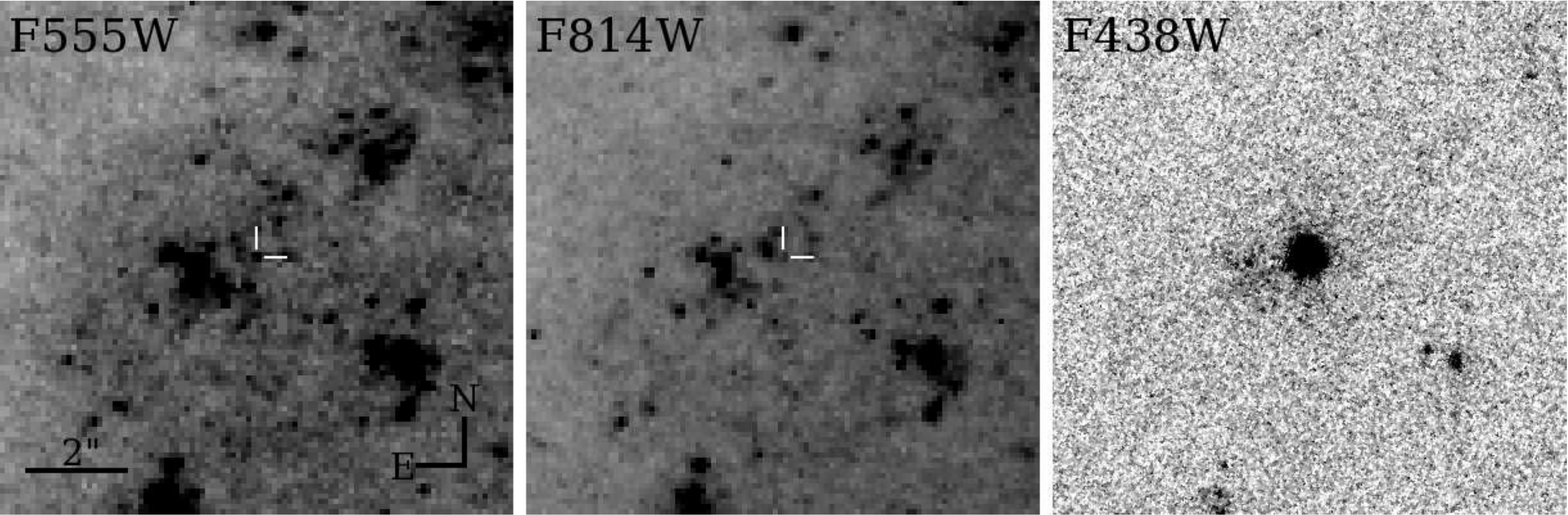}
\end{minipage}

\begin{minipage}{1.0\linewidth}
\centering
\includegraphics[width=0.4\linewidth]{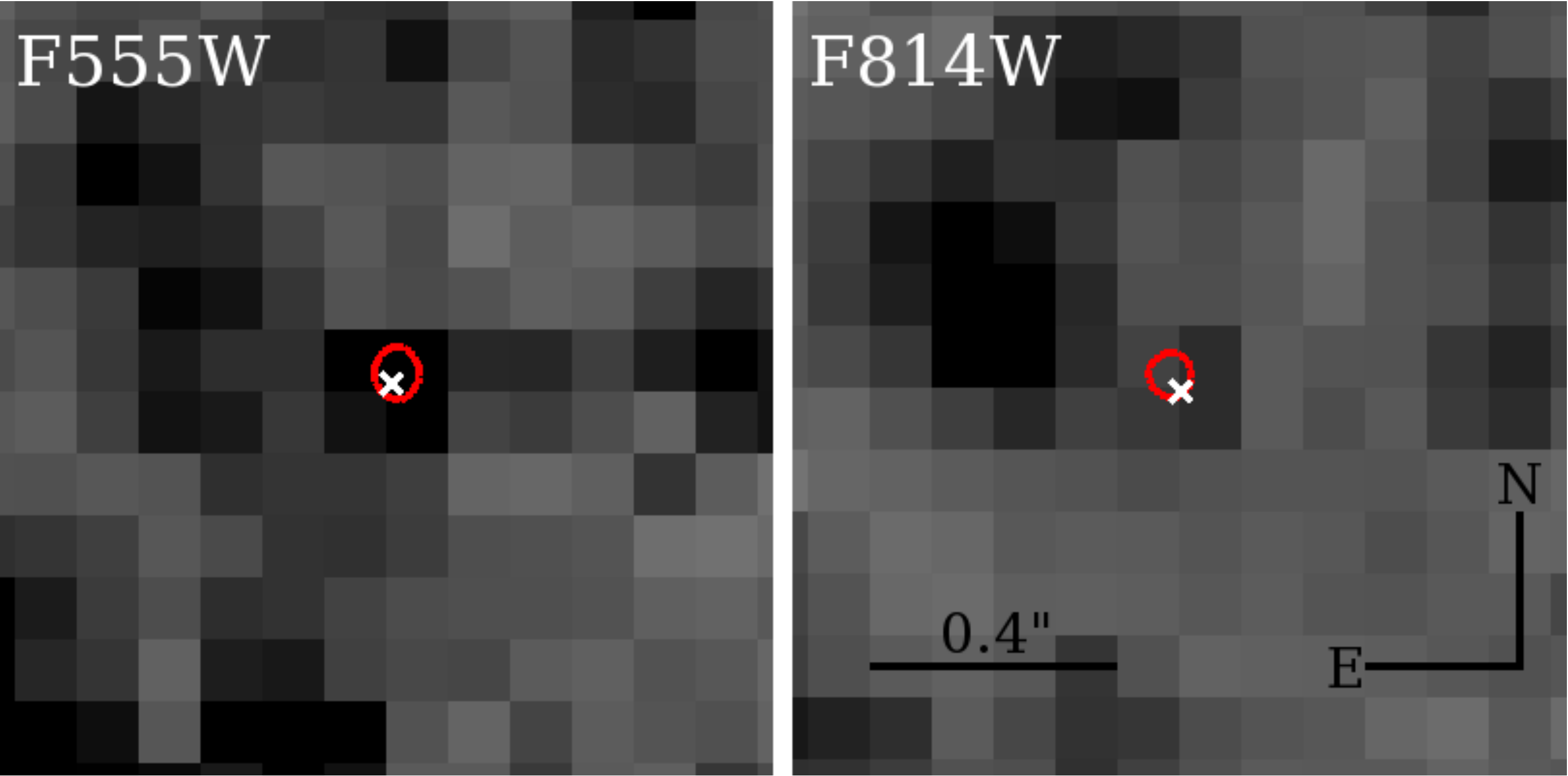}
\end{minipage}
\caption{\textit{Upper panel}: The regions around \mbox{SN~2017ein} showing on the pre-discovery HST F555W (left) and F814W (middle) images, and on the post-explosion HST F438W image (right). The white ticks mark the location of the progenitor star. \textit{Lower panel}: the zoomed-in pre-explosion HST images centering around the progenitor candidate in F555W (left) and F814W (right) filters, with the white crosses marking the center positions of the identified progenitor star. The center of red ellipses shows the SN position, with the radius of the ellipse showing the size of the error. All images are aligned. \label{fig:hstimg}}
\end{figure*}

The pre- and post-discovery HST images around the SN position are shown in Figure \ref{fig:hstimg}, where the zoomed-in regions centered at the SN site are also shown.
There is clearly a point-like source near the SN position in both the F555W and F814W images.
The position of this point-like source is obtained by averaging the measured values with four different methods (centroid, Gauss, $o$-filter center algorithm of IRAF task phot, and IRAF task Daophot).
On all pre-explosion images, the separation between the position transformed from the post-explosion SN site and the center of the nearby point-like source is within the uncertainty of the transformed position, suggesting that this nearby point source is the progenitor of \mbox{SN~2017ein}.
The coordinates of the \mbox{SN} on the post-explosion F438 image are found to be (539.302$\pm$0.002, 564.516$\pm$0.003) with different methods (centroid, Gauss and o-filter center algorithm of IRAF task phot, IRAF task imexamine, and SExtractor).
And are converted to (798.70, 1766.77) and (798.38, 1766.67) on the pre-explosion F555W and F814W images, respectively, as the coordinates of the progenitor.

To get accurate positions of \mbox{SN~2017ein} on the pre-discovery images, we first chose several stars that appeared on all the post- and pre-explosion images and then got their positions on each image using \mbox{SExtractor}.
A second-order polynomial geometric transformation function is applied using the IRAF geomap task to convert their coordinates on the post-explosion image to those on the pre-explosion images.
Based on the \mbox{IRAF} geoxytran task, this established the transformation \mbox{relationship} between the coordinates of \mbox{SN~2017ein} on the post-explosion image and those on the pre-explosion images.
The uncertainties of the transformed coordinates are a combination of the uncertainties in the SN position and the geometric transformation.
The transformed \mbox{SN} and progenitor positions and their uncertainties are listed in Table \ref{tab:pos}.
As one can see, the location of the SN center and the point-like source on the pre-explosion F555W and F814W images have small offsets of $0^{\prime\prime}.020$ and $0^{\prime\prime}.030$, respectively, which are smaller than the uncertainty of the location of the supernova.

We use DOLPHOT\footnote{\url{http://americano.dolphinsim.com/dolphot/}} to get photometry of the progenitor on the pre-explosion images.
The photometry is performed on raw C0M FITS images obtained from the \mbox{MAST} archive.
Magnitudes and their uncertainties of the progenitor candidate are extracted from the output of DOLPHOT.
We find that the progenitor star has magnitudes of 24.46$\pm$0.11$\ \mathrm{mag}$ and 24.39$\pm$0.17$\ \mathrm{mag}$ in F555W band and F814W band, respectively.
Note that our measurements are brighter by about 0.1 mag in F555W and 0.2 mag in F814W, respectively, compared with the results recently reported by \cite{2018ApJ...860...90V}.

To further constrain the properties of the identified point-like source at the SN position, we also studied the photometric properties of the surrounding stars within 5 arc-second, corresponding to a radius of about 400 pc, around the progenitor.
The measured PSF FWHMs of these stars are in a range of $0^{\prime\prime}.15\sim0^{\prime\prime}.20$ in both F555W and F814W bands, which are consistent with that derived for the progenitor (i.e. $0^{\prime\prime}.21$ in F814W).
While these FWHMs of the stars correspond to a radius of $\sim16$pc, which is possible for a compact star cluster.
So we can not be sure about whether the progenitor candidate is a single star, a multi-system, or a star cluster.

\begin{deluxetable*}{ccccccl}
\tablecaption{Pre- and post-explosion HST images on the site of \mbox{SN~2017ein}.\label{tab:hst-obs}}
\tablehead{
\colhead{Instrument} &\colhead{Filter} &\colhead{Obs. date} &\colhead{Exposure time(s)} &\colhead{Proposal ID} &\colhead{PI} &\colhead{Drizzled image}
}
\startdata
WFPC2   & F555W & 2007-12-11 & $230 \times 2$  & 10877  & Li      & hst\_10877\_10\_wfpc2\_f555w\_wf\_drz.fits \\
WFPC2   & F814W & 2007-12-11 & $350 \times 2$  & 10877  & Li      & hst\_10877\_10\_wfpc2\_f814w\_wf\_drz.fits \\
WFC3/UVIS & F438W & 2017-06-12 & $10  \times 27$ & 14645 & van Dyk & id9703010\_drc.fits \\
\enddata
\end{deluxetable*}

\begin{deluxetable}{ccc}
\tablecaption{Transformed SN positions on pre-explosion HST images\label{tab:pos}}
\tablehead{
\colhead{image} &\colhead{F555W} &\colhead{F814W}
}
\startdata
$s_\mathrm{pre}/(\mathrm{arcsec \cdot pixel^{-1}})$\tablenotemark{a} & 0.10 &0.10 \\
$N_\mathrm{stars}$\tablenotemark{b} & 14 & 16 \\
$\sigma_\mathrm{geomap}/\mathrm{pixel}$\tablenotemark{c} & 0.36/0.43 & 0.34/0.36 \\
$\sigma_\mathrm{geomap}/\mathrm{mas}$\tablenotemark{d} & 36/43 & 34/36 \\
$s_\mathrm{post}/(\mathrm{arcsec \cdot pixel^{-1}})$\tablenotemark{e} & \multicolumn{2}{c}{0.04} \\
$\sigma_\mathrm{SN}/\mathrm{mas}$\tablenotemark{f} & \multicolumn{2}{c}{0.1/0.1} \\
$\sigma_\mathrm{total}/\mathrm{mas}$\tablenotemark{g} & 35.8/42.7 & 34.1/35.8 \\
$(x, y)_\mathrm{SN}$\tablenotemark{h} & (798.70, 1766.77) & (798.38, 1766.67) \\
$(x, y)^*$\tablenotemark{i} & (798.60, 1766.60) & (798.56, 1766.43) \\
$\sigma_{*}/\mathrm{mas}$\tablenotemark{j} & 9/5 & 17/7 \\
$d/\mathrm{mas}$\tablenotemark{k} & 20 & 30 \\
\enddata
\tablenotetext{a}{Pixel scale of the pre-explosion image.}
\tablenotetext{b}{Number of stars used in the geometric transformation.}
\tablenotetext{c}{Uncertainty of the geometric transformation in pixel.}
\tablenotetext{d}{Uncertainty of the geometric transformation in milliarcseconds.}
\tablenotetext{e}{Pixel scale of the post-explosion image.}
\tablenotetext{f}{Uncertainty of SN position on the post-explosion image.}
\tablenotetext{g}{Uncertainty of the transformed SN position on pre-explosion image.}
\tablenotetext{h}{The position of the SN transformed on the pre-explosion image.}
\tablenotetext{i}{The position of the progenitor star.}
\tablenotetext{j}{Position uncertainty of the progenitor star.}
\tablenotetext{k}{Separation of the SN from the progenitor star.}
\end{deluxetable}

\subsection{Progenitor properties: a single WR star?}\label{sec:prog-hst}
From the HST photometry and the reddening derived in Section \ref{sec:dist_extinc}, we found that the progenitor star has a very high luminosity with M$_{\mathrm{F555W}}$ = $-$8.17$\pm$0.37 mag and a very blue color of $V-I$=$-$0.47$\pm$0.27 mag.
This indicates that \mbox{SN~2017ein} may have had a blue and \mbox{luminous} progenitor star. Despite the complexity of the progenitor candidate, for simplicity, we first consider it as a single star.

Metallicity is another important parameter to constrain the properties of the progenitor star.
From the measurements of \cite{2016MNRAS.462.1642K}, the metallicity at the site of \mbox{SN~2017ein} is found to be \mbox{$\mathrm{log}(Z/Z_{\odot})\approx0.22$} or 12$+\mathrm{log(O/H)\approx8.91}$\footnote{We adopt the average value of metallicity at the SN location measured by various methods in the literature \citep{2016MNRAS.462.1642K}, and the solar metallicity as 12+log(O/H)=8.69 \citep{2009ARA&A..47..481A}.}.
Such high metallicity for the progenitor is consistent with the general trend that normal SNe Ic tend to be metal rich compared to the subclass of SNe Ic-BL or GRB-SNe \citep{2003ApJ...591..288H}.
Considering the high luminosity and blue color, we assume that it is a WR star.
The lack of H features in the supernova spectra indicates a H poor progenitor.
Radiative transfer models including non-local thermodynamical equilibrium effects of SE SNe \citep[see][and references therein]{2012MNRAS.422...70H} suggest that only $\sim$0.1 M$_{\odot}$ of helium is needed in the progenitor star to produce an observable signature in the spectra.
As we do not see any significant helium features in \mbox{SN~2017ein}, its progenitor must also be helium poor.
Thus, we use the model spectra grids of Milky Way-like WC-type WR stars \citep{2012A&A...540A.144S} to fit the photometry data of the progenitor star.
The model spectra are characterized by two parameters: the effective temperature $T_{\mathrm{eff}}$ and the ``transformed radius'' $R_t$, which is defined by
\begin{equation}
R_t=R_*[\frac{v_{\infty}}{\mathrm{km\ s^{-1}}}/\frac{\dot{M}\sqrt{D}}{10^{-4}M_{\odot}\mathrm{yr}^{-1}}]^{2/3}
\end{equation}
Where $\dot{M}$ is the mass loss rate of the star, $v_{\infty}$ is the terminal wind velocity ($v_{\infty}=2000\ \mathrm{km\ s^{-1}}$ for WC \mbox{stars}), and $D$ is the so-called density contrast, which is set to be 10 for WC stars.
The flux of model spectra is normalized so that the luminosity of the star is $10^{5.3}$ times that of the sun.

Due to the fact that we only have observation in two bands, we use a MC simulation to better estimate the range of the stellar parameters.
We created a sample of fluxes in F555W and F814W bands using the measured magnitudes and extinction to the SN.
Then we fit the color index F555W-F814W to that of the model spectra to determine the effective temperature.
The radius and luminosity of the star are then determined using the scale factor between the apparent flux in F555W and F814W bands and that of the model SED.
Finally, we get distributions of effective temperature, luminosity and radius of the progenitor candidate.
By taking the full-width at half maximum (FWHM) of the distributions of the parameters, this analysis gives estimates of an effective temperature range of $\sim$70,800--15,900 K, a luminosity of $6.1\lesssim\mathrm{log}L/L_{odot}\lesssim7.3$, and a stellar radius of \mbox{6$\sim$14 R$_{\odot}$}.
We caution, however, that current estimates of the progenitor parameters contain considerable uncertainties due to large errors associated with the photometry, distance, and extinction of the progenitor star.
Moreover, for WR stars with very high temperature, the radiation in the optical is only a small proportion of \mbox{its} total luminosity \citep{2007ARA&A..45..177C}, and this will also give rise to a large uncertainty in determining the temperature from only two optical bands like F555W and F814W.

\begin{figure}[ht!]
\centering
\includegraphics[width=1.0\linewidth]{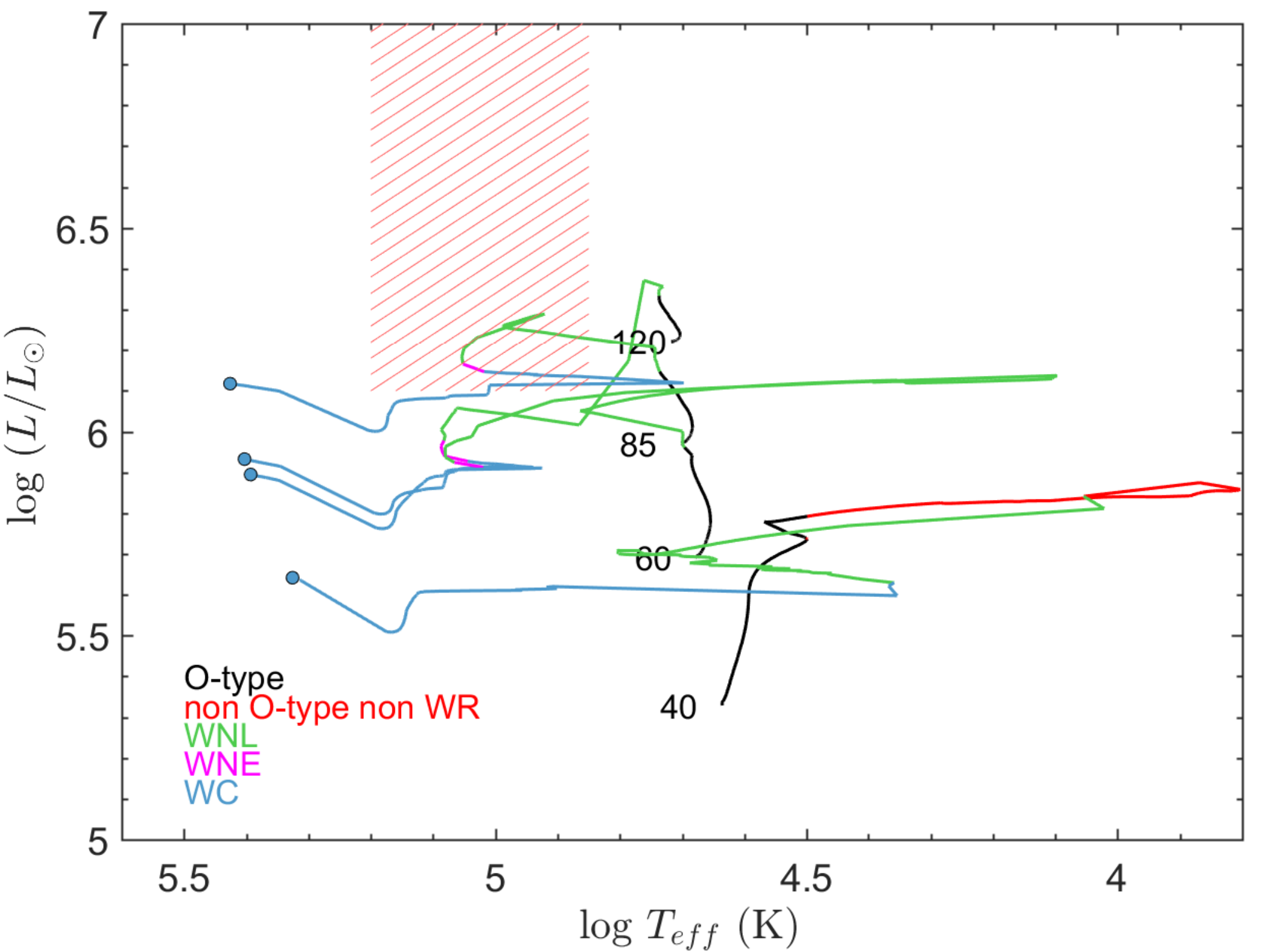}
\caption{Hertzsprung-Russel diagram showing the locus of the progenitor of \mbox{SN~2017ein}. The evolutionary tracks of massive stars at solar metallicity from \cite{2012A&A...542A..29G} are displayed for comparison. The numbers in the plot mark the corresponding ZAMS mass in unit of solar mass. The endpoints of the tracks are indicated by dots. Different colors on the lines represent different evolutionary stages of the stars.}
\label{fig:evol}
\end{figure}

Figure \ref{fig:evol} shows the stellar evolution tracks of massive stars at solar metallicity \citep{2012A&A...542A..29G}, with the locus of the progenitor of \mbox{SN~2017ein} overplotted.
These evolution models take rotational effects into account, with an initial rate of $v_{ini}/v_{crit}=0.4$ \citep[see also the model descriptions in][]{2012A&A...537A.146E}.
Different colors of tracks indicate different stellar types, which are determined following the definitions proposed by \cite{2003A&A...404..975M} and \cite{1991A&A...241...77S}.
Inspecting Figure \ref{fig:evol}, we propose that the initial mass of the progenitor star of \mbox{SN~2017ein} was larger than 60 M$_{\odot}$.
According to these stellar evolution models, the final CO core mass can be $\sim$17.5 M$_{\odot}$ for massive stars with initial mass of 60 M$_{\odot}$, and the explosion as SN Ic can produce an ejecta mass of \mbox{$>$10 M$_{\odot}$}, with a black hole remnant.
This expected ejecta mass is rather large compared to the value we inferred for \mbox{SN~2017ein}.
However, this mass estimate may be too large, although the progenitor radius that we get from the SED fitting is not far from the value given by the shock cooling fitting.
Given the uncertainty in deriving the bolometric luminosity, it is likely that the progenitor luminosity is somewhat overestimated due to the lack of multiband photometry data especially at shorter wavelengths.

\subsection{Binary, star cluster and low ejecta mass}\label{sec:discussionprog}
\begin{figure}[ht!]
\centering
\includegraphics[width=1.1\linewidth]{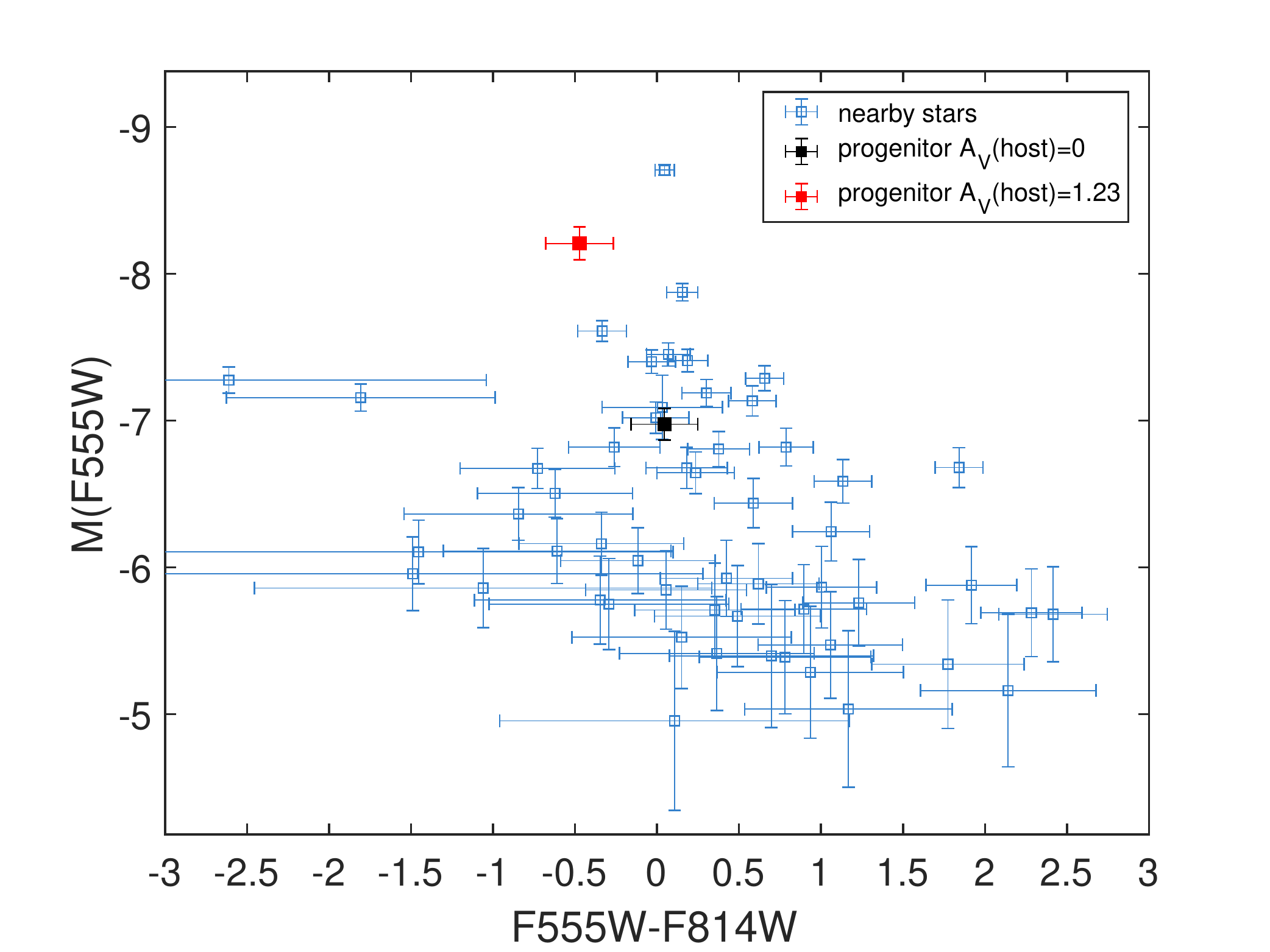}
\caption{A color--absolute magnitude diagram of the progenitor candidate of \mbox{SN~2017ein} and its surrounding stars at a distance within 5 arcseconds. The absolute magnitudes and color index of the surrounding stars are corrected for Galactic extinction but not the host extinction. A distance modulus of $\mu=$31.38 mag is adopted. }
\label{fig:hr-diagram}
\end{figure}

As an alternative, the high luminosity of the progenitor may be related to the contribution of a similarly luminous companion star.
According to observations of WR stars in the Milky Way and Large Magellanic Cloud (LMC), one finds that more than 80\% of luminous stars, i.e., M$_{V}$ $<$ $-$5.5 mag, are found to be binaries, while this fraction drops to about 50\% for the relatively faint counterparts \citep{2014A&A...565A..27H, 2006A&A...458..453V}.
After examining the surrounding stars within a radius of 5 arc-seconds from the progenitor of \mbox{SN~2017ein} on the pre-explosion images, we find that the progenitor \mbox{star} does not stand out from them but clearly belongs to the relatively blue and luminous subgroup before subtracting host extinction, as shown in Figure \ref{fig:hr-diagram}.
This, perhaps, favors  a multi-system or star cluster  as the  source.
In this case, we  expect to see a bright star at the SN position when the supernova fades away in a few years.

Binary interaction can result in more efficient mass loss for massive stars, resulting in lower ejecta masses in supernovae.
\cite{2016MNRAS.457..328L} found that explosions of stars with initial mass larger than $28$ M$_{\odot}$ will produce ejecta of mass not lower than 5 M$_{\odot}$, and even stars with initial mass $\geqslant$ 20 M$_{\odot}$ in binaries will produce ejecta of mass higher than 5 M$_{\odot}$.
Thus, lower-mass binary models are in better agreement with the distribution of ejecta mass of SNe Ib/c.
Considering the lower ejecta mass of $\sim$1 M$_{\odot}$ for \mbox{SN~2017ein}, it is more likely that its progenitor star is a lower mass star in a binary system rather than a single star with such a high initial mass.
As discussed in \cite{2018ApJ...860...90V}, the progenitor is consistent with massive binary stars.
\cite{2018MNRAS.480.2072K} has came to similar conclusions,  prefering a binary model of 80$+$48 M$_{\odot}$.
However, the expected \mbox{ejecta} mass is still higher than 5 M$_{\odot}$.

These results have demonstrated a connection between \mbox{SN~2017ein} and very massive progenitor stars, conflicting with the low ejecta mass.
This may be related to mass loss in massive stars.
The mass loss of massive \mbox{stars} dominates their evolution and determines their fate \citep{2014ARA&A..52..487S}. However, the mass loss of massive stars is quite complicated.
In addition to mass loss through strong and relatively mild stellar wind as WR stars, massive \mbox{stars} may experience eruptive mass loss, e.g. Luminous Blue Variables (LBVs).
And some stars undergo explosive mass loss shortly (a few years) before explosion \citep{2014ApJ...789..104O, 2014Natur.509..471G, 2017NatPh..13..510Y}.
Circumstellar dust can give rise to large uncertainty in observations of massive stars.
Some WR systems reveal persistent or episodic dust formation within months or years, resulting in strong optical or IR variations with an amplitude of a few magnitudes of the WR binaries, due to the newly formatted dust or the asymmetry of the dust distribution \citep[eg.][]{2009MNRAS.395.1749W}. This dust will eventually be ejected to the interstellar medium (ISM).

On the other hand, as also proposed by \cite{2018ApJ...860...90V}, the supernova explosion can also disrupt and vaporize circumstellar dust.
These facts demonstrate that the extinction derived from the Na {\footnotesize I} D lines in the SN spectra, which mostly represent extinction from the interstellar dust, rather than the dust surrounding the star, can not be directly applied to the progenitor.
The extinction to the SN should be a lower limit of the true extinction to the progenitor.
As a result, the progenitor could be even bluer and brighter.
Combining the effects of dust formation in massive stars and destruction by the SN explosion, the real extinction to the progenitor can be very complicated.
Detailed discussions on the mass loss and dust formation and disruption in \mbox{SN~2017ein} is beyond the scope of this paper.

\begin{figure*}[htbp]
    \centering
    \includegraphics[width=0.8\linewidth]{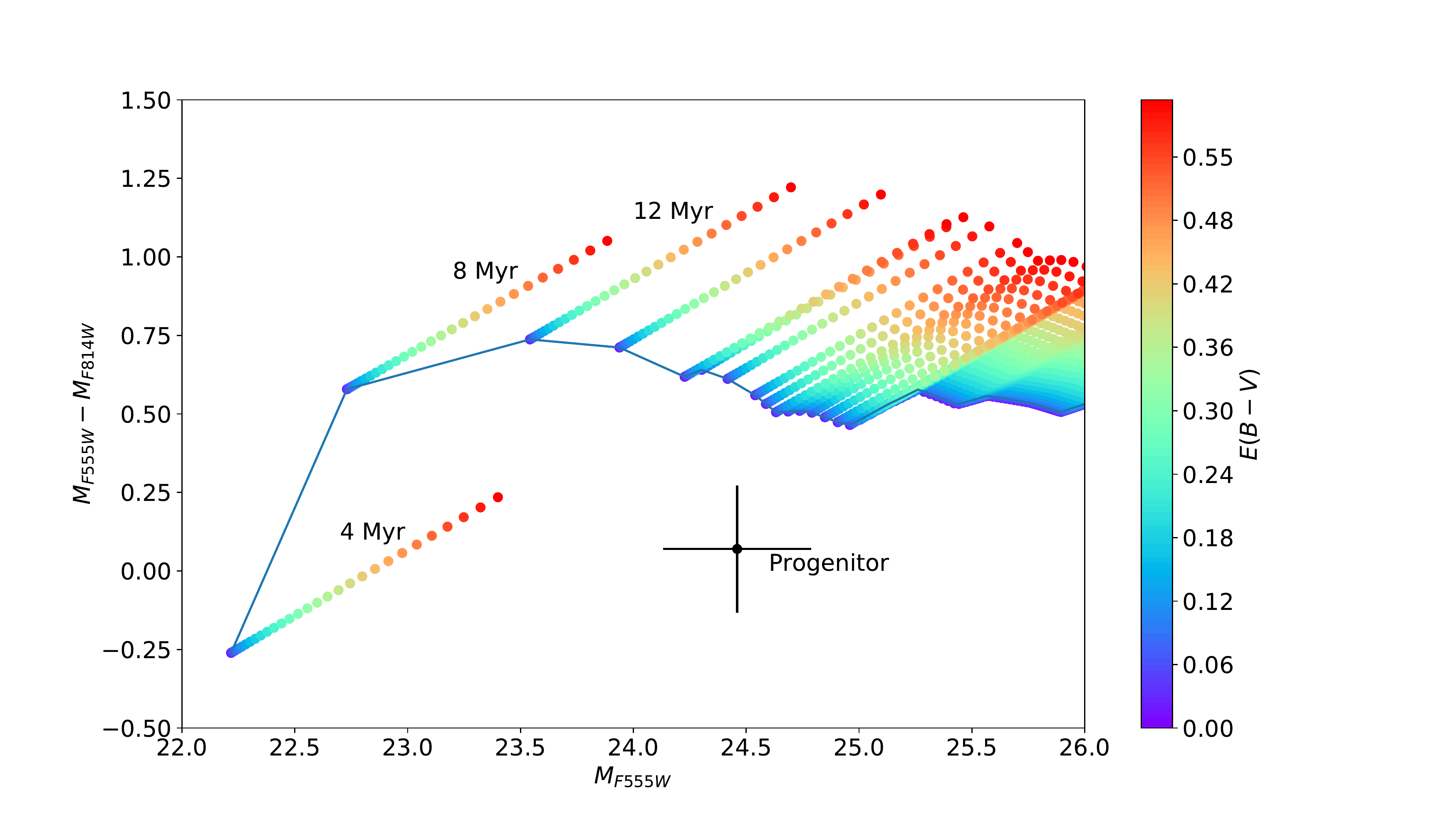}
    \caption{A set of cluster models from GALEV, for a $1\times10^4$ M$_{\odot}$ cluster at solar metallicty. The color of each point denotes the assumed total reddening, while the line connects the $E(B-V)=0$ cluster models to guide the eye. All tracks have placed at the distance of SN~2017ein. The observed progenitor color and magnitude is indicated.}
    \label{fig:cluster}
\end{figure*}
On the other hand, the progenitor candidate could also be a young and blue star cluster, as also pointed out by \cite{2018ApJ...860...90V}.
As noted previously, a $\lesssim$16 pc cluster would be unresolved in the pre-explosion data.
To explore this possibility, we created a set of models for a $10^{4}$ M$_{\odot}$ cluster with the GALEV code \citep{2009MNRAS.396..462K}.
We assume a single stellar population at solar metallicity, formed in a exponentially declining burst.
In Figure \ref{fig:cluster}, we compare the colour and magnitude of such a cluster to the source observed to be coincident with the SN.
The relatively blue color of the source implies that the progenitor must be $\sim4$ Myr old.
The absolute magnitude of the observed source is much fainter than the model cluster, implying that it must be substantially less massive.
Naively scaling our model magnitudes to the observed value, we find that the cluster would have a mass $\sim 2.5 \times10^{3}$ M$_{\odot}$.
As a test of our results, we also compared to the colours of a set of BPASS models \cite[v 2.2.1,][]{2018MNRAS.479...75S}, and found similar results.

Such a young cluster age implies that in this case the progenitor would have to be relatively massive.
Taking the lifetimes of single massive stars from the STARS code \citep{2008MNRAS.384.1109E}, we find that the progenitor would of necessity have a zero age main sequence mass of $\gtrsim$40 M$_{\odot}$.
Such a star would in fact lose its H envelope through stellar winds, and explode as a single WR star.
For a Salpeter initial mass function between 0.1 and 300 M$_{\odot}$, and assuming a cluster of 2500 M$_{\odot}$, then only around 160 M$_{\odot}$ of the cluster mass will be in stars more massive than 40 M$_{\odot}$.
So, if this were indeed the case, then the progenitor was one of the most massive members of the cluster.

The preceding analysis must be taken with considerable caution: with only two broadband filters it is extremely difficult to draw firm conclusions as to the properties of any possible cluster, especially when young clusters of hot massive stars will emit much of their flux at bluer wavelengths than F555W.
We also caution that we cannot even be certain as to whether we are observing a single population -- even a small 16 pc region could plausibly play host to multiple overlapping stellar populations.
Future observations, and in particular in the UV, can help resolve this uncertainty.

\section{Conclusions}\label{sec:summary}
We present the optical/ultraviolet observations of the nearby type Ic supernova \mbox{SN~2017ein} and analysis of the properties of the supernova and its progenitor, by taking advantages of both our extremely early-time photometry/spectroscopy and the public HST pre-explosion data.
\mbox{SN~2017ein} is similar to SN 2007gr and SN 2013ge in overall spectral evolution, all characterized by prominent carbon features in the early phase, while \mbox{SN~2017ein} clearly shows some distinct features from the latter two.
The expansion velocities of different lines of \mbox{SN~2017ein} are similar to the other two, but the Na {\footnotesize I} and Si {\footnotesize II} expansion velocities of \mbox{SN~2017ein} are higher than those of SN 2007gr, but comparable to those of \mbox{SN 2013ge}.
Another interesting feature is that the absorption lines of the Ca{\footnotesize II} NIR triplet split into two narrow lines at late phases, which was not seen in other SNe Ib/c including SNe 2007gr and 2013ge.
This probably suggests that the line forming region of \mbox{SN~2017ein} is confined into a relatively narrow shell.
Moreover, the late time light curves of \mbox{SN~2017ein} are fast declining, which is not seen in other SNe Ib/c.

From the very early photometry and non detection limit, we were able to constrain the first-light time of \mbox{SN~2017ein} to be within one day after explosion, representing one of the tightest constraints for SNe Ib/c.
Most interestingly, the early data show a possible excess emission relative to the Ni decay model at one day after explosion, which can be attributed to the shock cooling tail.
By fitting the $BVR$-band light curves, using a radioactive decay model and a shock cooling model, we find that the supernova synthesized $^{56}$Ni mass of \mbox{$\sim$0.13 M$_{\odot}$}, and the stellar envelope has a radius of \mbox{8$\pm$4 R$_{\odot}$} and a mass of \mbox{$\sim$0.02 M$_{\odot}$}.
Note that a smaller ejecta mass of $\sim$1 M$_{\odot}$ is found for \mbox{SN~2017ein}.
This indicates that the progenitor star has lost most of its H/He envelope before collapse, consistent with a compact Wolf-Rayet star or a massive binary.
\mbox{SN~2017ein} perhaps represents the first normal SN Ic observed to have a signature of shock breakout cooling from the stellar envelope. \mbox{Undoubtedly}, it provides a very good sample to test theories of \mbox{SN Ic} explosions.

By analyzing the pre-explosion HST images, we find a point-like source with an absolute magnitude of $M_V=-8.00\pm0.34$ mag and $M_I=-7.53\pm0.35$ mag coincides with the position of \mbox{SN~2017ein}.
Assuming that the source is a single star, it is found to have an effective temperature of $\sim$70,800--159,000 K, a luminosity of $\mathrm{log}(L/L_{\odot})\gtrsim6.1$, and a radius of $\sim$6--14 R$_{\odot}$, roughly consistent with a hot and luminous WR star.
Comparing with stellar evolutionary tracks, we found that the initial mass of this star could exceed 60 M$_{\odot}$. Such a hot and luminous star is very rare and we did not find a similar star in current WR star catalogs.
However, such massive star is inconsistent with the low ejecta mass found for SN 2017ein.
On the other hand, it is also possible that the point source identified at the SN position could be a binary or a young, blue star cluster.
The low ejecta mass inferred for \mbox{SN~2017ein} seems to favor for a low mass binary scenario.
By comparing the observational data and SSP star cluster models at solar metallicity, we find that the source is consistent with star clusters with age of 4 Myr and a low mass of $\sim2.5\times10^3$ M$_{\odot}$.
And the progenitor of \mbox{SN~2017ein} should be among the most massive members of the cluster.
Due to the low resolution of the HST images and the lack of data in shorter wavelengths, it is difficult to decide which case is favored.
Further distinguishing between different scenarios needs to wait for another visit of this galaxy with HST when the SN light dims out.
SN 2017ein provides a rare opportunity that  helps to throw light on massive star evolution and supernova explosion theories.

\acknowledgments

We acknowledge the support of the staff of the \mbox{Xinglong} 2.16~m and Lijiang 2.4~m telescope. This work is supported by the National Natural Science Foundation of China (NSFC grants 11178003, 11325313, and 11633002), the National Program on Key Research and Development Project (grant no. 2016YFA0400803), and the Tsinghua University Initiative Scientific Research Program. J.-J. Zhang is supported by the NSFC (grants 11403096, 11773067), the Key Research Program of the CAS (Grant NO. KJZD-EW- M06), the Youth Innovation Promotion Association of the CAS (grants 2018081), and the CAS ``Light of West China'' Program. T.-M. Zhang is supported by the NSFC (grants 11203034). This work was also partially Supported by the Open Project Program of the Key Laboratory of Optical Astronomy, National Astronomical Observatories, Chinese Academy of Sciences. Funding for the LJT has been provided by Chinese Academy of Sciences and the People's Government of Yunnan Province. The LJT is jointly operated and administrated by Yunnan Observatories and Center for Astronomical Mega-Science, CAS.

We thank the staff of AZT and ATLAS for their observations and allowance of the use of the data.
ATLAS is funded by NASA grant NN12AR55G under the Near Earth Objects Observations program (NEOO).
L.J.W. acknowledges the support of the National Program on Key Research and Development Project of China (Grant No. 2016YFA0400801).
This work makes use of \mbox{data} from the Las Cumbres Observatory network and the Global Supernova Project. DAH, CM, and GH are supported by the US National Science Foundation (NSF) grant AST-1313484. EB and JD are supported in part by NASA Grant NNX16AB256. MF is supported by a Royal Society - Science Foundation Ireland University Research Fellowship. J.V. is supported by the project ``Transient Astrophysical Objects'' GINOP 2.3.2-15-2016-00033 of the National Research, Development and Innovation Office (NKFIH), Hungary, funded by the European Union. K. Nomoto, A. Tolstove, and S. Blinnikov are supported by the World Premier International Research Center Initiative (WPI Initiative), MEXT, Japan, and JSPS KAKENHI (grant numbers JP16K17658, JP26400222, JP16H02168, JP17K05382). S.B. acknowledges funding from project RSF 18-12-00522.
\software{ZrutyPhot (Mo et al. in prep.), \mbox{SYNAPPS} \citep{2011PASP..123..237T}, IRAF \citep{1993ASPC...52..173T, 1986SPIE..627..733T}, DOLPHOT \citep{2016ascl.soft08013D}, SExtractor \citep{1996A&AS..117..393B}, GALEV \citep{2009MNRAS.396..462K}, STARS \citep{2008MNRAS.384.1109E}.}

\bibliography{ref}
\input{BVRI}
\input{atlas}
\input{swift}

\end{document}

%% file: BVRI.tex
\startlongtable
\begin{deluxetable*}{lccccc}
\tablecaption{Optical \textit{BVRI} photometry of SN 2017ein.\label{tab:lcBVRI}}
\tablehead{\colhead{MJD} &\colhead{\textit{B}}&\colhead{\textit{V}}&\colhead{\textit{R}}&\colhead{\textit{I}}&\colhead{Telescope}}
\startdata
57904.605   &16.438$\pm$0.026   &15.885$\pm$0.013   &15.571$\pm$0.030   &15.245$\pm$0.017   &TNT\\
57905.604   &16.270$\pm$0.044   &15.715$\pm$0.022   &15.390$\pm$0.030   &15.076$\pm$0.020   &TNT\\
57906.598   &16.083$\pm$0.026   &15.506$\pm$0.016   &15.208$\pm$0.032   &14.912$\pm$0.021   &TNT\\
57907.641   &...                &...                &15.165$\pm$0.016   &14.835$\pm$0.032   &TNT\\
57908.574   &15.929$\pm$0.028   &15.347$\pm$0.014   &15.057$\pm$0.029   &14.735$\pm$0.015   &TNT\\
57911.612   &15.975$\pm$0.019   &15.221$\pm$0.011   &14.881$\pm$0.028   &14.531$\pm$0.015   &TNT\\
57912.592   &16.038$\pm$0.022   &15.209$\pm$0.012   &14.852$\pm$0.030   &14.496$\pm$0.014   &TNT\\
57917.607   &16.483$\pm$0.030   &15.312$\pm$0.012   &14.851$\pm$0.029   &14.457$\pm$0.016   &TNT\\
57919.573   &16.644$\pm$0.029   &15.436$\pm$0.016   &14.928$\pm$0.033   &14.490$\pm$0.016   &TNT\\
57921.607   &16.910$\pm$0.055   &15.542$\pm$0.019   &15.034$\pm$0.032   &14.547$\pm$0.026   &TNT\\
57930.549   &17.678$\pm$0.035   &16.280$\pm$0.018   &15.626$\pm$0.037   &14.982$\pm$0.020   &TNT\\
57932.560   &...                &...                &15.773$\pm$0.079   &...                &TNT\\
57933.566   &...                &16.591$\pm$0.134   &15.855$\pm$0.053   &15.258$\pm$0.071   &TNT\\
57936.713   &18.190$\pm$0.021   &16.790$\pm$0.016   &16.084$\pm$0.012   &15.338$\pm$0.013   &AZT\\
57937.712   &18.184$\pm$0.026   &16.804$\pm$0.016   &16.127$\pm$0.008   &...                &AZT\\
57938.726   &18.150$\pm$0.029   &16.831$\pm$0.017   &16.182$\pm$0.008   &15.460$\pm$0.013   &AZT\\
57940.694   &18.334$\pm$0.047   &16.921$\pm$0.020   &16.257$\pm$0.010   &15.512$\pm$0.019   &AZT\\
57941.711   &18.344$\pm$0.021   &16.958$\pm$0.017   &16.287$\pm$0.008   &15.557$\pm$0.014   &AZT\\
57942.723   &18.325$\pm$0.040   &16.944$\pm$0.023   &16.330$\pm$0.010   &15.562$\pm$0.013   &AZT\\
57945.685   &18.121$\pm$0.116   &17.044$\pm$0.051   &16.365$\pm$0.011   &15.598$\pm$0.013   &AZT\\
57946.721   &18.606$\pm$0.110   &17.001$\pm$0.022   &16.364$\pm$0.011   &15.600$\pm$0.013   &AZT\\
57949.687   &18.335$\pm$0.072   &17.152$\pm$0.021   &16.488$\pm$0.013   &15.683$\pm$0.021   &AZT\\
57950.695   &18.556$\pm$0.028   &17.093$\pm$0.017   &16.503$\pm$0.011   &15.678$\pm$0.014   &AZT\\
57951.695   &18.431$\pm$0.018   &17.171$\pm$0.015   &16.551$\pm$0.008   &15.735$\pm$0.022   &AZT\\
57952.684   &18.500$\pm$0.094   &17.210$\pm$0.020   &16.544$\pm$0.012   &15.732$\pm$0.019   &AZT\\
57953.678   &18.524$\pm$0.030   &17.227$\pm$0.024   &16.609$\pm$0.016   &15.761$\pm$0.015   &AZT\\
57954.686   &18.600$\pm$0.018   &17.239$\pm$0.019   &16.621$\pm$0.015   &15.788$\pm$0.018   &AZT\\
57955.686   &18.684$\pm$0.102   &17.277$\pm$0.016   &16.634$\pm$0.010   &15.804$\pm$0.016   &AZT\\
57956.693   &18.800$\pm$0.120   &17.269$\pm$0.026   &16.615$\pm$0.017   &15.796$\pm$0.017   &AZT\\
57957.674   &18.793$\pm$0.109   &17.324$\pm$0.033   &16.686$\pm$0.020   &15.881$\pm$0.014   &AZT\\
57958.683   &18.715$\pm$0.049   &17.338$\pm$0.024   &16.715$\pm$0.009   &15.888$\pm$0.013   &AZT\\
57960.683   &18.642$\pm$0.166   &17.296$\pm$0.017   &16.696$\pm$0.022   &15.852$\pm$0.013   &AZT\\
57963.670   &18.460$\pm$0.039   &17.449$\pm$0.046   &16.786$\pm$0.011   &15.908$\pm$0.016   &AZT\\
57964.667   &18.615$\pm$0.150   &17.367$\pm$0.020   &16.820$\pm$0.008   &15.921$\pm$0.017   &AZT\\
57965.667   &18.820$\pm$0.028   &17.410$\pm$0.034   &16.827$\pm$0.015   &15.969$\pm$0.016   &AZT\\
58107.022   &21.858$\pm$1.412   &20.151$\pm$0.098   &19.639$\pm$0.045   &19.456$\pm$0.239   &AZT\\
58124.882   &21.636$\pm$1.494   &22.635$\pm$1.440   &19.852$\pm$0.108   &21.230$\pm$1.004   &AZT\\
58133.075   &20.097$\pm$1.340   &19.113$\pm$0.391   &19.432$\pm$0.128   &19.826$\pm$0.242   &AZT\\
58139.040   &20.039$\pm$0.322   &22.644$\pm$1.036   &21.158$\pm$0.104   &20.422$\pm$0.098   &AZT\\
58139.962   &20.581$\pm$0.195   &...                &21.487$\pm$1.077   &20.267$\pm$0.519   &AZT\\
58140.954   &...                &21.599$\pm$0.905   &21.372$\pm$1.064   &20.900$\pm$0.605   &AZT\\
58143.896   &19.676$\pm$0.611   &23.596$\pm$1.241   &20.507$\pm$0.286   &20.018$\pm$0.346   &AZT\\
58150.919   &20.158$\pm$1.441   &21.834$\pm$0.605   &19.399$\pm$0.246   &21.078$\pm$0.933   &AZT\\
58152.898   &...                &21.061$\pm$0.673   &20.245$\pm$0.280   &...                &AZT\\
58154.958   &21.784$\pm$0.727   &21.371$\pm$0.196   &20.674$\pm$0.194   &...                &AZT\\
58156.966   &...                &20.820$\pm$0.124   &20.259$\pm$0.082   &20.318$\pm$0.293   &AZT\\
58167.949   &20.121$\pm$0.756   &20.633$\pm$0.149   &21.456$\pm$0.336   &19.844$\pm$0.425   &AZT\\
58170.917   &19.208$\pm$0.393   &24.017$\pm$1.374   &20.768$\pm$0.668   &21.959$\pm$0.550   &AZT\\
58171.904   &20.520$\pm$0.840   &21.682$\pm$0.310   &21.252$\pm$0.136   &20.347$\pm$0.227   &AZT\\
\enddata
\end{deluxetable*}

%% file: atlas.tex
\startlongtable
\begin{deluxetable*}{lccc|lccc}
\tablecaption{ATLAS photometry of SN 2017ein \label{tab:atlas}}
\tablehead{
\colhead{MJD} &\colhead{AB mag} &\colhead{err} &\colhead{Filter} &\colhead{MJD} &\colhead{AB mag} &\colhead{err} &\colhead{Filter}}
\startdata
57869.506  &>20.06 &1.35  &c &57907.281  &15.17  &0.01  &o\\
57869.520  &>20.01 &0.53  &c &57907.296  &15.16  &0.01  &o\\
57869.535  &>19.87 &2.18  &c &57908.269  &15.29  &0.11  &o\\
57869.550  &>19.67 &0.00  &c &57908.284  &15.25  &0.05  &o\\
57870.425  &>20.13 &0.00  &c &57908.294  &15.17  &0.04  &o\\
57870.437  &>20.19 &0.00  &c &57911.261  &14.84  &0.01  &o\\
57870.444  &>20.21 &0.45  &c &57911.284  &14.84  &0.01  &o\\
57870.459  &>20.11 &0.00  &c &57911.300  &14.84  &0.01  &o\\
57870.482  &>20.12 &0.56  &c &57912.260  &14.83  &0.01  &o\\
57882.354  &>19.34 &0.61  &o &57912.262  &14.84  &0.01  &o\\
57882.380  &>19.52 &0.00  &o &57912.274  &14.80  &0.01  &o\\
57882.391  &>19.54 &0.43  &o &57912.291  &14.79  &0.01  &o\\
57882.415  &>19.44 &2.92  &o &57912.291  &14.80  &0.01  &o\\
57887.280  &>19.83 &0.00  &o &57912.304  &14.85  &0.01  &o\\
57887.296  &>19.80 &0.73  &o &57912.305  &14.83  &0.01  &o\\
57887.307  &>19.47 &0.00  &o &57914.269  &14.93  &0.01  &o\\
57887.314  &>19.64 &0.00  &o &57914.271  &14.93  &0.01  &o\\
57887.320  &>19.33 &0.00  &o &57914.290  &14.88  &0.01  &o\\
57887.335  &>19.59 &0.97  &o &57914.299  &14.90  &0.01  &o\\
57887.339  &>19.64 &0.68  &o &57914.302  &14.92  &0.01  &o\\
57887.354  &>19.62 &1.02  &o &57914.304  &14.90  &0.01  &o\\
57892.322  &>19.95 &0.81  &o &57914.318  &14.94  &0.01  &o\\
57892.324  &>19.72 &0.00  &o &57914.327  &14.89  &0.01  &o\\
57892.337  &>19.75 &1.85  &o &57914.335  &14.91  &0.01  &o\\
57894.333  &>20.24 &0.00  &c &57915.310  &14.83  &0.01  &o\\
57894.356  &>20.02 &0.00  &c &57915.333  &14.81  &0.01  &o\\
57894.370  &>20.09 &0.97  &c &57915.355  &14.82  &0.01  &o\\
57894.390  &>20.07 &1.40  &c &57918.269  &14.86  &0.01  &o\\
57894.404  &>20.11 &0.00  &c &57918.280  &14.87  &0.01  &o\\
57896.252  &>20.04 &1.41  &o &57922.260  &15.00  &0.01  &o\\
57896.267  &>20.07 &0.00  &o &57922.270  &15.03  &0.02  &o\\
57896.283  &>19.61 &0.66  &o &57925.250  &15.15  &0.01  &o\\
57898.336  &18.43  &0.10  &c &57925.260  &15.16  &0.01  &o\\
57898.354  &18.69  &0.14  &c &57925.278  &15.17  &0.01  &o\\
57898.362  &18.53  &0.11  &c &57925.283  &15.15  &0.01  &o\\
57898.384  &18.35  &0.10  &c &57927.278  &15.38  &0.01  &o\\
57898.390  &18.25  &0.10  &c &57927.287  &15.40  &0.01  &o\\
57900.255  &16.98  &0.05  &o &57929.259  &15.55  &0.01  &o\\
57900.262  &16.92  &0.04  &o &57929.262  &15.53  &0.01  &o\\
57900.267  &17.07  &0.05  &o &57929.277  &15.52  &0.01  &o\\
57900.280  &16.83  &0.04  &o &57929.286  &15.52  &0.01  &o\\
57900.284  &17.02  &0.05  &o &57933.259  &15.75  &0.01  &o\\
57900.313  &17.05  &0.05  &o &57933.262  &15.80  &0.02  &o\\
57903.256  &15.96  &0.02  &o &57933.276  &15.74  &0.01  &o\\
57903.270  &15.88  &0.02  &o &57933.287  &15.79  &0.02  &o\\
57903.285  &15.81  &0.02  &o &57937.258  &16.00  &0.02  &o\\
57903.299  &15.81  &0.02  &o &57937.263  &16.01  &0.02  &o\\
57907.259  &15.17  &0.01  &o &57937.273  &16.04  &0.02  &o\\
57907.264  &15.18  &0.01  &o &57937.288  &16.03  &0.02  &o\\
\enddata
\end{deluxetable*}

%% file: swift.tex
\begin{deluxetable*}{lcccc}
\tablecaption{\textit{Swift}-UVOT photometry of SN 2017ein.\label{tab:swift}}
\tablehead{\colhead{MJD} &\colhead{uvw1} &\colhead{\textit{u}} &\colhead{\textit{b}} &\colhead{\textit{v}}}
\startdata
57900.510   &...               &17.619$\pm$0.213   &17.943$\pm$0.182   &17.133$\pm$0.181 \\
57911.922   &17.674$\pm$0.176  &15.980$\pm$0.084   &15.966$\pm$0.069   &15.228$\pm$0.070 \\
57915.465   &17.713$\pm$0.189  &16.812$\pm$0.123   &16.188$\pm$0.075   &15.221$\pm$0.072 \\
57919.892   &...               &17.473$\pm$0.171   &16.643$\pm$0.084   &15.401$\pm$0.074 \\
57923.103   &...               &18.238$\pm$0.288   &17.094$\pm$0.099   &15.737$\pm$0.080 \\
57927.849   &...               &18.500$\pm$0.368   &17.775$\pm$0.149   &16.292$\pm$0.105 \\
57933.992   &...               &...                &18.114$\pm$0.224   &16.764$\pm$0.162 \\
57937.877   &...               &...                &18.518$\pm$0.256   &16.796$\pm$0.138 \\
57942.460   &...               &...                &18.409$\pm$0.240   &16.928$\pm$0.154 \\
57945.483   &...               &...                &18.424$\pm$0.250   &17.078$\pm$0.174 \\
57949.869   &...               &...                &18.382$\pm$0.346   &16.699$\pm$0.196 \\
57953.332   &...               &...                &18.350$\pm$0.248   &16.950$\pm$0.173 \\
57955.423   &...               &...                &18.253$\pm$0.205   &17.057$\pm$0.161 \\
57960.206   &...               &...                &18.568$\pm$0.285   &17.015$\pm$0.175 \\
\enddata
\end{deluxetable*}